\documentclass{emulateapj}
\usepackage{apjfonts}

\usepackage{graphics,graphicx,rotating}
\usepackage{natbib}
\usepackage{xspace}
\usepackage{amssymb}
\usepackage{amsmath}
\usepackage{subfigure}

\shorttitle{Constraining Intra-cluster Gas Models with \AMIBAW}
\shortauthors{Molnar, et al.}

\newcommand{\simless} 
     {\ensuremath{\lower 3pt\hbox{$\rlap{\raise5pt\hbox{$\char'074$}}\mathchar"7218$}}}
\newcommand{\simgreat}
     {\ensuremath{\lower 3pt\hbox{$\rlap{\raise5pt\hbox{$\char'076$}}\mathchar"7218$}}}

\newcommand{\simgt}{\lower.5ex\hbox{$\; \buildrel > \over \sim \;$}}
\newcommand{\simlt}{\lower.5ex\hbox{$\; \buildrel < \over \sim \;$}}

\newcommand{\nop}{\noindent}

\newcommand{\AMIBA}{{\sc AM}i{\sc BA}}
\newcommand{\AMIBAW}{{\sc AM}i{\sc BA13}}
\newcommand{\AMIBAS}{{\sc AM}i{\sc BA7}}
\newcommand{\ENZO}{{\sc ENZO}}
\newcommand{\WMAP}{{\sc WMAP}}

\newcommand{\SUZAKU}{{\sc Suzaku}}
\newcommand{\CHANDRA}{{\sc Chandra}}

\newcommand{\rmsub}[1]{\ensuremath{_{\rm #1}}}
\newcommand{\MSUN}{{\ensuremath{\mbox{\rm M}_{\odot}}}}
\newcommand{\RVIR}{\ensuremath{R\rmsub{vir}}}
\newcommand{\MVIR}{{\ensuremath{M\rmsub{vir}}}}
\newcommand{\RCORE}{{\ensuremath{r\rmsub{core}}}}
\newcommand{\RT}{{\ensuremath{r\rmsub{T}}}}
\newcommand{\RCTR}{{\ensuremath{r\rmsub{c}}}}
\newcommand{\TCTR}{{\ensuremath{a\rmsub{c}}}}
\newcommand{\TCMB}{{\ensuremath{T\rmsub{CMB}}}}

\newcommand{\ICMB}{{\ensuremath{I\rmsub{CMB}}}}
\newcommand{\VCL}{{\ensuremath{V\rmsub{CL}}}}
\newcommand{\VSOURCE}{{\ensuremath{V\rmsub{source}}}}
\newcommand{\VCMB}{{\ensuremath{V\rmsub{CMB}}}}
\newcommand{\VCMBP}{{\ensuremath{V^\prime\rmsub{CMB}}}}
\newcommand{\VSZ}{{\ensuremath{V\rmsub{SZ}}}}
\newcommand{\VBKND}{{\ensuremath{V\rmsub{bknd}}}}
\newcommand{\VNOISE}{{\ensuremath{V\rmsub{noise}}}}
\newcommand{\VNOISEP}{{\ensuremath{V^\prime\rmsub{noise}}}}
\newcommand{\KBOLTZ}{{\ensuremath{k\rmsub{B}}}}

\begin{document}

\title{Constraining Intra-cluster Gas Models with \AMIBAW}

\author{Sandor M. Molnar\altaffilmark{1}, Keiichi Umetsu\altaffilmark{1}, Mark Birkinshaw\altaffilmark{2}, 
       Greg Bryan\altaffilmark{3}, Zolt\'an Haiman\altaffilmark{3}, Nathan Hearn\altaffilmark{4}, \\
       Cien Shang\altaffilmark{3}, Paul T.P. Ho\altaffilmark{1,5}, Chih-Wei Locutus Huang\altaffilmark{6}, 
       Patrick M. Koch\altaffilmark{1}, Yu-Wei Victor Liao\altaffilmark{6}, \\
       Kai-Yang Lin\altaffilmark{1}, Guo-Chin Liu\altaffilmark{1,7}, 
       Hiroaki Nishioka\altaffilmark{1}, Fu-Cheng Wang\altaffilmark{6}, Jiun-Huei Proty Wu\altaffilmark{6}
       }
\email{sandor@asiaa.sinica.edu.tw}

\altaffiltext{1}{Institute of Astronomy and Astrophysics, Academia Sinica, 
                      P.O. Box 23-141, Taipei 106, Taiwan, ROC.;
                      sandor@asiaa.sinica.edu.tw}

\altaffiltext{2}{H. H. Wills Physics Laboratory, University of Bristol, Tyndall Ave, Bristol BS8 1TL, UK}

\altaffiltext{3}{Department of Astronomy, Columbia University, 550 West 120th Street, 
                      New York, NY 10027}

\altaffiltext{4}{ASC/Alliances Center for Astrophysical Thermonuclear Flashes, 
                      University of Chicago, Chicago IL 60637}

\altaffiltext{5}{Harvard-Smithsonian Center for Astrophysics, 60 Garden Street, Cambridge, MA 02138}

\altaffiltext{6}{Department of Physics, Institute of Astrophysics, and Center
                      for Theoretical Sciences, National Taiwan University, Taipei 10617, Taiwan, ROC.}

\altaffiltext{7}{Department of Physics, Tamkang University, 251-37, 
                      Tamsui, Taipei County, Taiwan, ROC.}

\begin{abstract}
Clusters of galaxies have been used extensively to determine cosmological 
parameters. A major difficulty in making best use of Sunyaev--Zel$'$dovich 
(SZ) and X-ray observations of clusters for cosmology is that using X--ray 
observations it is difficult to measure the temperature distribution and therefore 
determine the density distribution in individual clusters of galaxies out to the 
virial radius. Observations with the new generation of SZ instruments are a 
promising alternative approach. We use clusters of galaxies drawn from 
high-resolution adaptive mesh refinement (AMR) cosmological simulations to 
study how well we should be able to constrain the large-scale distribution of 
the intra-cluster gas (ICG) in individual massive relaxed clusters using \AMIBA\
in its configuration with 13 1.2-m diameter dishes (\AMIBAW) along with X-ray 
observations. We show that non-isothermal $\beta$ models provide a good 
description of the ICG in our simulated relaxed clusters. We use simulated 
X-ray observations to estimate the quality of constraints on the distribution of 
gas density, and simulated SZ visibilities (\AMIBAW\ observations) for constraints 
on the large-scale temperature distribution of the ICG. We find that \AMIBAW\ 
visibilities should constrain the scale radius of the temperature distribution to 
about 50\% accuracy. We conclude that the upgraded \AMIBA, \AMIBAW, 
should be a powerful instrument to constrain the large-scale distribution of the 
ICG.
\end{abstract}


\section{Introduction}
\label{S:Intro}

According to our standard structure formation scenarios based on the dark matter (DM) models, 
clusters of galaxies, the most massive virialized objects in the Universe, form from the largest 
positive density fluctuations, thus their formation and evolution are sensitive to the underlying 
cosmological model.
Taking advantage of this feature, clusters have been used extensively to determine
cosmological parameters (e.g., Henry 2000; Allen et al. 2004; Ettori 2004;
Vikhlinin et al. 2008; for recent reviews see Voit 2005 and Borgani 2006).
Prospects of determining cosmological parameters using much larger samples of clusters of galaxies
from next generation surveys were discussed in detail by e.g.,
Haiman, Mohr \& Holder (2001); Holder, Haiman \& Mohr (2001); \cite{MolnBirkMush02} and \cite{Molnet04}.

While theory predicts the mass function of clusters of galaxies and the distribution of mass, 
gas density and temperature within individual clusters, observations directly measure 
the X-ray luminosity and intra-cluster gas temperature functions, luminosity functions based
on the Sunyaev-Zel$'$dovich (SZ) effect, and the projected distribution of X-ray emissivity and 
electron pressure.
To connect theory and observation it is crucial to understand the physics of clusters 
out to their virial radii and beyond.
The observed large-scale distribution of the intra-cluster gas (ICG) and its evolution can be
directly compared to predictions of cosmological structure formation models
and so constrain them.
Also, when using the X-ray/SZ method to derive distances to clusters directly,
and thus determining cosmological parameters, the large scale distribution
of the ICG has to be known well since incorrect ICG models lead to bias
in the determination of the distance and thus in cosmological parameters
(e.g., \citealt{Kawaet08}; for a summary of systematic errors see \citealt{MolnBirkMush02}).

In this paper we focus on what qualitatively new constraints on the 
large scale distribution of the ICG we can expect from analyzing data to be 
taken with the Yuan-Tseh Lee Array for Microwave Background Anisotropy 
(\AMIBA; Ho et al. 2009; Wu et al. 2009) interferometer operating at 94 GHz 
with the planned upgrade to 13 antennas (\AMIBAW; Koch et al., in preparation).
Thus we carry out a feasibility study to estimate how well we should be able to 
constrain the large-scale distribution of the ICG using \AMIBAW.
We first summarize the presently available observational
constraints on the large-scale distribution of the ICG (\S\ref{S:LargeScale}).
In \S\ref{S:AMR} we derive a family of models for the ICG from our high-resolution
cosmological simulations. We then present our methods of generating SZ and X-ray images of
simulated clusters of galaxies in \S\ref{S:IMAGES}.
Our method to simulate visibilities for mock \AMIBAW\ observations is described in
\S\ref{S:AMIBAVIS}.
Model fitting and the results are presented in \S\ref{S:AMIBA}.
Finally, in \S\ref{S:Discussion}, we discuss our results for the constraints on the 
shape parameters of our ICG models from mock \AMIBAW\ observations.
We quote all errors at 68\% confidence levels (CLs).

Our companion papers describe the details of the design, performance, 
and the science results from the first observational season of \AMIBA\
with the first configuration (\AMIBAS).
\cite{Ho__et09} describe the design concepts and specifications of the
\AMIBA\ telescope. Technical aspects of the instruments are
described by \cite{Chenet09} and \cite{Kochet09}. 
Details of the first SZ observations and data analysis are presented
by \cite{Wu__et08}. \cite{Nishet08} assess the integrity of \AMIBAS\ data 
performing several statistical tests. 
\cite{Linet09} discuss the system performance and verification.
Contamination from foreground sources and the primary 
cosmic microwave background (CMB) fluctuations is estimated by Liu et al. (2009).
\cite{Kochet10} present a measurement of the Hubble constant using \AMIBAS\ and 
archival X-ray data. 
\cite{Umetet09} determine gas mass fractions using gravitational lensing and 
\AMIBAS\ observations of galaxy clusters.
\cite{Huanet10} discuss cluster scaling relations between \AMIBAS\ and X-ray data.

%
%
\begin{figure*}
\centerline{
\includegraphics[width=17cm]{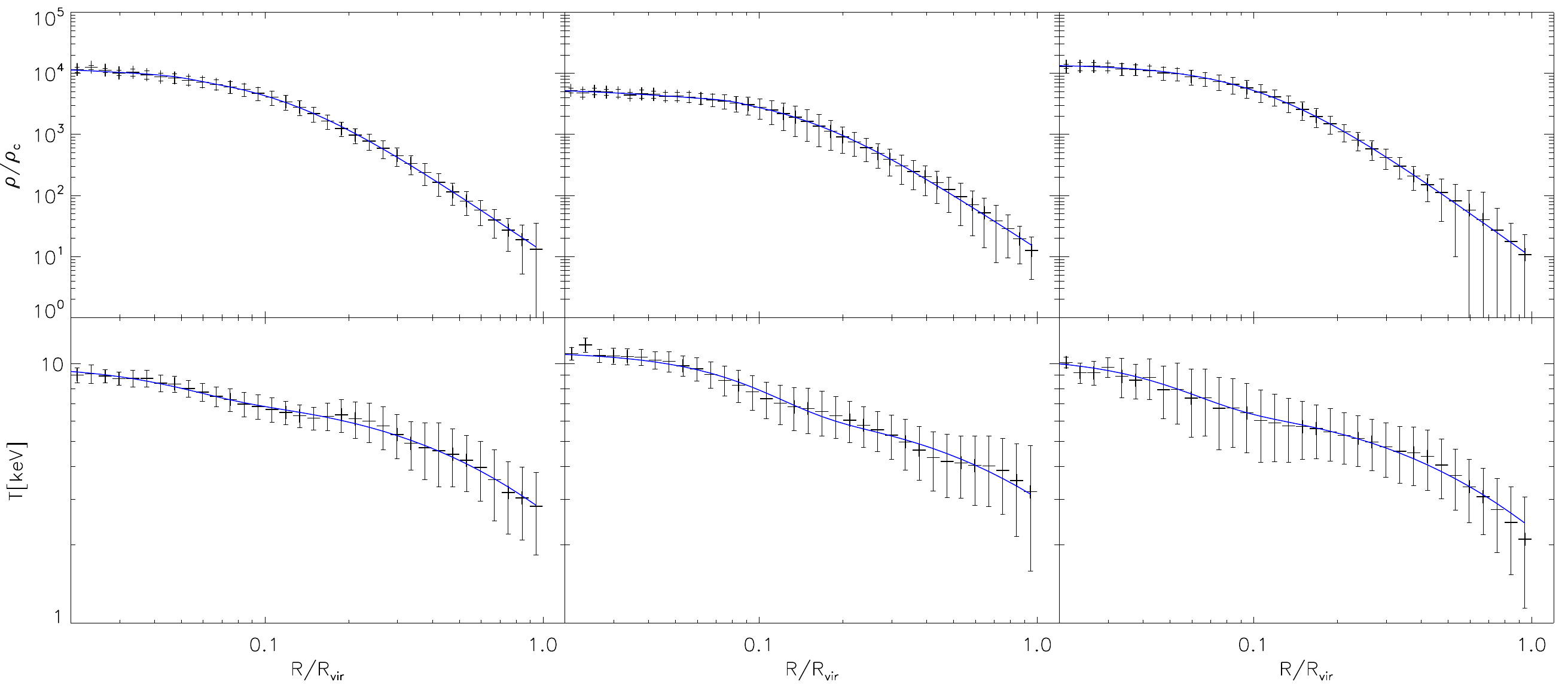}
}
\caption{
         Spherically averaged gas density, $\rho$, (in units of the critical density,
         $\rho_c$) and temperature, $T$ (in keV) distributions of the relaxed massive 
         clusters CL1 (left panels) and CL2 (middle panels) and non-relaxed cluster CL3 (right panels)
         as a function of radius in units of the virial radius, \RVIR\ (dashed lines).
         The error bars represent {\it rms} variations due to spherical averaging.
         The solid lines show the best-fit density and temperature models to the respective 3D distributions. 
}
\label{F:f1}
\end{figure*}

\section{Constraints on the large-scale Distribution of the Intra-Cluster Gas}
\label{S:LargeScale}

Thermal bremsstrahlung is generated by the scattering of two particles 
(an electron and an ion) in the ICG, thus the X-ray emission in massive (hot)
clusters, where this is the dominant emission process, 
is proportional to the square of the electron density,
since the ion density is proportional to the electron density (e.g., Sarazin 1988).
   The SZ effect, the inverse-Compton scattering of cold photons of the
   CMB by electrons in the hot ICG, is proportional to the electron density 
(Sunyaev \& Zel'dovich 1980; for recent reviews see 
Rephaeli 1995; Birkinshaw 1999; and Carlstrom, Holder \& Reese 2002).
As a consequence, X-ray observations are more sensitive to the inner parts of 
clusters, while SZ observations are relatively more sensitive to the outer regions.
The X-ray flux is dominated by signal from regions with
$0.1 \simlt r / \RVIR \simlt 0.4$, where the virial radius, \RVIR, 
is defined here according to the usage in Bryan \& Norman (1998), 
and the total SZ decrement is dominated by signal from regions near \RVIR\ 
(see Figure 10 of Fang \& Haiman 2008).

We measure the SZ signal, which is proportional to the line-of-sight (LOS)
integral of the electron pressure, and the X-ray surface brightness,
which is proportional to the LOS integral of the X-ray
emissivity.
  The projections cause the observables to depend on the
  LOS size of the cluster. This size can be estimated based
  on the angular size from the X-ray or SZ image and the
  angular diameter distance to the cluster, $D_A$. 
  The X-ray and SZ brightnesses of the cluster then provide
  two equations from which we can estimate two unknowns, $D_A$ and
  some characteristic electron density in the cluster.
Observing a sample of clusters, we can derive $D_A$ as a function of the 
redshift, $z$, and thus constrain cosmological parameters. This 
is usually called the SZ--X--ray (SZX) method (e.g., Birkinshaw 1999; see
Koch et al. 2010 for an application of this method using \AMIBAS\ observations).

In practical implementations of the SZX method, we determine the spatial model
for the cluster from the higher signal-to-noise (SN) X-ray observations, which typically go 
out to about half of the virial radius.
The caveat of this method is that there is no guarantee that the ICG 
 distribution at large radii follows an extrapolation of the distribution 
 determined from X-ray data.
Also, due to projection effects, measurement errors, etc.,
the distribution of the ICG determined from X-ray measurements might be biased.
Models used to describe the X-ray observations of clusters are 
typically $\beta$ models for the density distribution, 
$\rho_g \propto (1 + r^2/ \RCORE^2)^{-3\beta/2}$, where the
spatial extent is determined by the core radius, \RCORE, and 
the fall off by the exponent, $\beta$ (Cavaliere \& Fusco-Femiano 1976);
with either constant temperature (isothermal $\beta$ models), 
or a gradually declining temperature as a function of the distance from the 
cluster center.
The resulting $\beta$ parameters are in the range of 0.5 - 0.8 typically.
Many relaxed clusters have $\beta \approx 2/3$, which provides a 
 shallow density profile, $\rho \propto r^{-2}$, at large radii
(e.g., Sarazin 1988; for recent results see Maughan et al. 2008 and references therein).

However, numerical simulations as well as SZ and X-ray observations suggest a 
much steeper fall off of the density at large radii.
\cite{Roncet06} used a sample of 9 clusters of galaxies in the mass range of 
$1.5 \times 10^{14}~\MSUN$ -- $3.4 \times 10^{15}~\MSUN$ 
from smoothed particle hydrodynamic simulations (SPH) to derive
gas density and temperature profiles in the outskirts of clusters. 
They used simulations with and without cooling, supernova feedback and
thermal heat conduction and found that the profiles steepen as a function of radius.
They also found that cooling and supernova feedback do not affect the 
density and temperature profiles at large radii significantly.
Their results support the theoretical expectation that the distribution of gas at large 
radii in clusters of galaxies is determined mainly by gravity.
\cite{Hallet07} fitted isothermal $\beta$ models to mock X-ray and SZ observations of 
simulated clusters drawn from adaptive mesh refinement (AMR) cosmological simulations.
They found that isothermal $\beta$ model fits to X-ray surface brightness distributions 
of simulated clusters are biased to low $\beta$ values relative to fits to SZ distributions, 
and that the fitted $\beta$ values depend on the projected outer cut off radii used. 
   When \cite{Hallet07} used a projected radius limit equal to \RVIR, 
   the $\beta$ parameters based on SZ structures scattered around $\beta = 1$.
Haugbolle, Sommer-Larsen \& Pedersen (2007) derived an empirical model
for the pressure distribution in clusters of galaxies based on high-resolution 
SPH simulations and observations.
They also found a steeper fall off of the pressure at large radii than that predicted by 
X-ray observations.

\cite{Afshet07} used WMAP 3 year data to stack images of 193 massive clusters of 
galaxies and detected the SZ effect statistically out to about $2~\RVIR$. 
Using a larger cluster sample, \cite{Atriet08} determined the average electron pressure profile 
in clusters by stacking 700 known clusters extracted from the 3-year WMAP data.
They showed that the isothermal $\beta$ model does not provide a good fit on large scales.
Both \cite{Afshet07} and \cite{Atriet08} concluded that an ICG model with 
a density profile with a fall off of $\rho_g \propto r^{-3}$ at large radii
and a temperature profile derived from hydrostatic equilibrium is a good description of their data.

The large-scale distribution of the ICG was studied in three individual clusters of
galaxies (Abell 1835, Abell 1914 and CL J1226.9+3332) by \cite{Mrocet09} 
using SZA observations at 30 GHz (and at 90 GHz for CL J1226.9+3332).
The SZA is an interferometer consisting of eight 3.5 meter diameter dishes
\citep{Muchet07}.
They used a parameterized pressure profile with five parameters based on 
cosmological numerical simulations of Nagai et al (2007).
 \cite{Mrocet09} fixed the three slope parameters at their values derived from simulations 
and X--ray observations, and fitted only for the amplitude and the pressure scale radius 
using their SZ data.
They used a density distribution derived from X--ray observations to determine
the temperature distribution based on the ideal gas law (temperature $\propto$ pressure/density).
\cite{Mrocet09} found that the SZ profiles drop more steeply than predicted by 
isothermal $\beta$ models, and, similarly to previous studies, that the profiles
drop more steeply than predicted by a $\beta = 2/3$ model even if the change in the 
temperature is taken into account (Figure 3 of \citealt{Mrocet09}).

\cite{Vikhet05} measured the temperature profile in 13 low redshift
relaxed clusters using \CHANDRA\ data. In three clusters the temperature
profiles were measured out to about 0.7\RVIR. 
At $r \simgt$ 0.1--0.2 \RVIR\
they found that the fall off of the temperature with radius is self similar
in relaxed clusters when scaled by \RVIR. 
Recently \cite{EttBal09} and \cite{Bautet09} studied the outer regions of galaxy 
clusters using X-ray observations.
\cite{EttBal09} used \CHANDRA\ observations of 11 clusters with SN ratio greater than 
2 out to  $r > 0.7$\RVIR.
The low count rate in the outer regions of galaxy clusters did not allow them to determine
the temperature distribution out to \RVIR. They derived the slope of the gas density
and temperature distribution at the virial radius assuming hydrostatic equilibrium.
Ettori \& Balestra found that the X-ray surface brightness distribution is steepening with larger radii,
implying an equivalent $\beta \approx 1$ (within errors) at \RVIR.
\cite{Bautet09} used \SUZAKU\ observations of relaxed cluster Abell 1795.
They mapped the X-ray surface brightness and temperature distribution out to about 0.9 \RVIR\
in two directions and found $\beta = 0.64$ within $r < 1$ Mpc.
At larger radii they found a steeper fall off of the density in the South ($\beta > 0.64$), but a rising
density profile towards the North with a maximum at 1.9 Mpc (1\RVIR).
The increase of the X-ray surface brightness in the North direction might be due to a contribution 
from a filament in the LOS.

\bigskip
\bigskip
\section{Models for the ICG from AMR Simulations}
\label{S:AMR}

We derive self-similar spherically symmetric models for the distribution of the 
ICG in relaxed clusters of galaxies using a sample of clusters drawn from cosmological 
AMR simulations performed with the cosmological code \ENZO\ (O'Shea et al. 2004) 
assuming a spatially flat cold DM model with cosmological parameters
($\Omega_m$, $\Omega_\Lambda$, $\Omega_b$, h, $\sigma_8$) = (0.3, 0.7, 0.047, 0.7, 0.92),
where $\Omega_m$, $\Omega_b$, $\Omega_\Lambda$ encode 
the current matter and baryon densities and the cosmological constant, 
$\sigma_8$ is the power spectrum normalization on 8 h$^{-1}$ Mpc scales, 
and the Hubble constant $H_0 = 100$~h km s$^{-1}$ Mpc$^{-1}$.
This cosmological model is close to the model implied by the \WMAP\ 5 year results except for 
$\sigma_8$, which is much larger \citep{Dunk08}.
The AMR simulations were adiabatic (in the sense that no heating, cooling, or feedback
were included). 
The box size of the original, low resolution, cosmological simulation was 300 h$^{-1}$ Mpc.
The clusters of galaxies in our sample were re-simulated with high resolution 
using the same technique as described in \cite{YounBria07}.

%
%
\begin{figure}
\centerline{
\includegraphics[width=.45\textwidth]{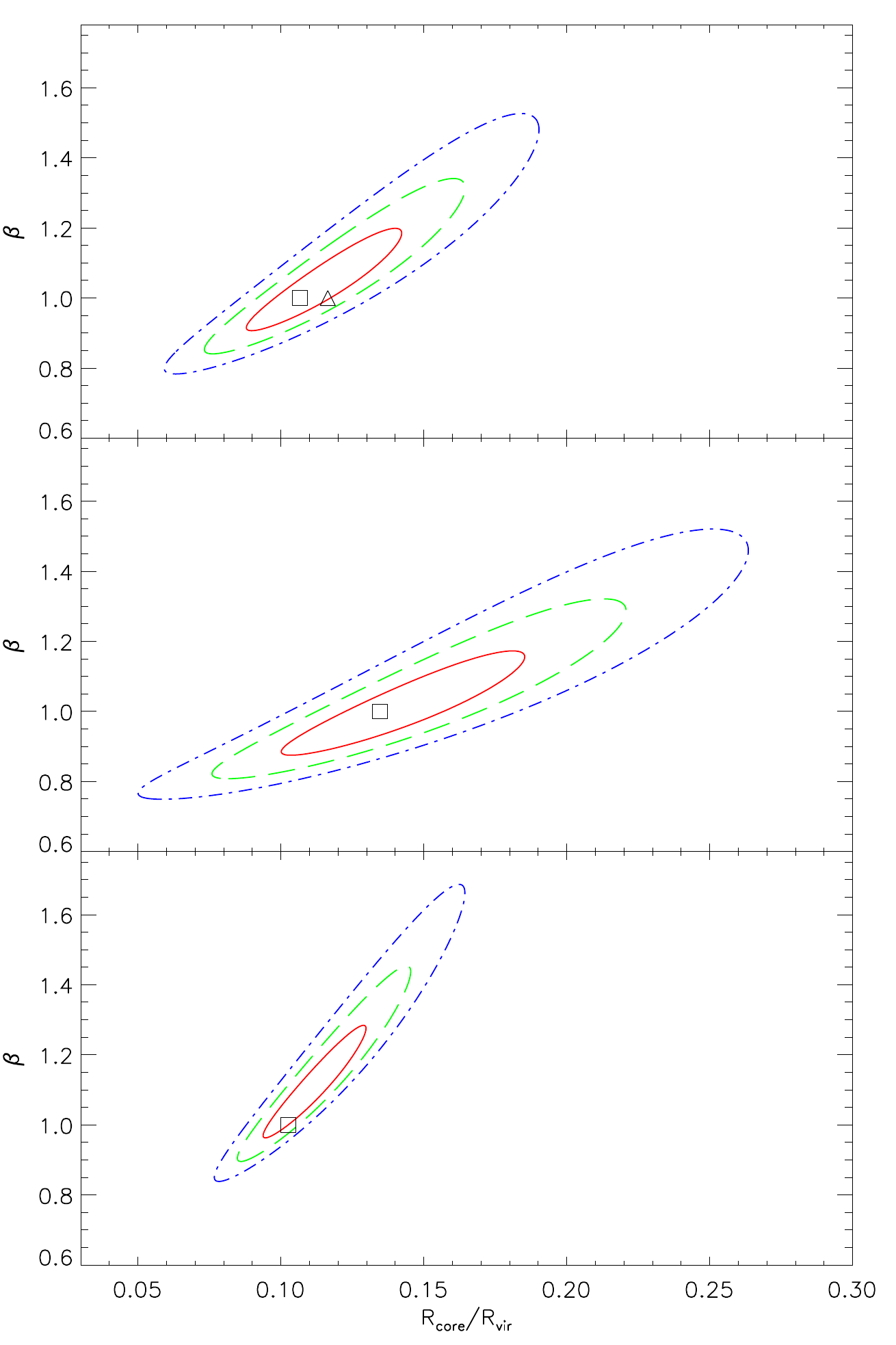}
}
\caption{Likelihood contours (68\%, 95.4\% and 99.7\% CLs, solid red, dashed green, and dash-dotted blue lines)
              as a function of the outer $\beta$ model parameters for fitting double $\beta$ models to the 3-dimensional 
              density distribution of CL1, 2, and CL3 (top to bottom).
              The square and triangle in the top figure and the squares in the middle and bottom panel
              represent best fit values from fitting non-isothermal double $\beta$ models to 
              simulated X-ray images in Projection X and Y of CL1, Projection Y of CL2 and Projection X of CL3.
\hspace*{0.1in}
}
\label{F:f2}
\end{figure}

The resolutions (minimum AMR cell size) of the high-resolution simulations 
at $R$ = 0, 1 and 4\RVIR\ were about 25 kpc, 80 kpc and 250 kpc.
The total virial masses of the 10 massive clusters in our cluster sample fell between
1 and $2 \times 10^{15}~\MSUN$.
Relaxed clusters were selected based on their density distribution: 
after the removal of filaments, we chose clusters with a smooth 
spherically averaged density profile with little angular variation, no sign
of recent major merger events, and a relaxed core 
(for more details about our simulated cluster sample and analysis see Molnar et al. 2009).
Out of a total of 10 clusters, two clusters satisfy our criteria for relaxed clusters
(CL1 and CL2).
We show spherically averaged gas density and temperature profiles
for our massive relaxed clusters and one massive cluster with a non-relaxed core,
CL3, in Figure~\ref{F:f1}.
We include CL3 to check if we could constrain the large scale distribution of the ICG 
in a cluster which has a non-relaxed core, but is otherwise relaxed.
The error bars represent the ${\it rms}$ of the density and temperature variations
due to spherical averaging.
  While the density distributions are similar in all relaxed
  clusters, showing only small deviations from radial averaging, 
  the temperature profiles show more variation.
This is due to the sensitivity of the temperature to shocks from merging and internal flows.
The solid curves in Figure~\ref{F:f1} show the density and
temperature profiles of the best-fit models for all simulated clusters (see below).
 The physical parameters of our selected clusters are summarized in Table~1.

We use spherically symmetric double $\beta$ models truncated at the virial radius, 
\RVIR, to describe the density distribution of the ICG in massive relaxed clusters.
We use the same functional form for the temperature distribution at large radii as \cite{Lokeet02},
and a gaussian to describe the core region.
The gas density and temperature models for massive clusters within the 
virial radius can be summarized as:

\begin{eqnarray}  \label{E:DENS}
     \rho(r) & = & \rho_1 \theta_1 (1 + r^{\,2} / r^{\,2}_1)^{-3\beta_1/2}  + 
                         \rho_2 \, \theta_2  (1 + r^{\,2} / r^{\,2}_{\rm core})^{-3\beta/2}
\nonumber\\                  & = & \rho_1 \, {\cal F}_1  + \rho_2 \, {\cal F}_2
\end{eqnarray}

\nop
and

\begin{equation}  \label{E:TEMP}
     T(r) = T_0 \bigl( a_c  \exp[-r^2/r_c^2] + 1 \bigr)   \bigl( 1 + r / r_{\rm T} \bigr)^{-\delta}
,
\end{equation} 

\smallskip
\nop
where the large scale distribution is described by \RCORE, $\beta$, \RT\ and $\delta$, 
and a possible extra component at the center is parameterized by $r_1$, $\beta_1$, \TCTR\ 
and \RCTR, $\rho_1$ and $\rho_2$ are the central densities, 
and the transition between the two $\beta$ models at $a_1$ is facilitated by 
$\theta_1 (r,a_1) = \theta(a_1 - r)$ and $\theta_2 (r, a_1) = \theta(r - a_1)$, where $\theta$ 
is the Heaviside step function.
We determine the best-fit parameters for each cluster by maximizing the likelihood 
functions

\begin{equation}  \label{E:LIKE3D}
       -2\, \ln {\cal L}_{F} =  \sum_{i}  { [ (F_O)_i -  (F_M)_i \}]^2  \over \sigma_{Fi}^2 } = \chi_F^2
,
\end{equation}

\nop
where $(F_O)_i$ is the median value of the 3-dimensional (3D) density or 
temperature ($F = \rho$ or $T$) of a simulated cluster in the $i$-th radial bin, 
$(F_M)_i$ is the corresponding value predicted by the model considered, 
and $\sigma_{Fi}$ is the corresponding standard deviation
(here we use an approximation and assume that the sum of the fluctuations due to substructure, 
asphericity, etc. are gaussian, which is a reasonable assumption since we excluded filaments with
large positive density fluctuations).
These functions provide good fits to the density and temperature profiles of our selected
clusters out to the virial radius:
the fits are well within the 1$\sigma$ error bars due to spherical averaging, 
except for one point where the deviation is 1$\sigma$ for a temperature profile
(see solid lines in Figure~\ref{F:f1}).
The best fit parameters are summarized in Table~1. 
In Figure~\ref{F:f2} we show the likelihood contours as a function of \RCORE\ and $\beta$
(the shape parameters of the second $\beta$ model, which describe the large scale
distribution of the ICG) to the 3D distribution of the density in the outer parts 
of our selected clusters (the other parameters marginalized). 
The contour levels were determined based on the standard $\chi^2$ statistic. 
The likelihood contours show how well constrained the parameters are, 
subject to our assumption of spherical symmetry, and allow us to estimate 
the level of degeneracies between parameters.
In carrying out fits to the 3D temperature distribution of simulated relaxed clusters
we noticed that the exponent of the temperature model, 
$\delta$, does not change much from cluster to cluster 
(in agreement with an analysis using more clusters by \citealt{Lokeet02}). 
Also, the best-fit models are not significantly better than models with fixed $\delta = 1.6$.
Thus we fix $\delta = 1.6$ in our fitting and in the rest of our analysis.
Since the best fit $\beta$ values are close to 1 in all selected clusters, 
we conclude that ICG models with $\beta = 1$ and $\delta = 1.6$ provide 
good fits to these clusters including CL3, which has a non-relaxed core.
This result verifies our assumption that the outer region of CL3 is relaxed.
Therefore, we find that the density distribution at large radii can be approximated
with a power law, $r^{-\alpha}$, with $\alpha = 3$, which is close to $\alpha = 3.4$ 
as found by \cite{Roncet06}.
Our results suggest that the pressure ($\propto \rho \, T$), at large radii can be
approximated with a power law with $\alpha = 3\beta + \delta = 4.6$, which is close to 
$\alpha = 5$, found by \cite{Nagaet07} and used by \cite{Mrocet09}.
 We conclude that the density and temperature functions (Equations~\ref{E:DENS} and \ref{E:TEMP})
 are adequate for relaxed clusters and provide a family of ICG models that can
 be fitted to observational data.

%
%
\begin{deluxetable*}{cccccccccccc}
\tablecolumns{12}
\tablecaption{
\label{tab:Table1}
 Fitted shape parameters to clusters of galaxies from AMR simulations
} 
\tablewidth{0pt} 
\tablehead{ 
 \multicolumn{1}{c}{ID\tablenotemark{a}}          &
 \multicolumn{1}{c}{\MVIR\tablenotemark{b}}    &
 \multicolumn{1}{c}{\RVIR\tablenotemark{c}}     &
 \multicolumn{1}{c}{$a_1$\tablenotemark{d}}    &
 \multicolumn{1}{c}{$r_1$\tablenotemark{d}}     &
 \multicolumn{1}{c}{$\beta_1$}                           &
 \multicolumn{1}{c}{\RCORE\tablenotemark{d}} &
 \multicolumn{1}{c}{$\beta$}                               &
 \multicolumn{1}{c}{\TCTR}                                 &
 \multicolumn{1}{c}{\RCTR\tablenotemark{d}}    &
 \multicolumn{1}{c}{\RT\tablenotemark{d}}         &
 \multicolumn{1}{c}{$\delta$\tablenotemark{e}}
}
\startdata  
   CL1  &   9.1E+14  &  2.0  & 0.05  & 0.013  &   0.10  &     0.113   &   1.033     &  0.26   &  0.052    &  1.07  &       1.6  \\
   CL2  &   1.2E+15  &  2.2  & 0.08  & 0.025  &  0.13   &     0.139   &   1.001     &  0.57   &  0.011    &  1.40  &       1.6  \\
   CL3\tablenotemark{f} &  1.1E+15  &  2.1  &   -- &  --  &  --  &  0.110   &   1.096    &  0.46   &  0.056     &  1.00  &  1.6
\enddata
\tablenotetext{a}{Galaxy Cluster ID}
\tablenotetext{b}{Virial mass in Solar Mass}
\tablenotetext{c}{Virial radius in Mpc}
\tablenotetext{d}{in units of \RVIR}
\tablenotetext{e}{fixed}
\tablenotetext{f}{single $\beta$ model}
\end{deluxetable*}

X-ray observations of clusters of galaxies show that most relaxed clusters have 
cool cores (e.g., Vikhlinin et al. 2006), suggesting that cool-core clusters are relaxed.
Cosmological numerical simulations suggest that early major 
mergers destroy the developing cool cores, but cool cores are strong 
enough to survive late major mergers, and thus cool cores are associated 
with cluster formation history and not necessarily with the dynamical state
of clusters (Burns et al. 2008). 
This conclusion seems to be supported by the observational result that
a substantial number of relaxed clusters do not posses a cool core at their center 
(e.g., HIFLUGS sample~\footnote{http://www.astro.virginia.edu/~cls7i/papers/HIFLUGCS\_CC.pdf.}).
In our sample of clusters we do not have cool core clusters, 
but our method would work with either cool-core or non-cool-core clusters
since we model the core of the cluster separately. In the case of cool-core
clusters the amplitudes of our central temperature model, $a_c$, would be negative.
Since outside of core gravitational physics dominates 
(Roncarelli et al. 2006), and the core region is modeled separately,
we conclude that our adiabatic cosmological 
simulations and cluster models are adequate for our purpose.

\bigskip
\smallskip
\section{SZ and X-ray Images of Simulated Clusters}
\label{S:IMAGES}

We derive the 2-dimensional SZ and X-ray surface brightness distributions 
for simulated clusters in projections along the X, Y and Z axes 
(projections to the YZ, XZ and XY planes, Projection X, Y and Z, hereafter).
We ignore relativistic effects, which is a good approximation in our case 
since the intra-cluster gas temperatures in our relaxed AMR clusters is less than about 10 keV.
We derive the SZ signal in Projection Z by integrating along the LOS 
($\ell$ which coincides with $z$ in this case) 
over the extent of the cluster (from $\ell_1$ to $\ell_2$) using

\begin{equation}  \label{E:CLSZ}
           \Delta T_{CL}(x,y) =  \Delta T_{CL0} \, N_{CLSZ}^{-1}\,
                                                     \int_{\ell_1}^{\ell_2} n_e(x,y,\ell) \, T_e(x,y,\ell) \, d\ell
,
\end{equation}

\bigskip

\nop
where $x$ and $y$ are spatial coordinates in the plane of the sky,
i.e. perpendicular to $\ell$; 
$n_e = f_g \rho_g / \mu_e m_P$ is the electron density, where $\mu_e$ 
is the mean molecular weight per electron and $m_P$ is the proton mass;
$\rho_g$ is the gas density; $f_g$ is the mass fraction of baryons 
in the cluster that are contained in the ICG (we adopt $f_g=0.9$), 
and we use the standard assumption that the electron temperature equals 
the gas temperature, $T_e = T$.
The frequency dependence is contained in 
$\Delta T_{CL0} =  p(x_{\nu}) \, \TCMB \, k_B \sigma_T / (m_e c^2)$, 
where the dimensionless frequency $x_{\,\nu} = h_P \nu / (\KBOLTZ \TCMB)$, 
where \TCMB\ is the monopole term of the CMB, 
$h_P$ and $k_B$ are the constants of Planck and Boltzmann, and the 
function $p(x_{\nu}) = x_{\nu} \coth (x_{\nu}/2) - 4$ 
(e.g., Birkinshaw 1999). The SZ normalization is 
\begin{equation}  \label{E:NCLSZ}
      N_{CLSZ} = \int_{\ell_1}^{\ell_2} n_e(0,0,\ell) \, T_e(0,0,\ell) \, d\ell
.
\end{equation}
In practice, we pixelize $x,y$ and $\ell$, and approximate the integral with a sum
over the LOS from $\ell_1 = -10$ Mpc to $\ell_2 = 10$ Mpc.
Similar expressions were used for Projections X and Y.

Liu et al. (2010) studied the contamination from CMB, galactic diffuse emission 
and point sources in six clusters of galaxies observed in the first year of AMIBA. 
They found that the contamination is dominated by CMB fluctuations.
The low level of contamination by point sources at around 90 GHz is due to the low angular 
resolution of AMIBA and the falling spectra of most radio sources.
Even though some sources have inverted spectra, theoretical
predictions based on VLA observations at lower frequencies suggest that only about 2\% 
of clusters are contaminated at a significant level by point sources at this frequency
(see Figure 13 of \citealt{Sehet10}).
Therefore we include CMB contamination in our SZ image simulations but not 
point sources, since we can select relaxed clusters with no significant point source 
contamination for structural studies. Thus we have 

\begin{equation}  \label{E:SZ}
           \Delta T(x,y) =  \Delta T_{CL}(x,y) + \Delta T_{CMB} (x,y)
,
\end{equation}

\nop
where the CMB contamination, $\Delta T_{CMB}$, is generated as in 
Umetsu et al. (2004).

We derive the X-ray surface brightness in Projection Z by integrating along the 
LOS ($\ell$) over the extent of the cluster (from $\ell_1$ to $\ell_2$) as

\begin{equation}  \label{E:SXCL}
  S_{CL}(x,y) = S_{CL0}\, N_{CLX}\, \int_{\ell_1}^{\ell_2} n_e^2(x,y,\ell) \, T_e^{1/2}(x,y,\ell) \, d\ell
,
\end{equation}

\nop
where $S_{CL0}$ is the central X-ray surface brightness and the normalization, $N_{X}$, is 

\begin{equation}  \label{E:NX0}
  N^{\,-1}_{CLX} = \int_{\ell_1}^{\ell_2} n_e^2(0,0,\ell) \, T_e^{1/2}(0,0,\ell) \, d\ell
,
\end{equation}

\nop
where we use the scaling $S_X \propto \sqrt{T_e}$ for thermal bremsstrahlung 
(similar expressions were used for Projections X and Y).  
Similarly to the SZ surface brightness, we pixelize $x,y$ and $\ell$, and approximate 
the integral with a sum over the LOS from $\ell_1 = -10$ Mpc to $\ell_2 = 10$ Mpc.

We simulate X-ray images of our relaxed AMR clusters assuming a field of view (FOV) of 
$16\arcmin \times 16\arcmin$ with a pixel size of  $2\arcsec \times 2\arcsec$.
We sample the Poisson distribution with expectation
values equal to the integrated flux per pixel as
\begin{equation}  \label{E:XRAY}
  F_X(x,y) = F_{CL}(x,y) + B_X
,
\end{equation}

\nop
where we calculate $F_{CL}(x,y)$ using Equation~\ref{E:SXCL} assuming 
$F_{CL0} = 7.5$ cnts/pixel ($S_{CL0}$ in units of integrated flux/pixel)
and a uniform background of $B_X = 0.2$ cnts/pixel
(typical parameters for X-ray observations of clusters of galaxies).

\begin{figure}
     \centering
     \subfigure{n
          \includegraphics[width=.15\textwidth]{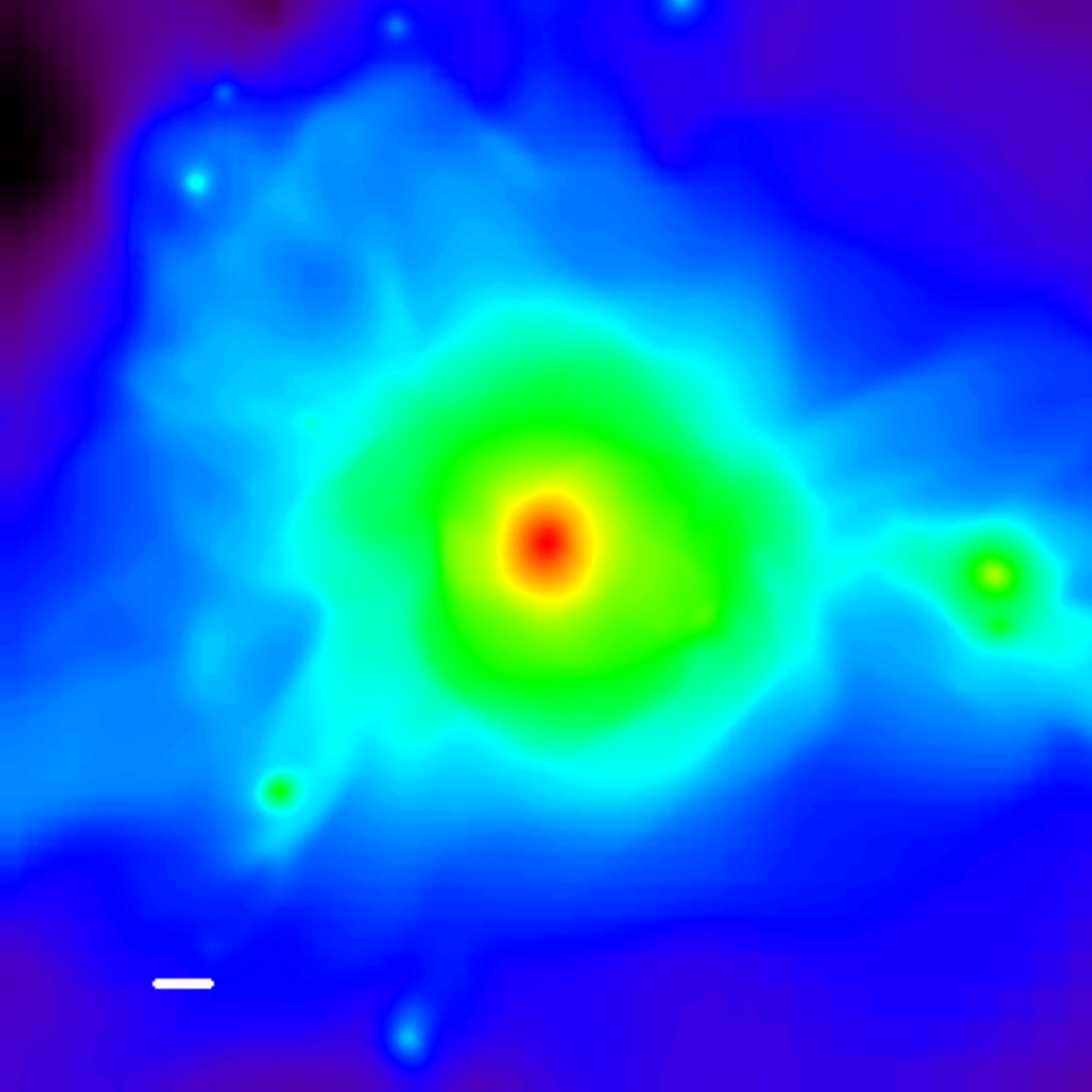}}
     \subfigure{
          \includegraphics[width=.15\textwidth]{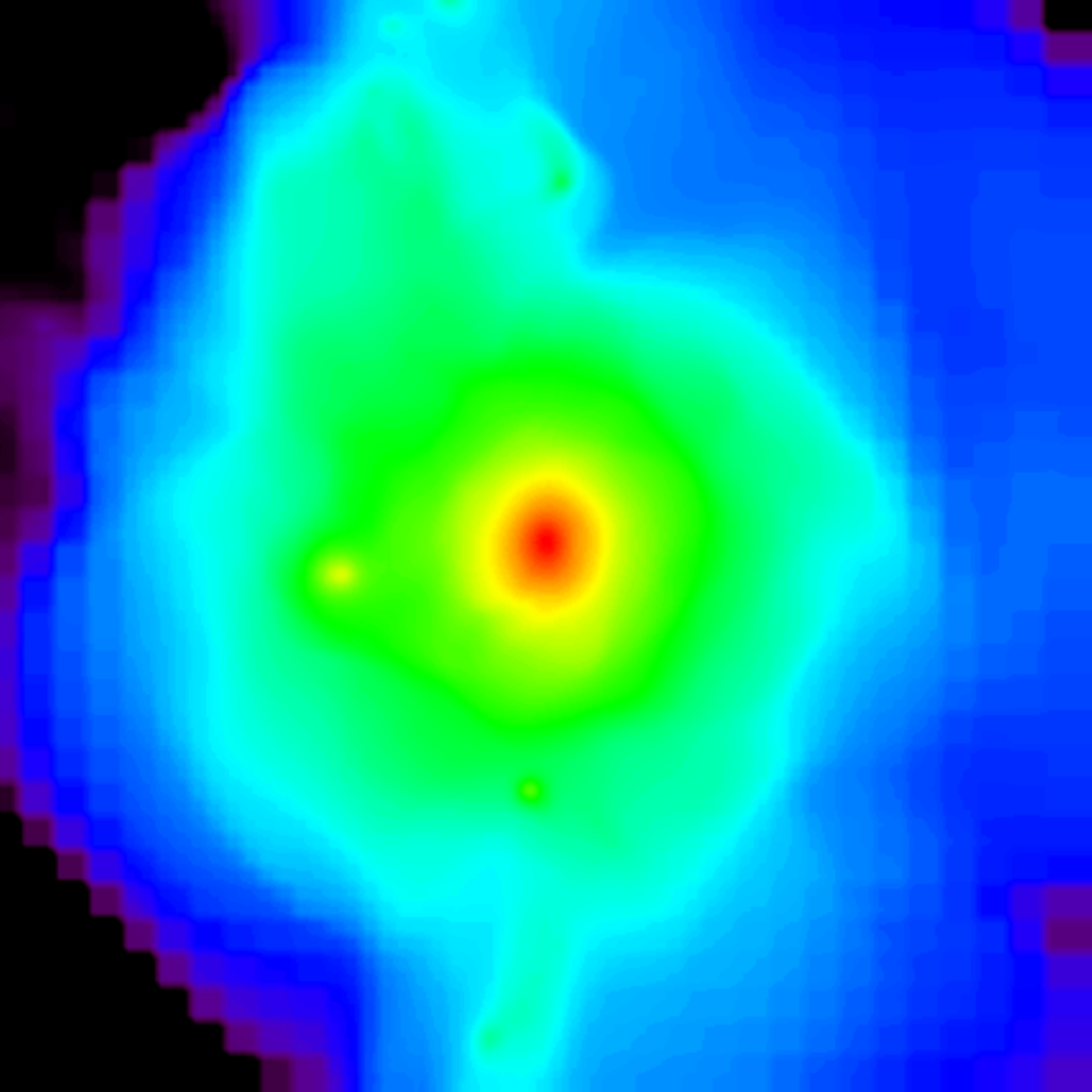}}
     \subfigure{
          \includegraphics[width=.15\textwidth]{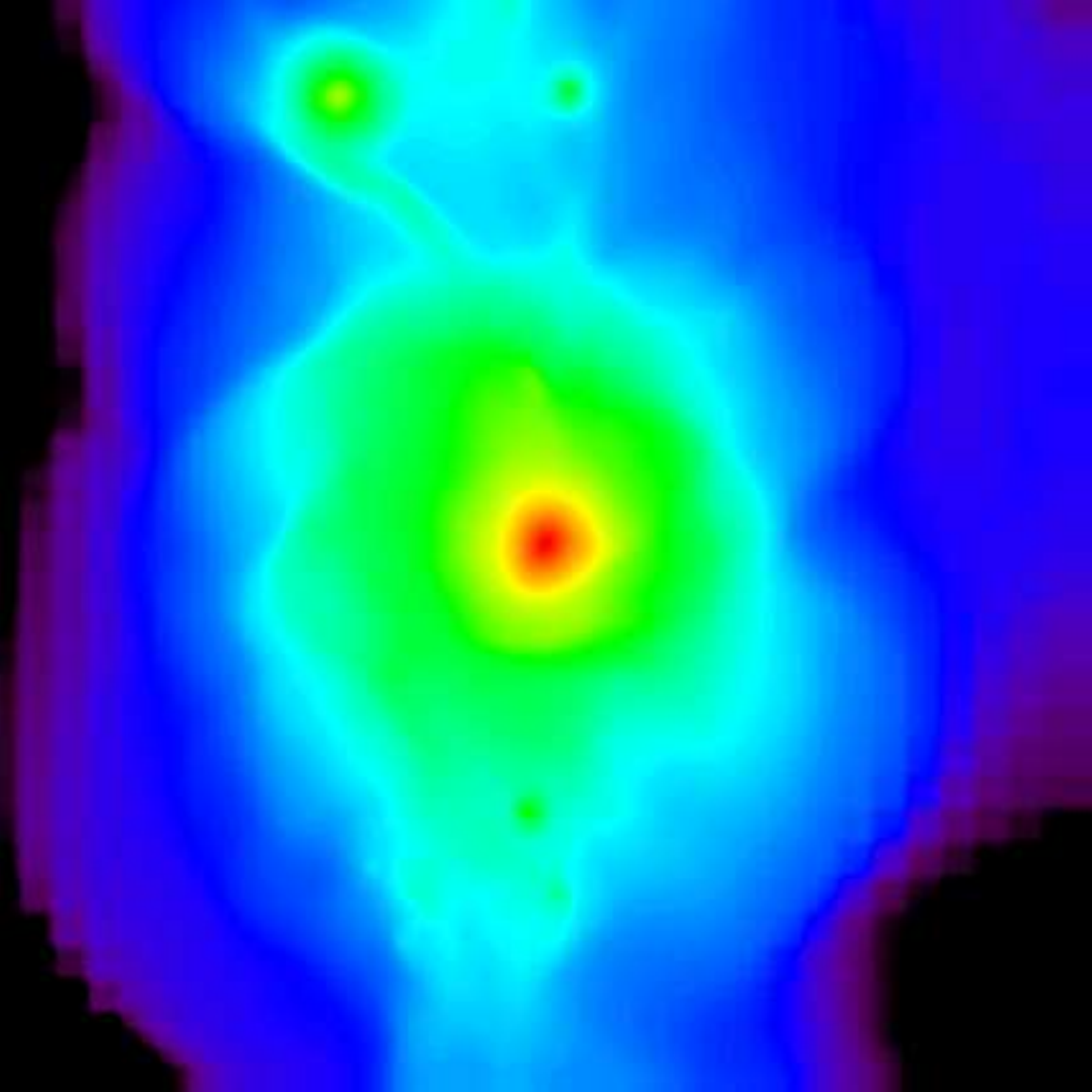}}
     \subfigure{
          \includegraphics[width=.15\textwidth]{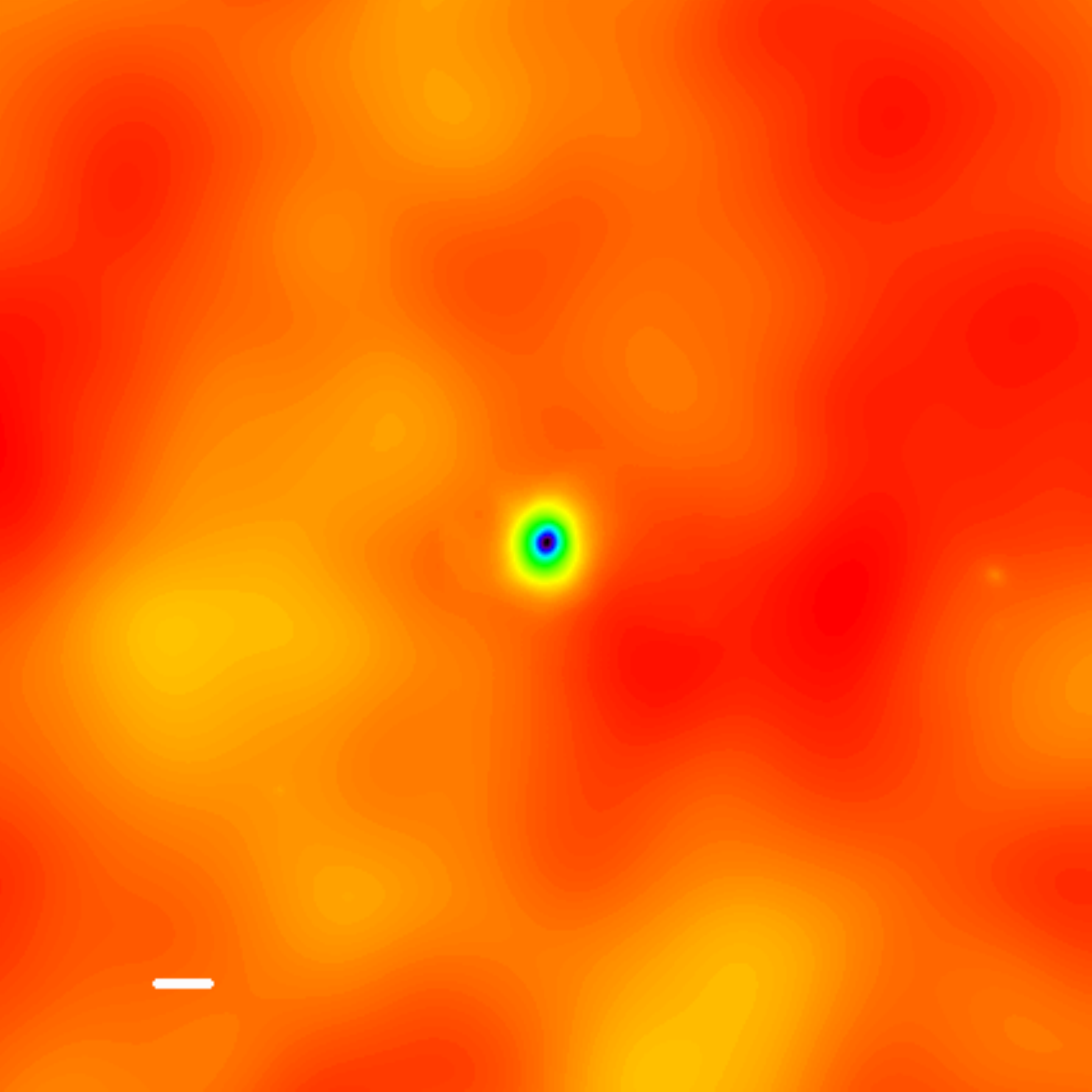}}
     \subfigure{
          \includegraphics[width=.15\textwidth]{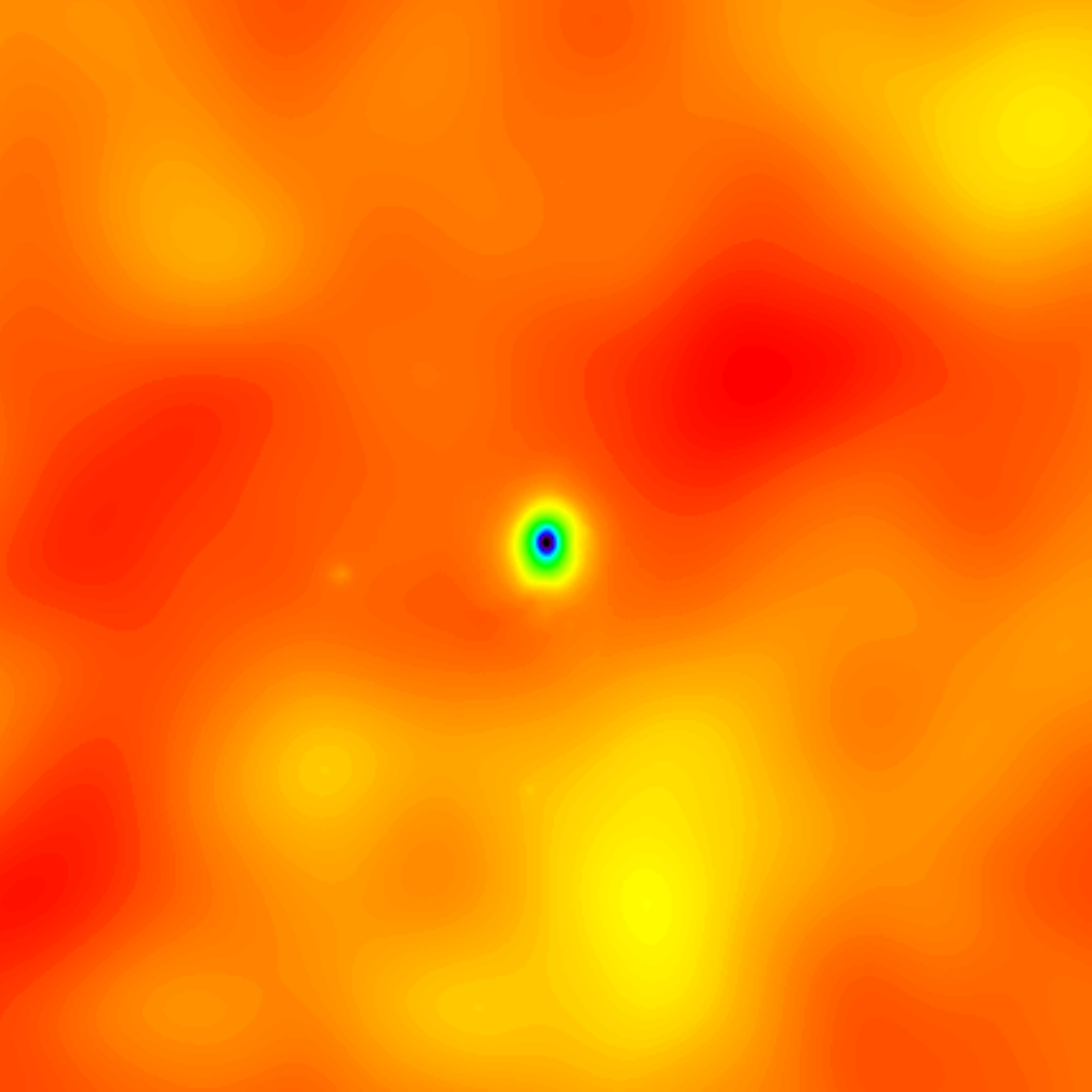}}
     \subfigure{
          \includegraphics[width=.15\textwidth]{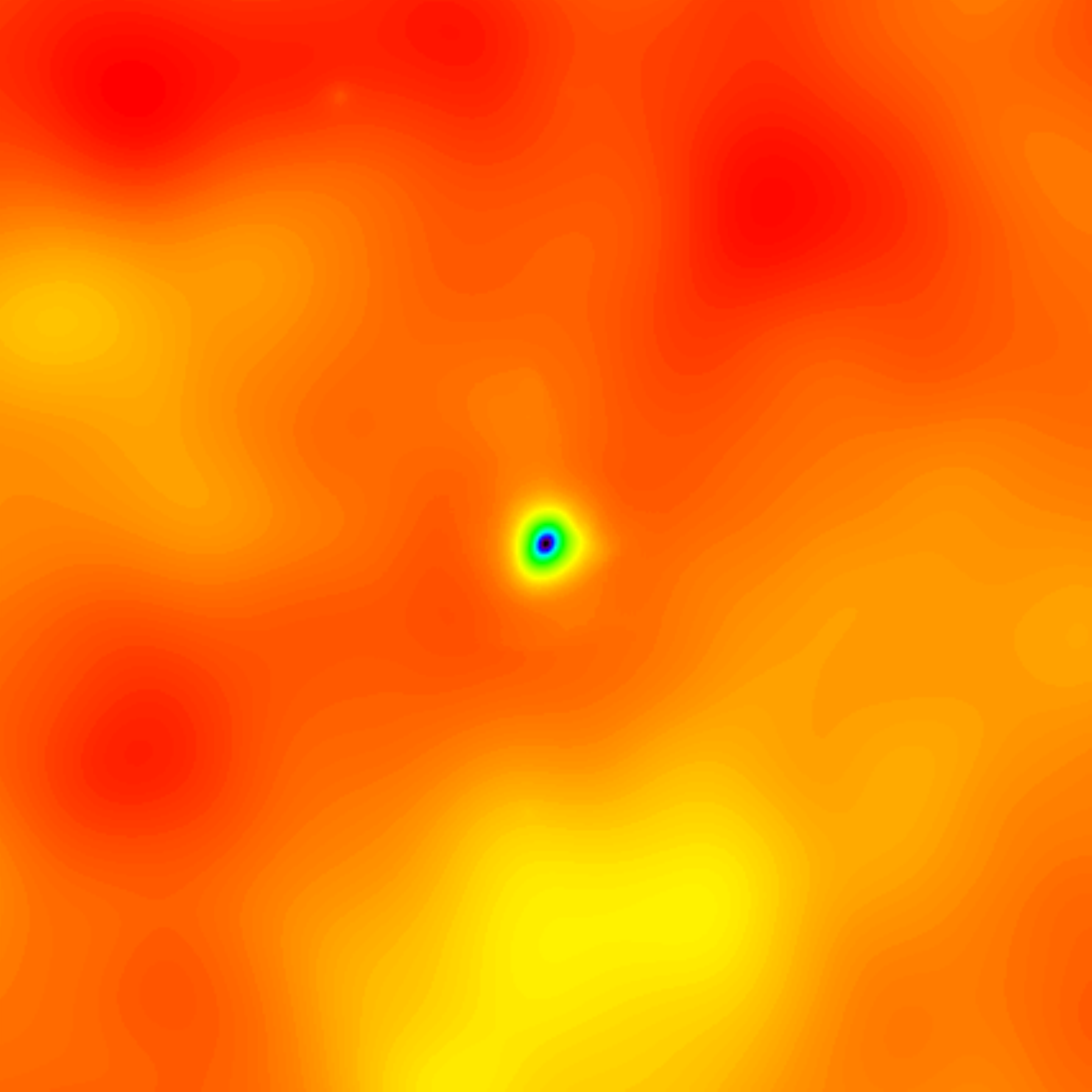}}
      \subfigure{
          \includegraphics[width=.15\textwidth]{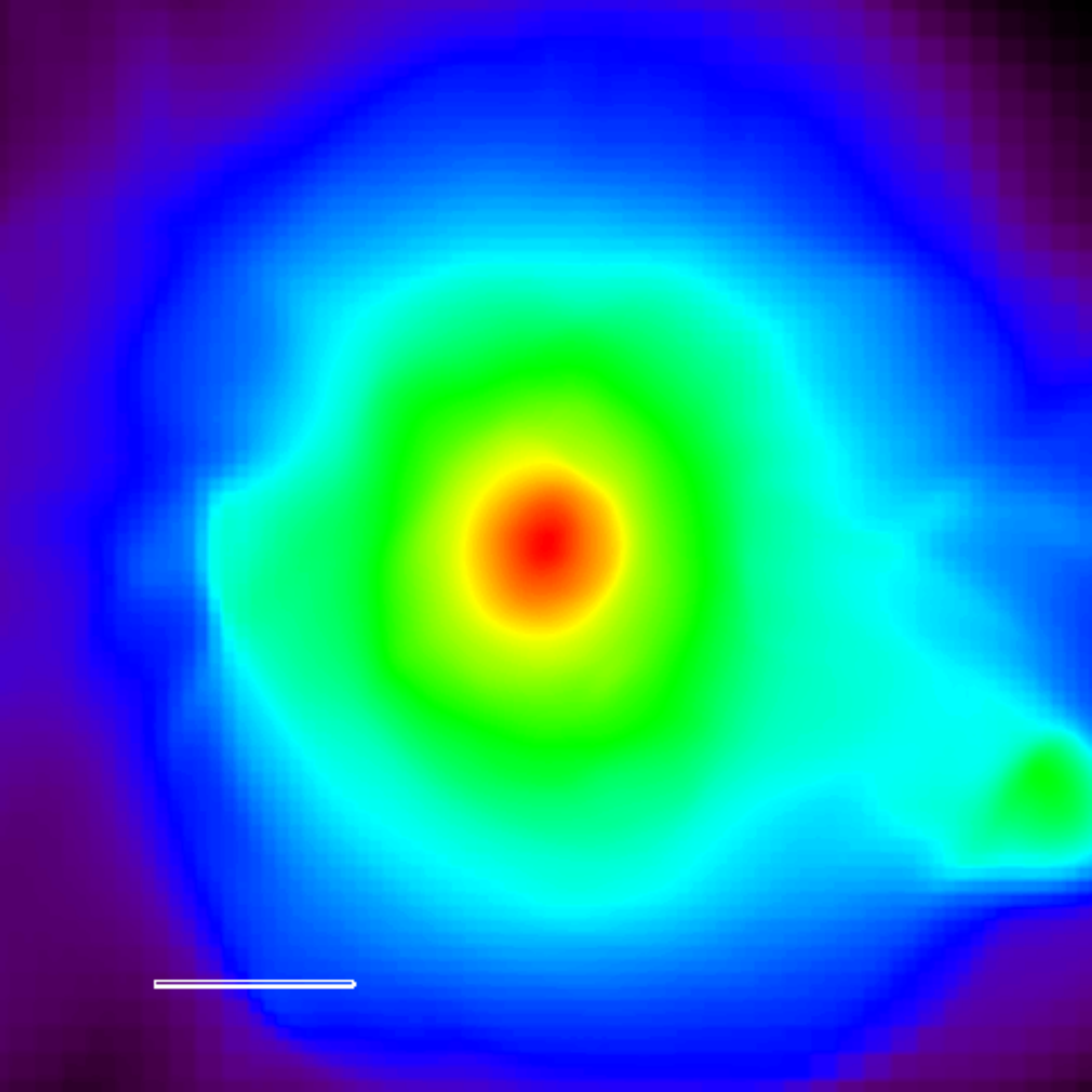}}
     \subfigure{
          \includegraphics[width=.15\textwidth]{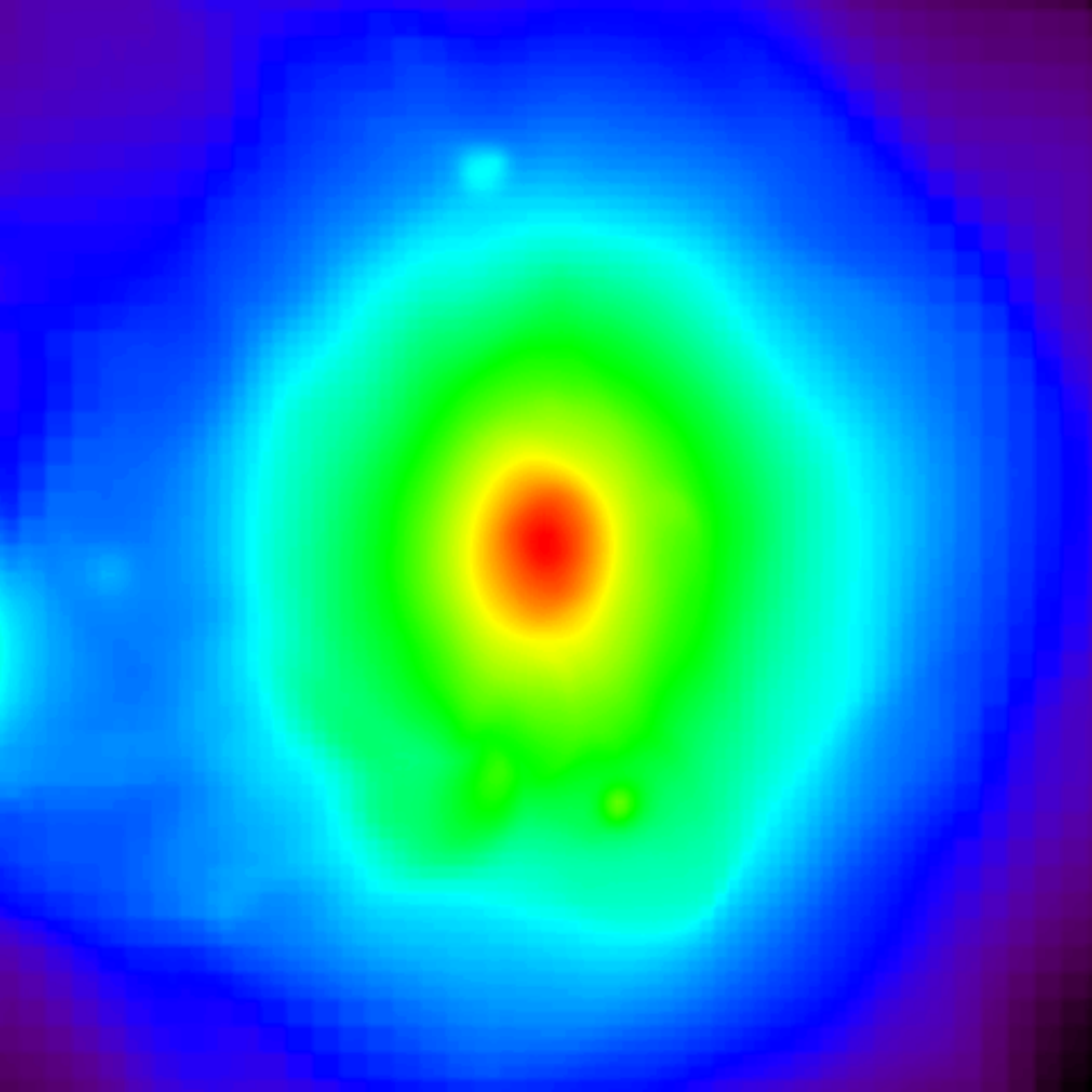}}
     \subfigure{
          \includegraphics[width=.15\textwidth]{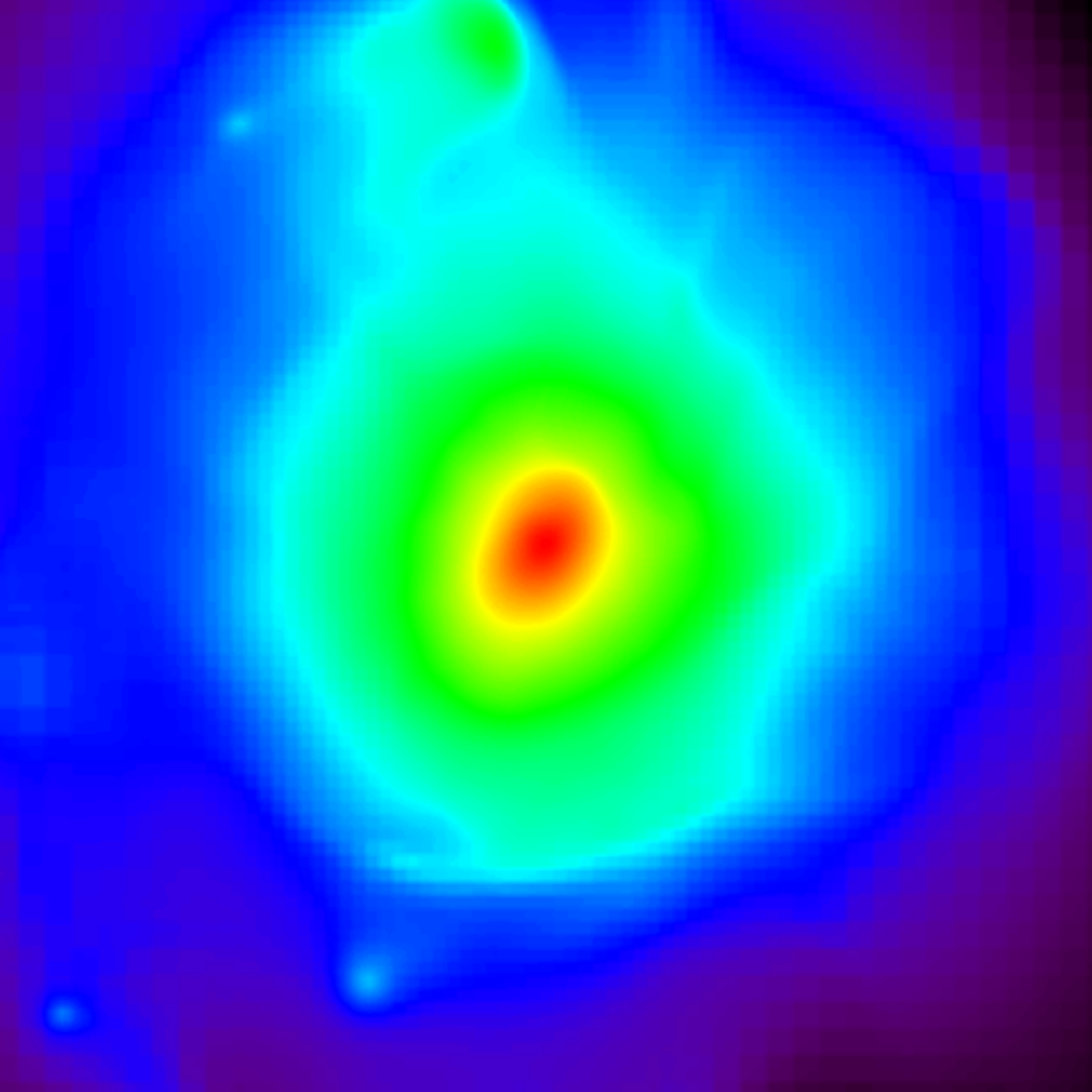}}
     \subfigure{
          \includegraphics[width=.15\textwidth]{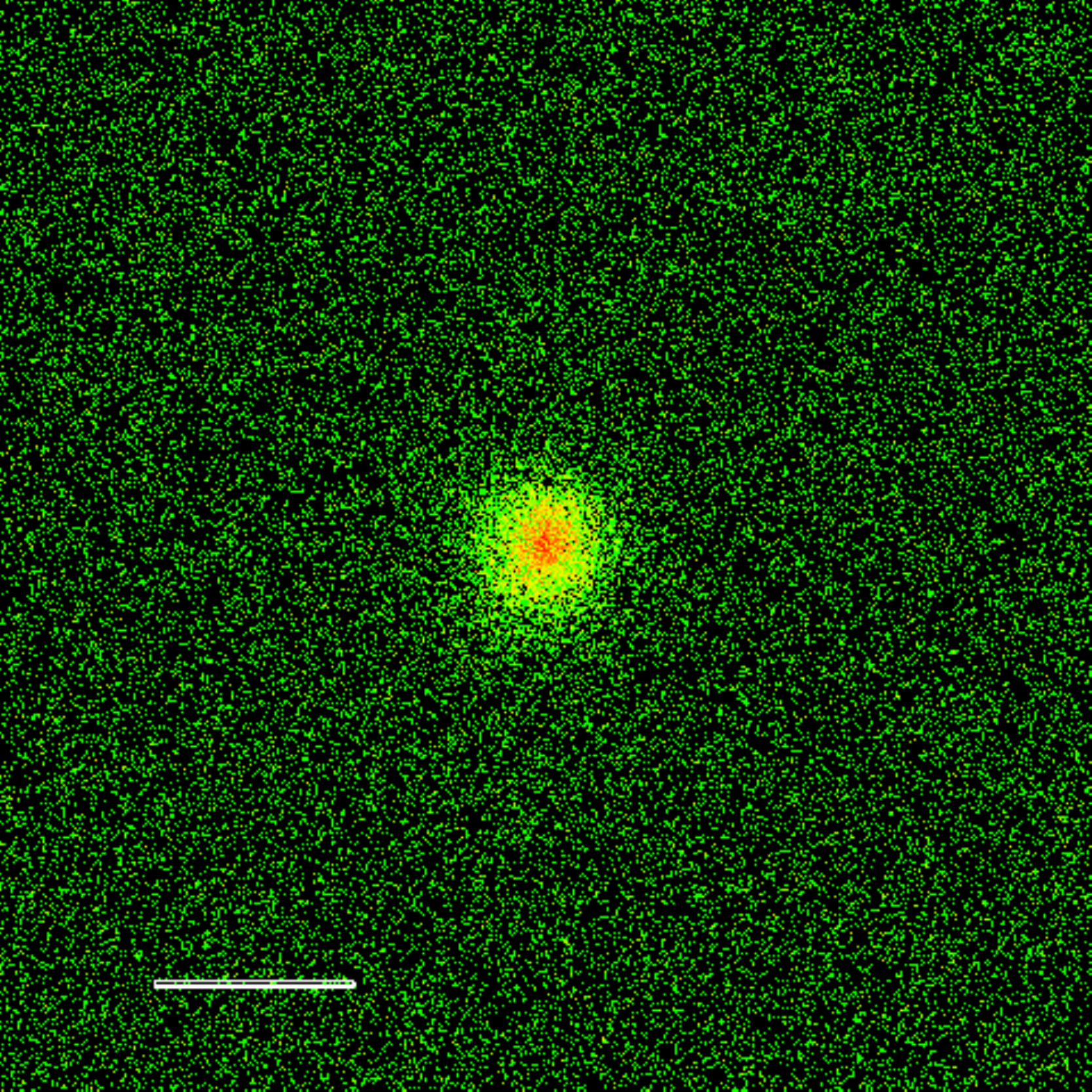}}
     \subfigure{
          \includegraphics[width=.15\textwidth]{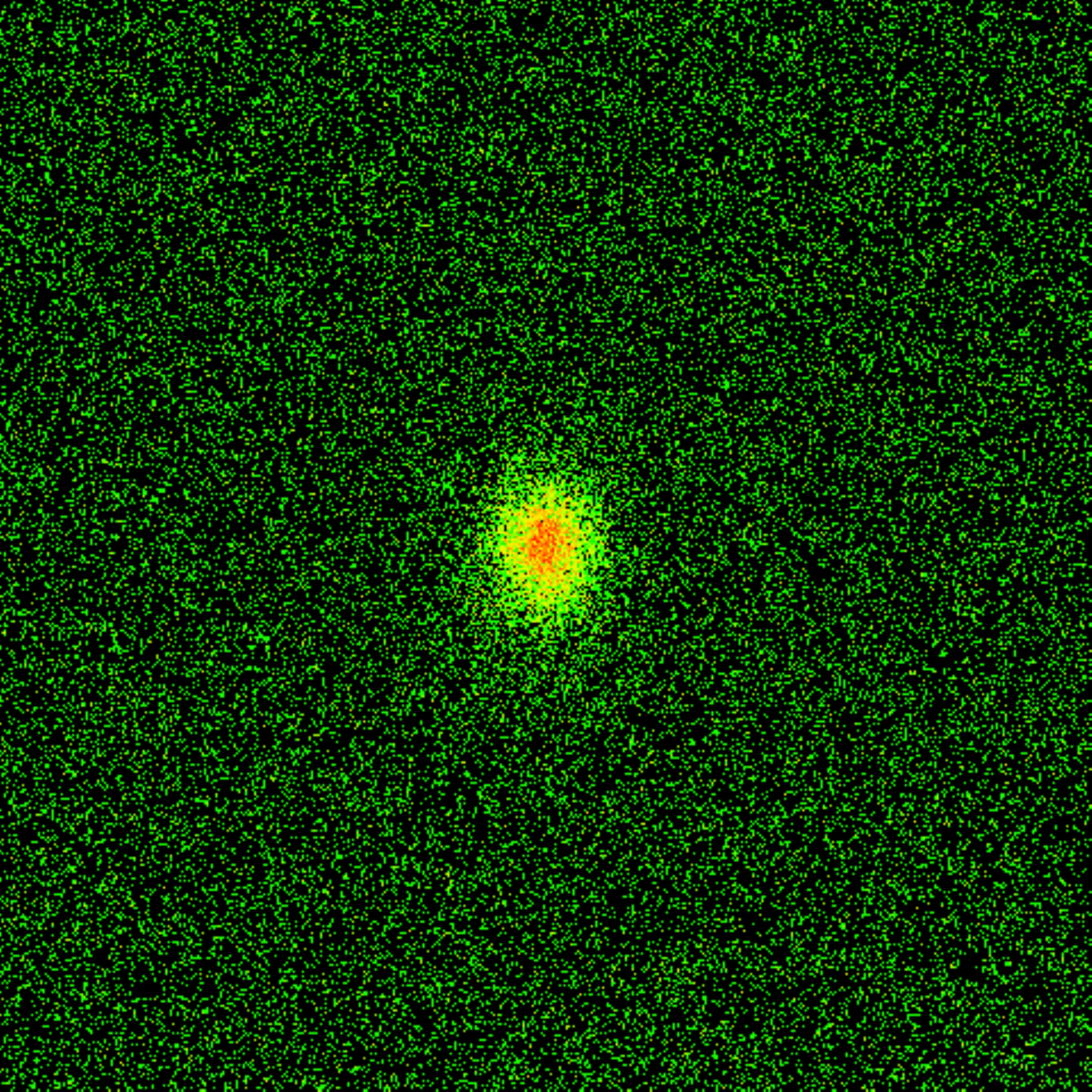}}
     \subfigure{
          \includegraphics[width=.15\textwidth]{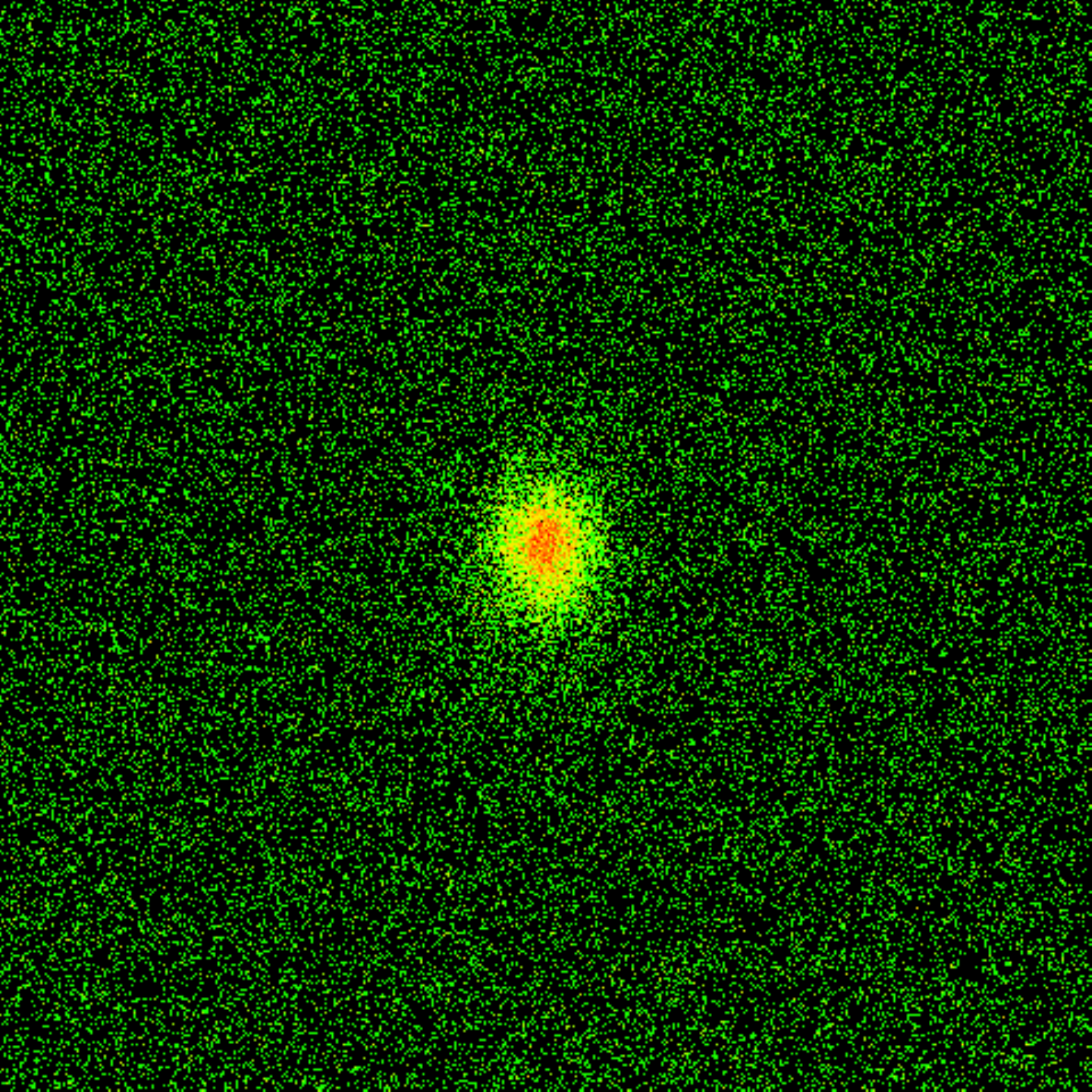}}
\caption{
\label{F:f3}
          Simulated images for Projections X, Y, and Z of relaxed cluster CL1
          (left, middle and right column). From top to bottom:
          Compton-$y$, $\Delta T$ (including the cluster SZ signal and CMB fluctuations 
          at 94 GHz), X-ray surface brightness and simulated X-ray image including
          background noise (logarithmic color scale except SZ map which is linear).
          The image size is $50^{\,\prime} \times 50^{\,\prime}$ for the SZ images
          (first two rows) and $16^{\,\prime} \times 16^{\,\prime}$ for the X-ray images 
          (last two rows). The physical scale of 1 Mpc, same within each raw, 
          is represented by horizontal bars. 
\bigskip  
}
\end{figure}

\begin{figure}
     \centering
     \subfigure{
          \includegraphics[width=.15\textwidth]{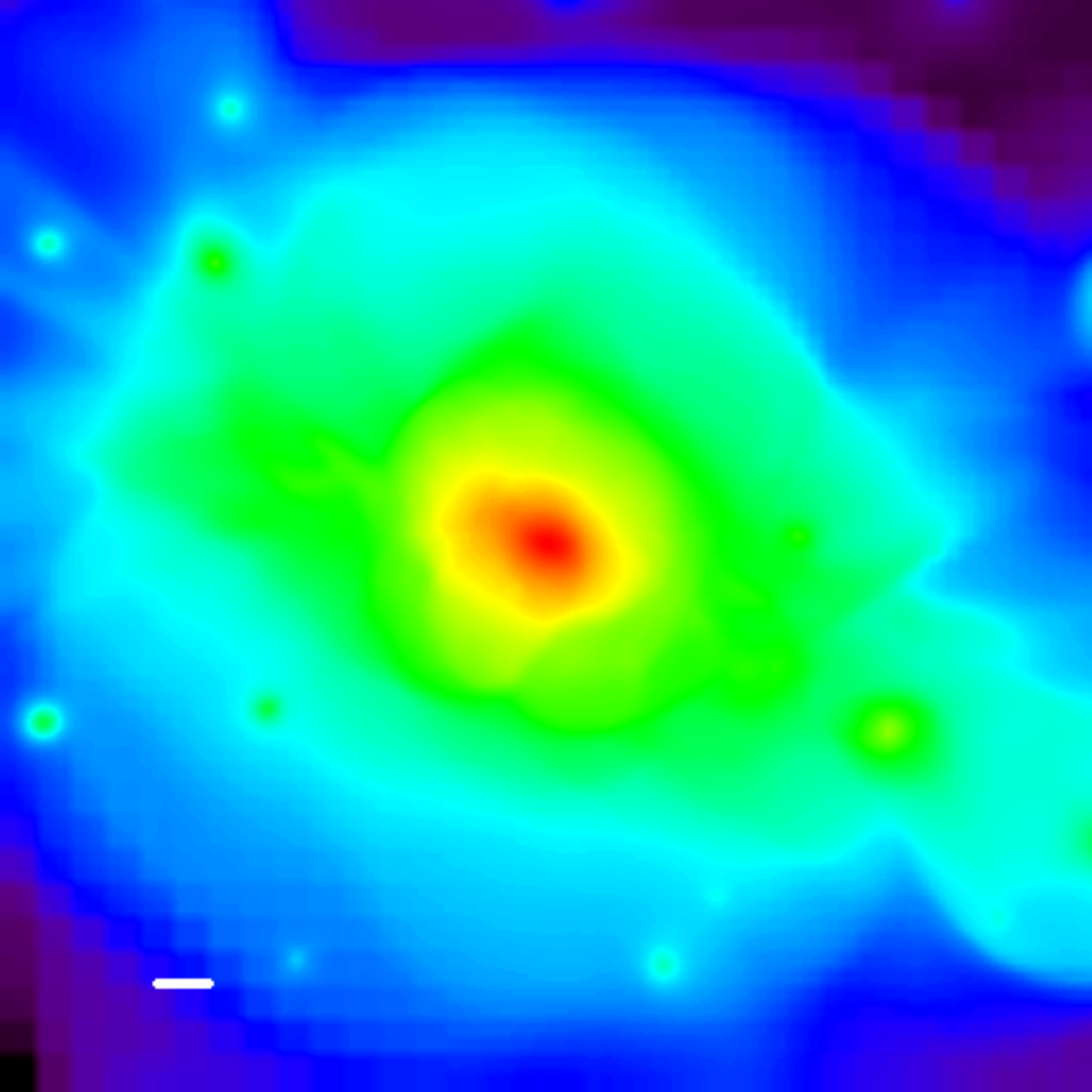}}
     \subfigure{
          \includegraphics[width=.15\textwidth]{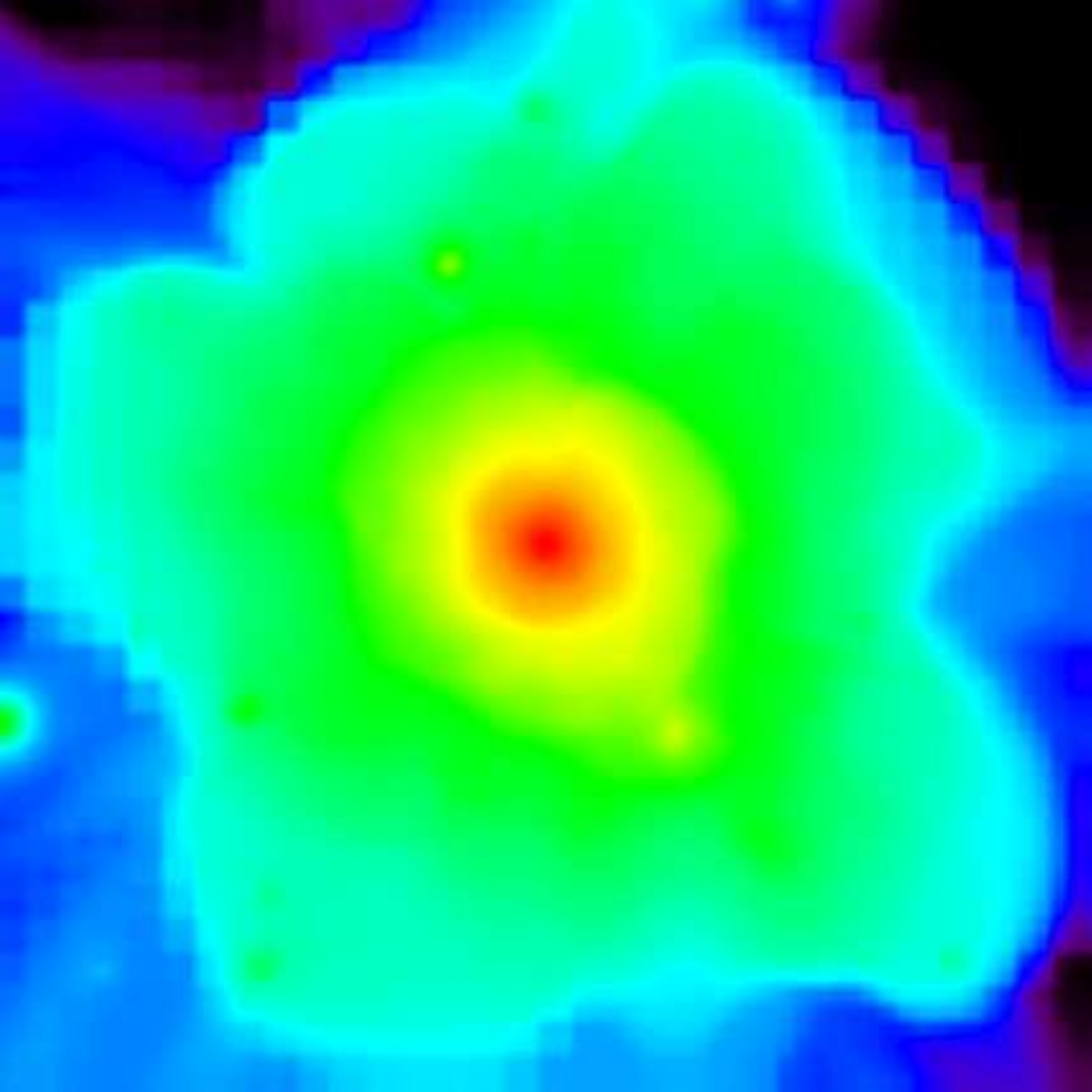}}
     \subfigure{
          \includegraphics[width=.15\textwidth]{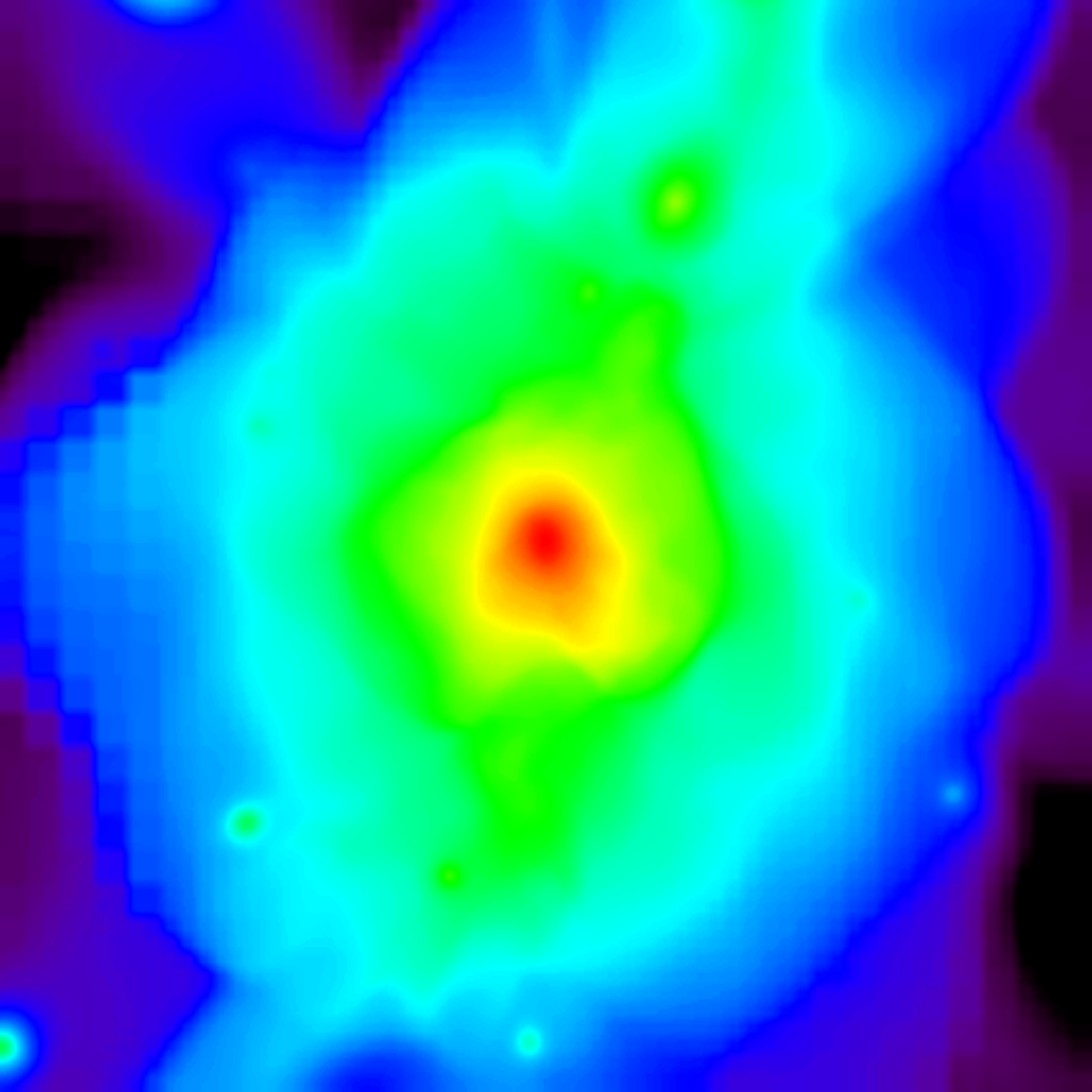}}
     \subfigure{
          \includegraphics[width=.15\textwidth]{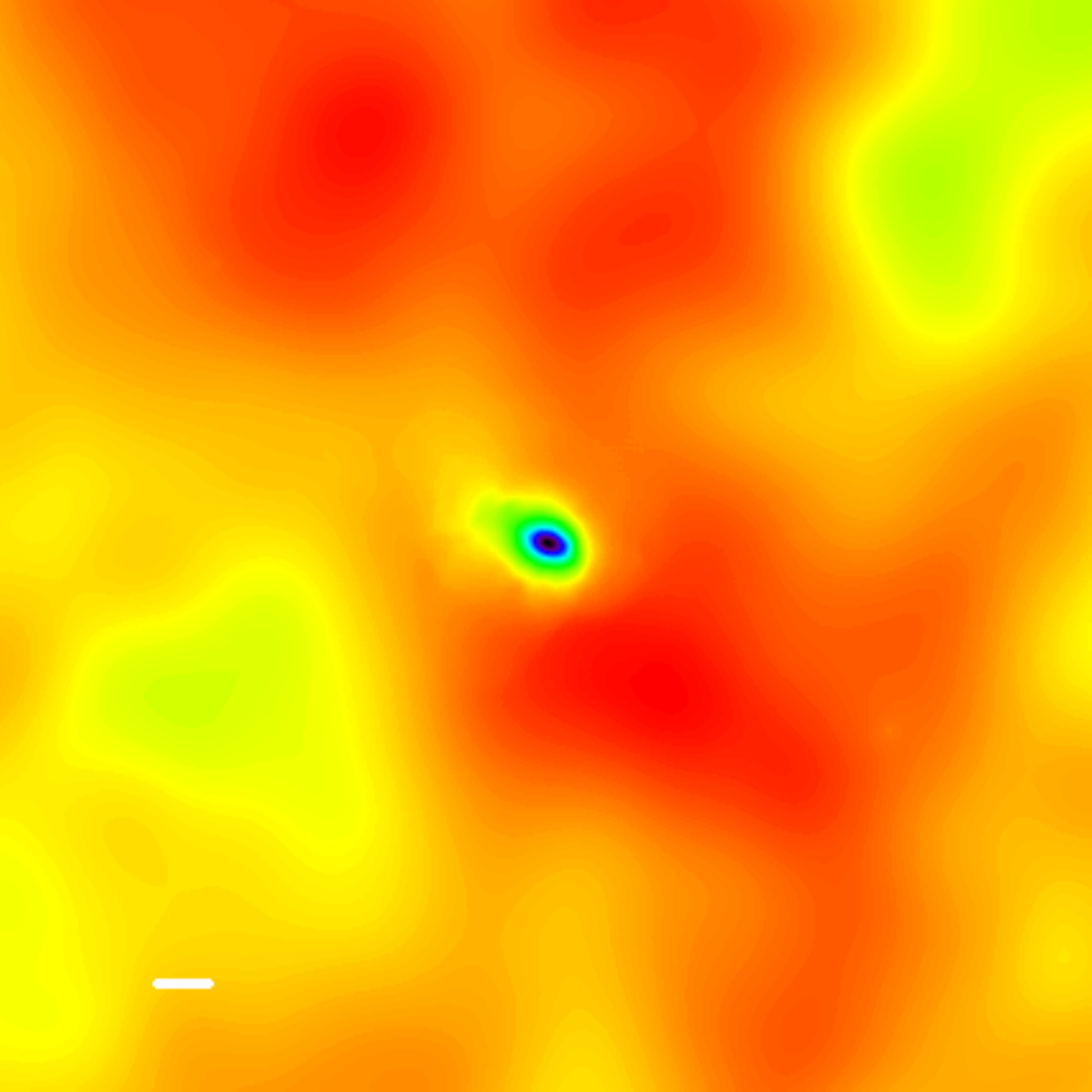}}
     \subfigure{
          \includegraphics[width=.15\textwidth]{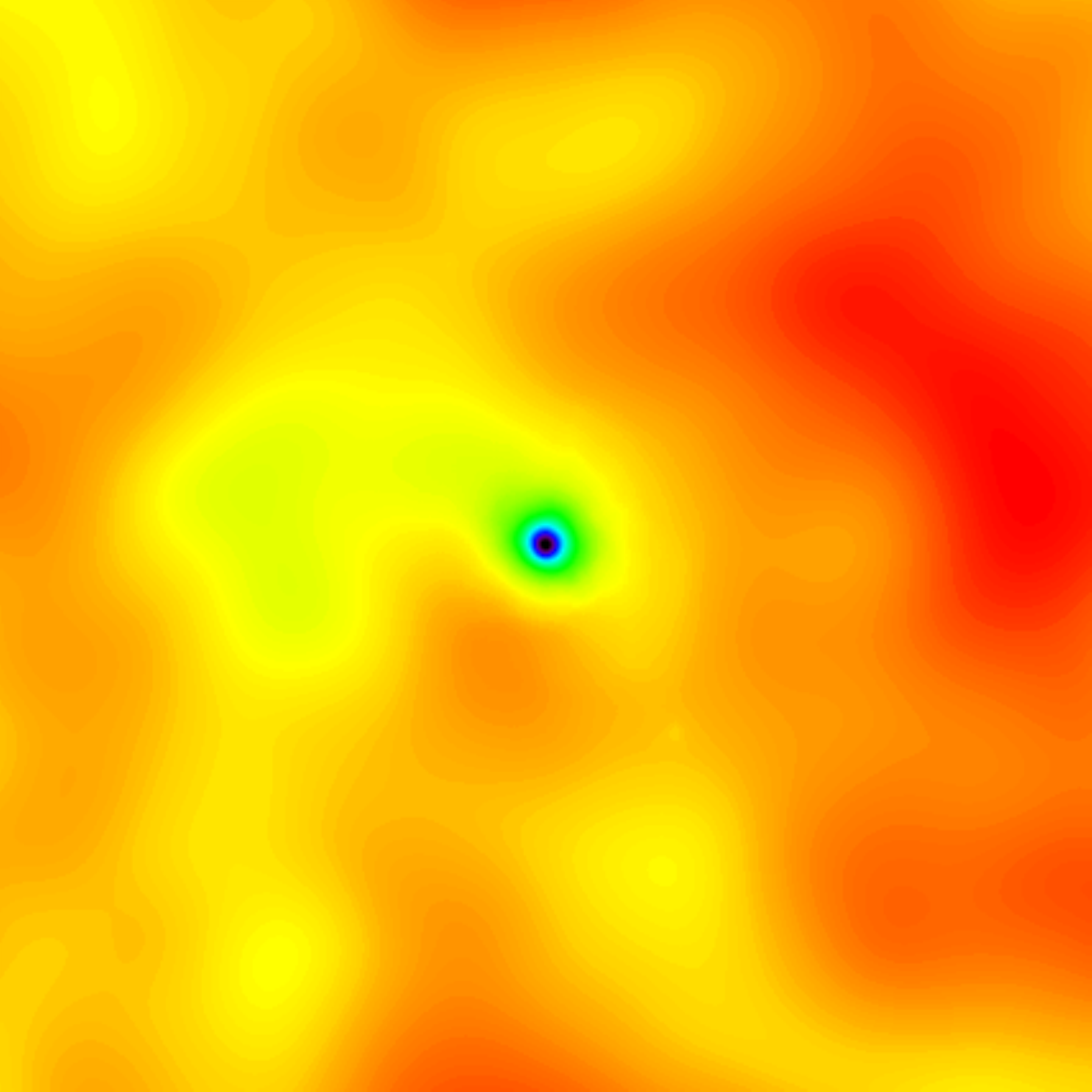}}
     \subfigure{
          \includegraphics[width=.15\textwidth]{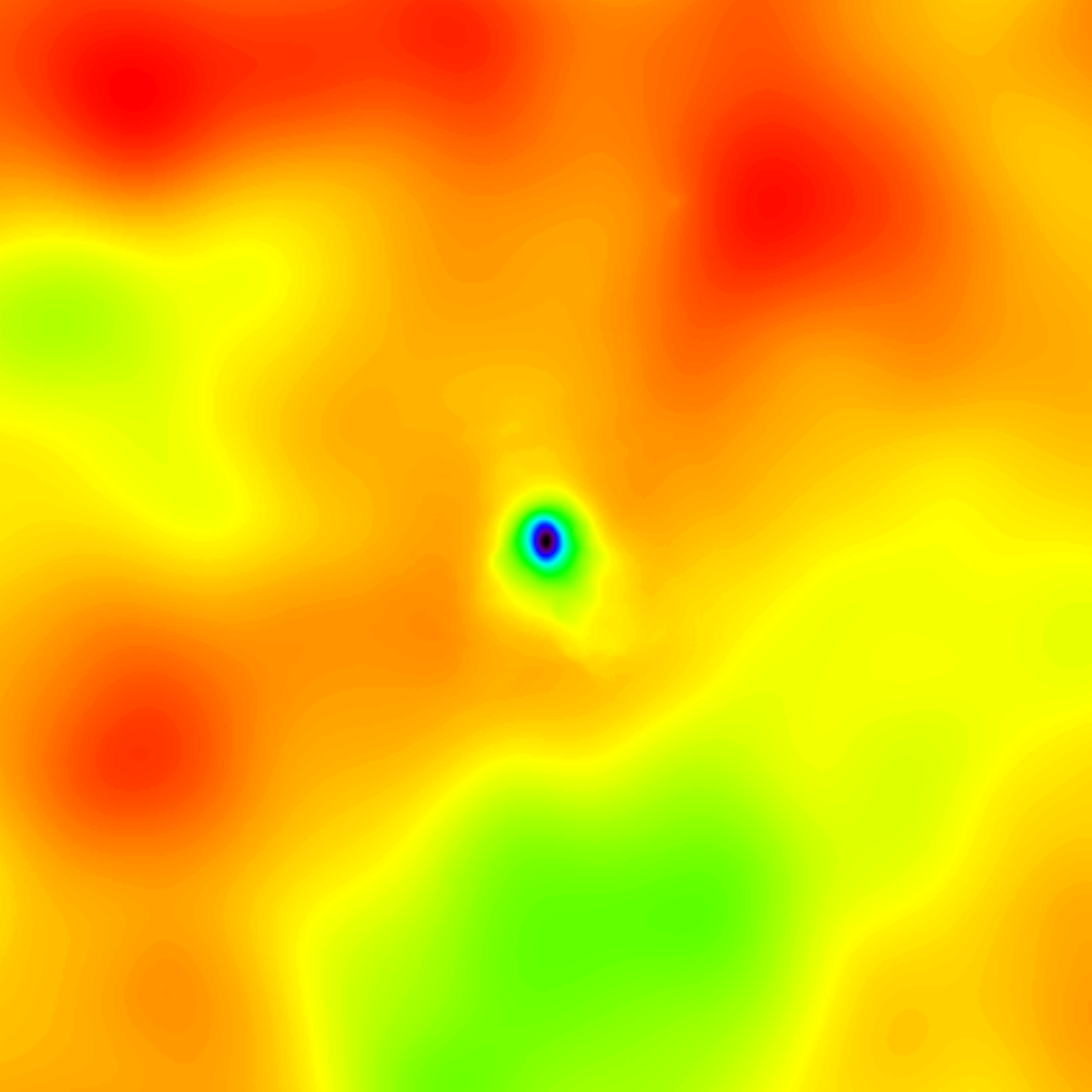}}
      \subfigure{
          \includegraphics[width=.15\textwidth]{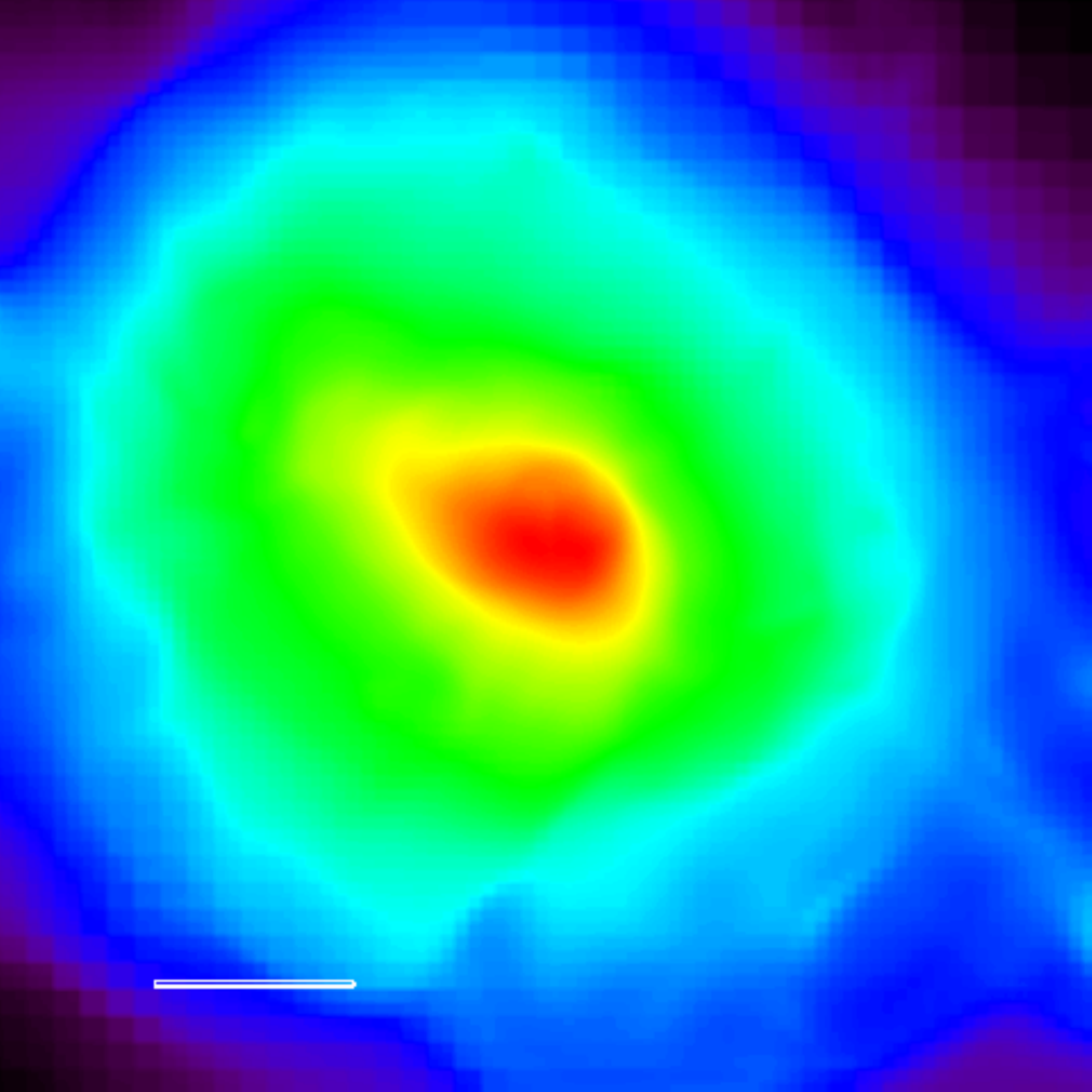}}
     \subfigure{
          \includegraphics[width=.15\textwidth]{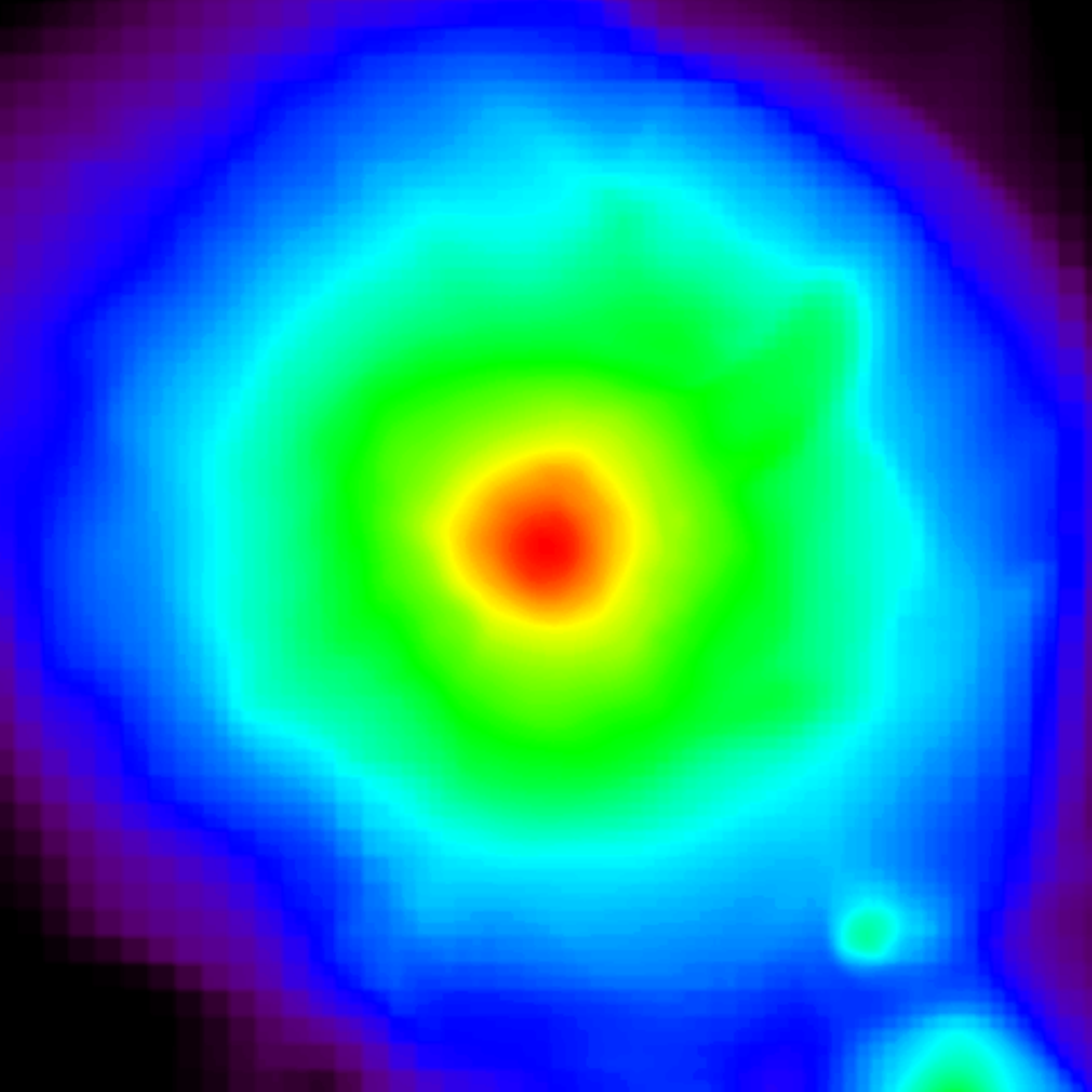}}
     \subfigure{
          \includegraphics[width=.15\textwidth]{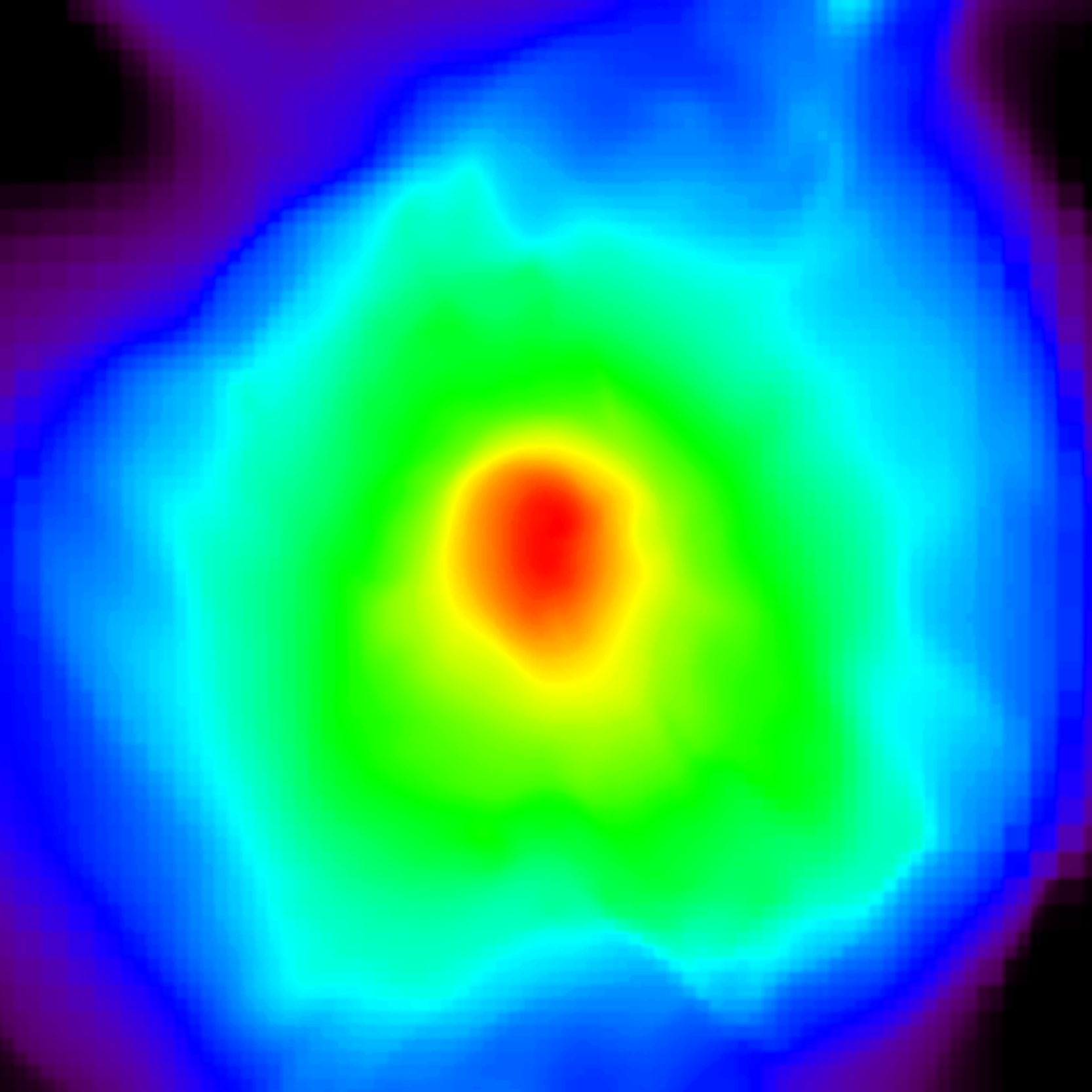}}
     \subfigure{
          \includegraphics[width=.15\textwidth]{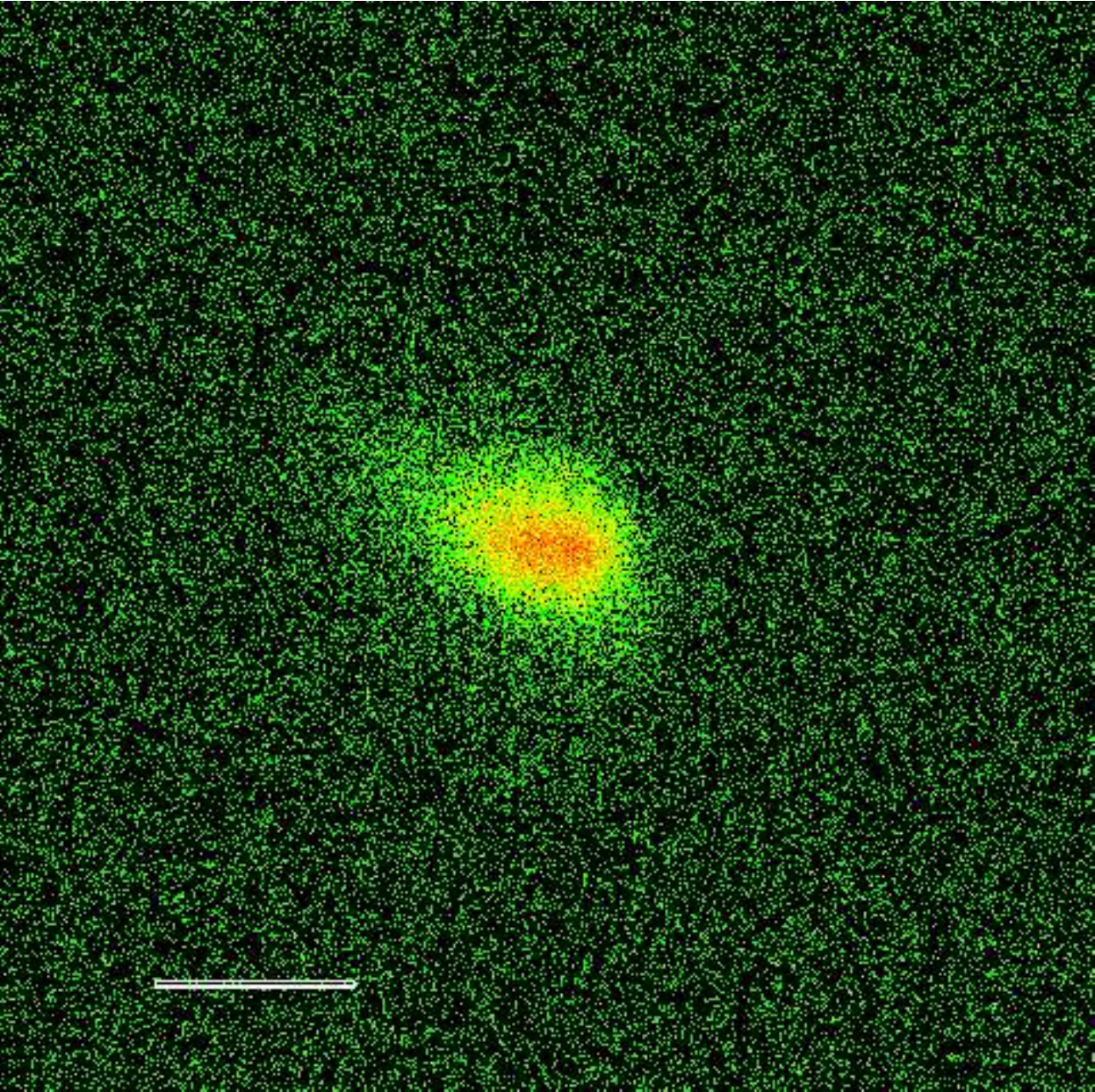}}
     \subfigure{
          \includegraphics[width=.15\textwidth]{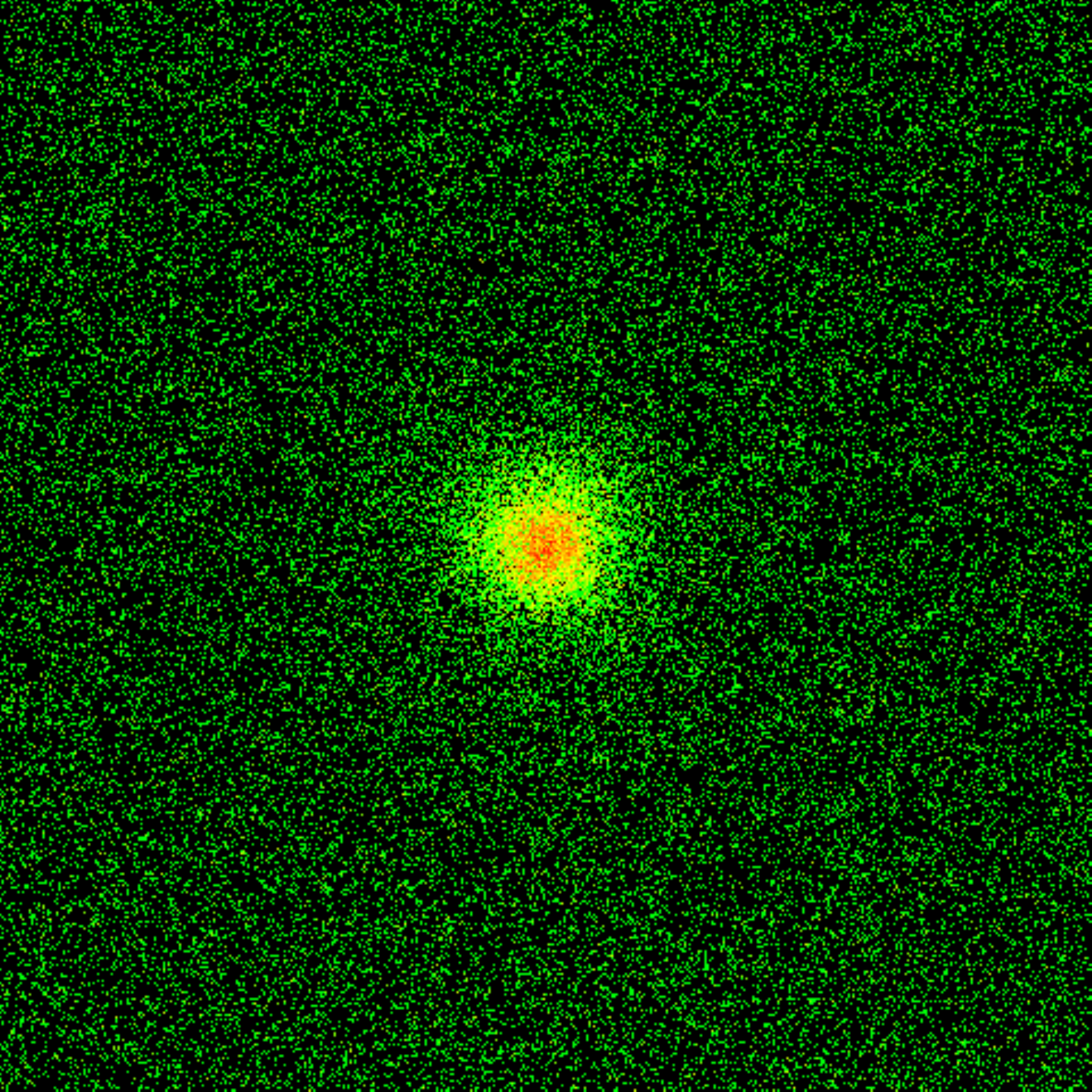}}
     \subfigure{
          \includegraphics[width=.15\textwidth]{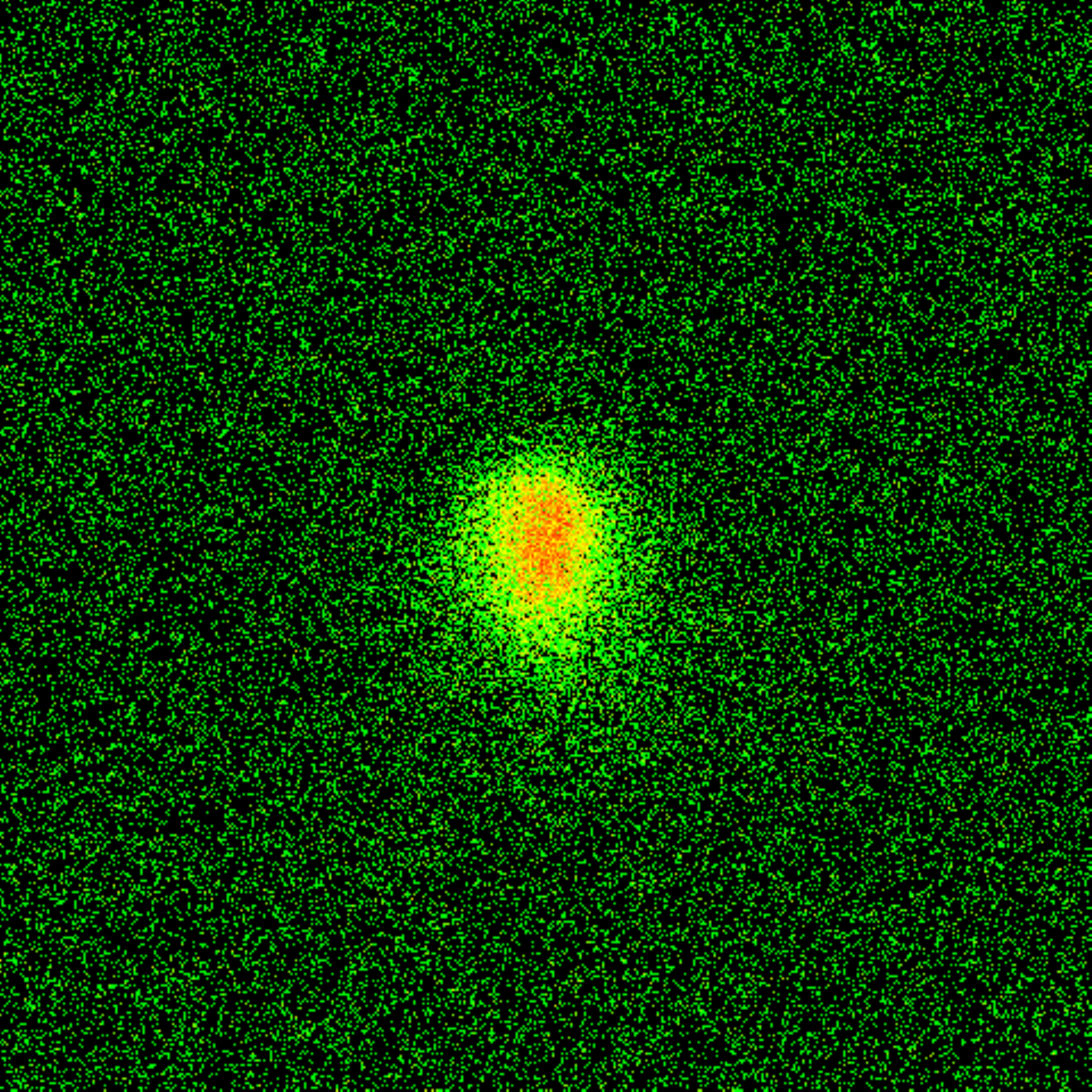}}
\caption{
\label{F:f4}
    Same as Figure~\ref{F:f3} but for simulated images of relaxed cluster CL2.
}
\end{figure}

The Compton--$y$ images, $y = \int d \tau \, (\KBOLTZ T_e) / (m_e c^2)$, 
where $\tau$ is the optical depth, the SZ images (cluster plus CMB 
at the \AMIBA\ frequency band, 94 GHz), X-ray surface brightness, 
and simulated X-ray images including background noise, 
of our two relaxed clusters (CL1 and CL2) and one non-relaxed cluster (CL3)
in Projections X, Y and Z assuming that they are located at a redshift of 0.3, 
are shown in Figures~\ref{F:f3}, \ref{F:f4} and \ref{F:f5}.
The Compton--$y$, X-ray surface brightness and 
simulated X-ray images are shown in logarithmic scale (first, third and fourth rows),
and the SZ images with CMB contamination (second row) are shown in linear scale.
The virial radii of our massive relaxed clusters are about 2 Mpc, 
which span about 8$\arcmin$ on the sky at this redshift.
The dark blue regions ($\approx -1.2$~mK) on the SZ images mark the cluster centers,
the yellow and red regions represent positive and negative CMB fluctuations 
with an amplitude of about $\pm 130\, \mu$K and an {\it rms} of about 90$\,\mu$K.
In the Compton-$y$ images we can follow the diffuse gas out to about 3\RVIR,
where the external shocks of massive clusters are found (see \citealt{Molnet09}), 
but in the SZ images the diffuse gas around clusters seems to extend out to
about \RVIR\ only, due to contamination from CMB fluctuations, which 
dominate the large-scale structure. 
However, the SZ images of relaxed clusters, in most projections,
show similar characteristics within the virial radius: 
a circularly-symmetric center and somewhat elongated outer regions.
The core regions in the X-ray and SZ images (for example regions with yellow 
color in rows 1 and 3), in a few projections (Projection Z of CL1 and 
Projections X and Z of CL2), show asymmetry due to asphericity of
the cluster and contamination by filaments in the LOS.
The core region of non-relaxed cluster CL3 seems to be disturbed in all projections.

%
%

%
%

\smallskip
\section{\AMIBAW\ Visibility Simulations}
\label{S:AMIBAVIS}

Interferometers measure visibilities, the Fourier transforms of the intensity distribution
multiplied by the primary beam of the telescope.
In the small-angle approximation, the visibilities can be written as

\begin{equation}  \label{E:VISIB}
           V_\nu (u,v) =   \int^{\infty}_{-\infty}  \int^{\infty}_{-\infty} B_\nu (x,y) \,  I_\nu (x,y) \, 
                                        e^{-i 2 \pi (ux + vy)}  \, dx \,dy
,
\end{equation}

\bigskip
\nop
where $V_\nu (u,v)$ is the visibility function in the $uv$ plane, 
which is the Fourier conjugate of the positions, $x$ and $y$ on the sky.
$B_\nu (x,y)$ and $I_\nu (x,y)$ are the primary telescope beam pattern and
source intensity at $x$ and $y$ at frequency $\nu$, 
and we ignore effects due to a finite bandwidth, finite time averaging, 
and other effects associated with practical interferometers.
We convert temperature differences to intensity units using

\begin{equation}  \label{E:DELI}
     {\Delta I \over \ICMB} = {x_{\,\nu} \, e^{\,x_\nu} \over (e^{\,x_\nu} -1)} \,  
                                                                                                {\Delta T \over \TCMB}
,
\end{equation}
\nop
where \ICMB\ is the intensity of the monopole term in the CMB.
The visibilities, \VCL\ and \VCMB\ at frequency $\nu$ are calculated from 
$\Delta T_{\rm CMB}$ and $\Delta T_{\rm CL}$ using Equations~\ref{E:VISIB} and \ref{E:DELI}.

We simulated visibilities for mock two-patch \AMIBAW\ observations of our 
relaxed clusters assuming a compact configuration for the 13 antennas 
(Figure~\ref{F:f6}; for a detailed description see Koch et al., in preparation).
In this configuration the many short baselines provide the highest sensitivity 
to the large-scale radio structure. 
The instrument noise and the contribution from the CMB were simulated in visibility space 
and added to the visibilities of the simulated clusters (see Umetsu et al. 2004). 
We assumed an observing time of 60 hours on-source. 
The errors in the azimuthally averaged visibilities for the simulated clusters, 
for the CMB and for the instrument noise are all comparable at this exposure time. 
Using longer exposure would reduce the errors for the instrument noise, 
but would not lower the errors either for our ICG models or for the CMB.
Therefore we would not be able to reduce the error bars significantly by
using longer observations.
Usually a subtraction scheme is used in cluster SZ observation with half of the time 
on-source, and half of the time off source to reduce systematics, ground pick up and 
sky background (e.g., Wu et al. 2009).
Therefore 60 hours of on-source integration involve 120 hour of total observing time.
This can be conveniently accommodated into a 1-month observing schedule.

%
%
\begin{figure}
     \centering
     \subfigure{
          \includegraphics[width=.15\textwidth]{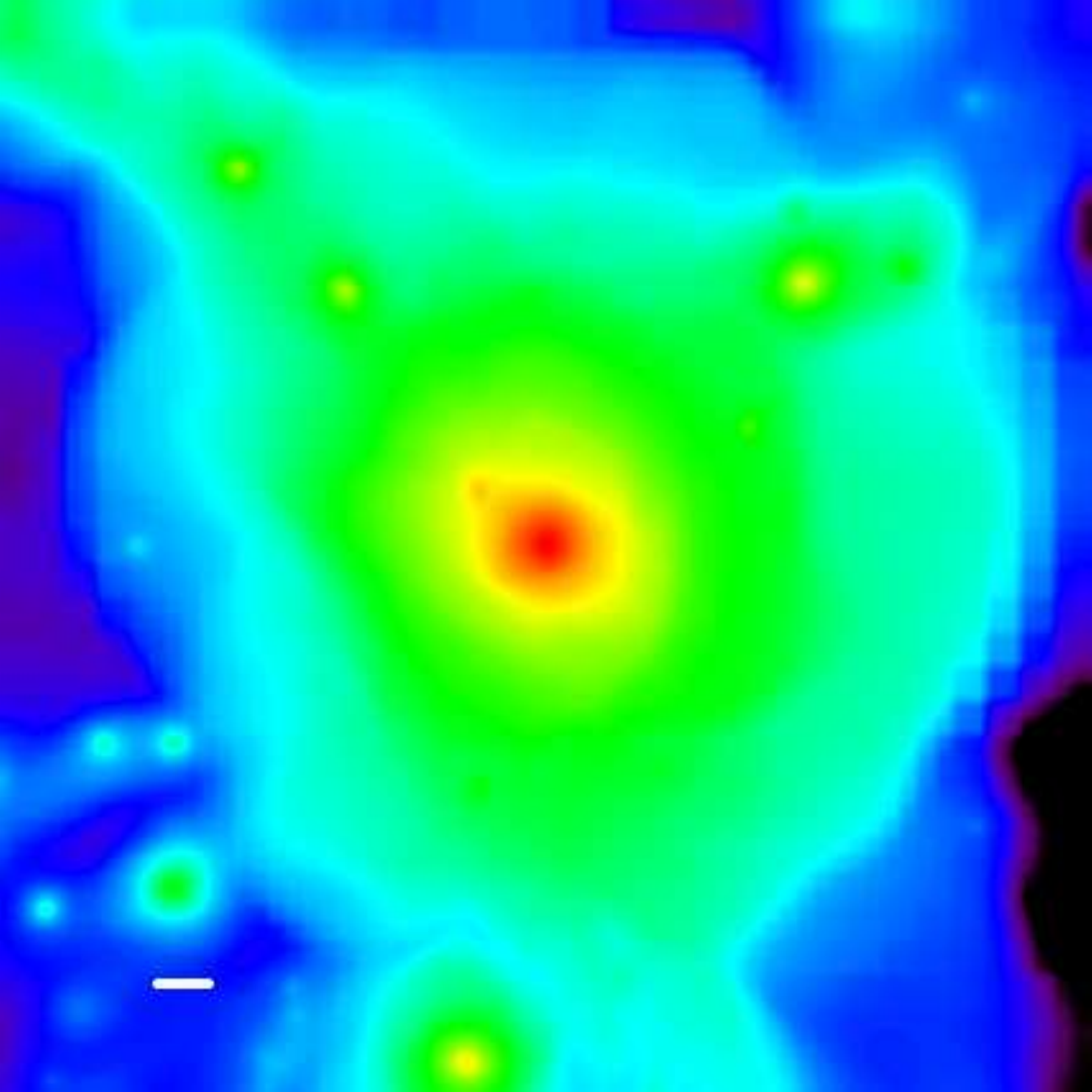}}
     \subfigure{
          \includegraphics[width=.15\textwidth]{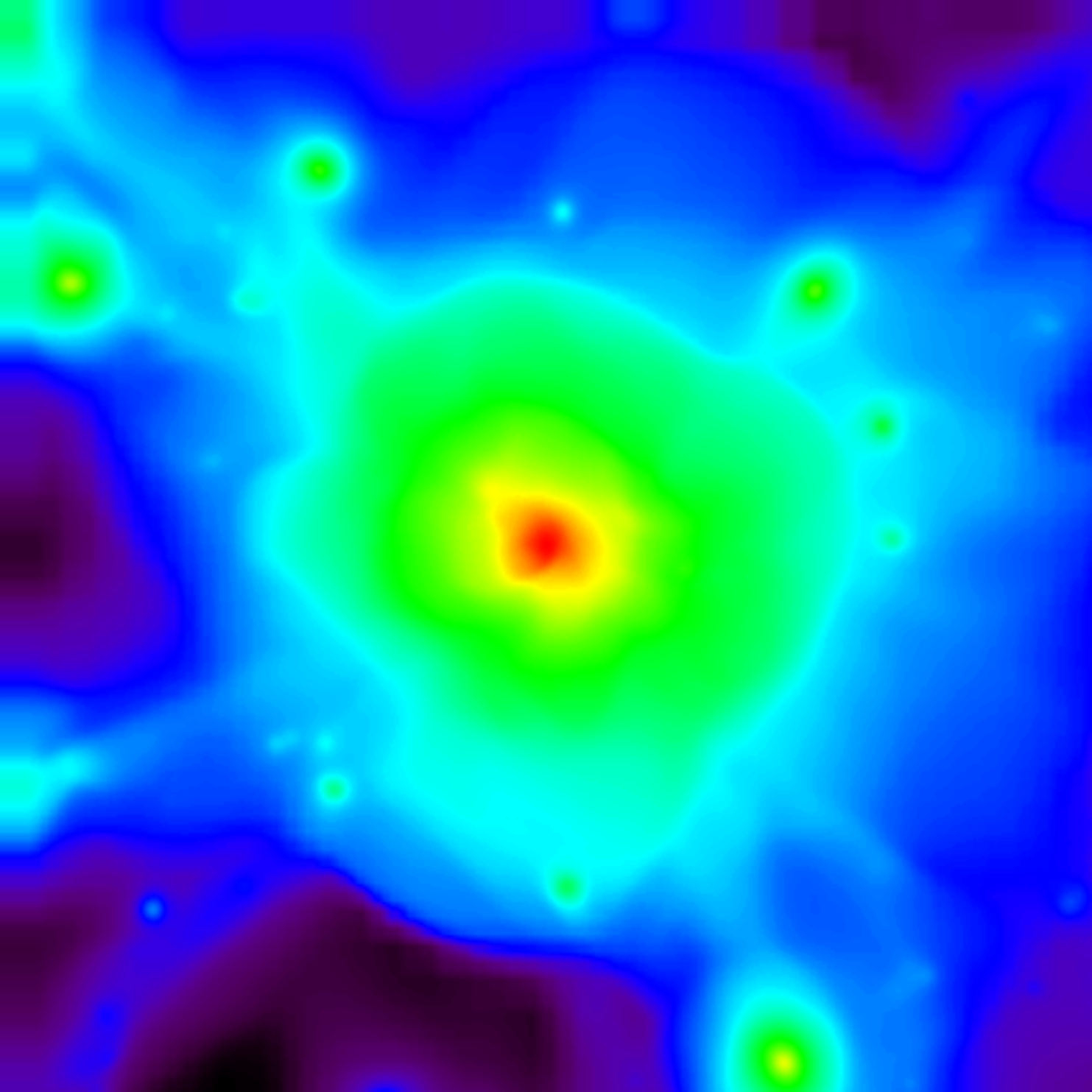}}
     \subfigure{
          \includegraphics[width=.15\textwidth]{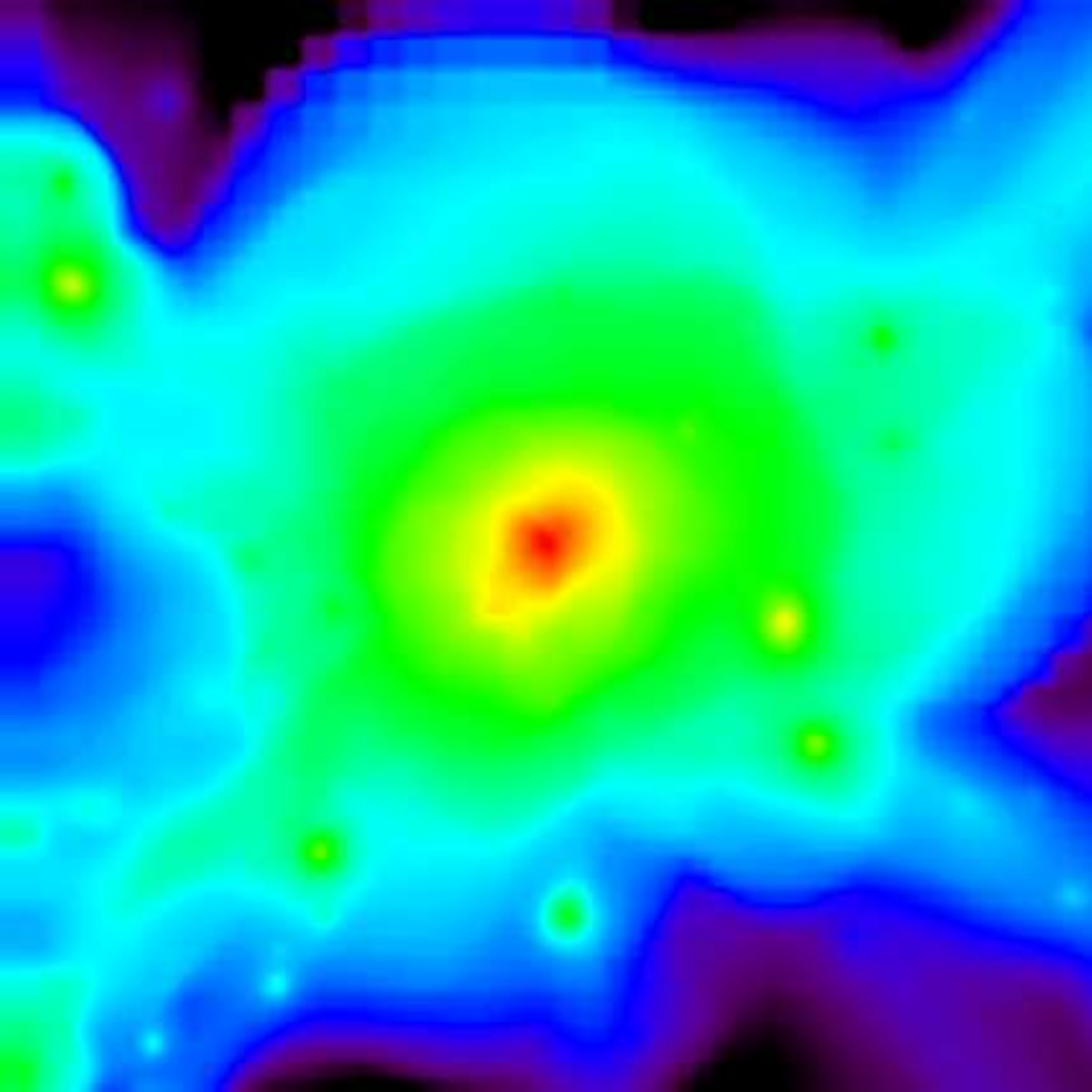}}
     \subfigure{
          \includegraphics[width=.15\textwidth]{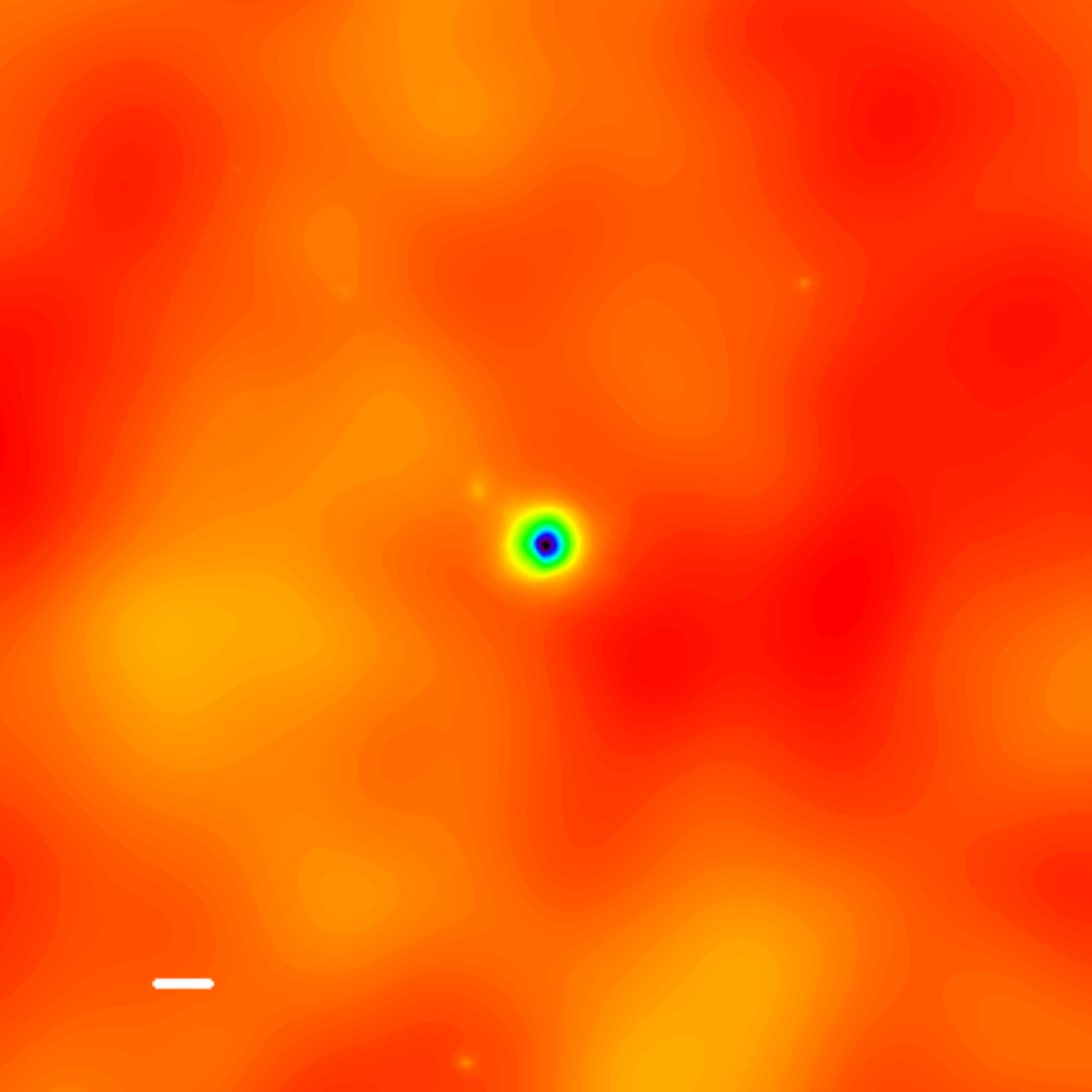}}
     \subfigure{
          \includegraphics[width=.15\textwidth]{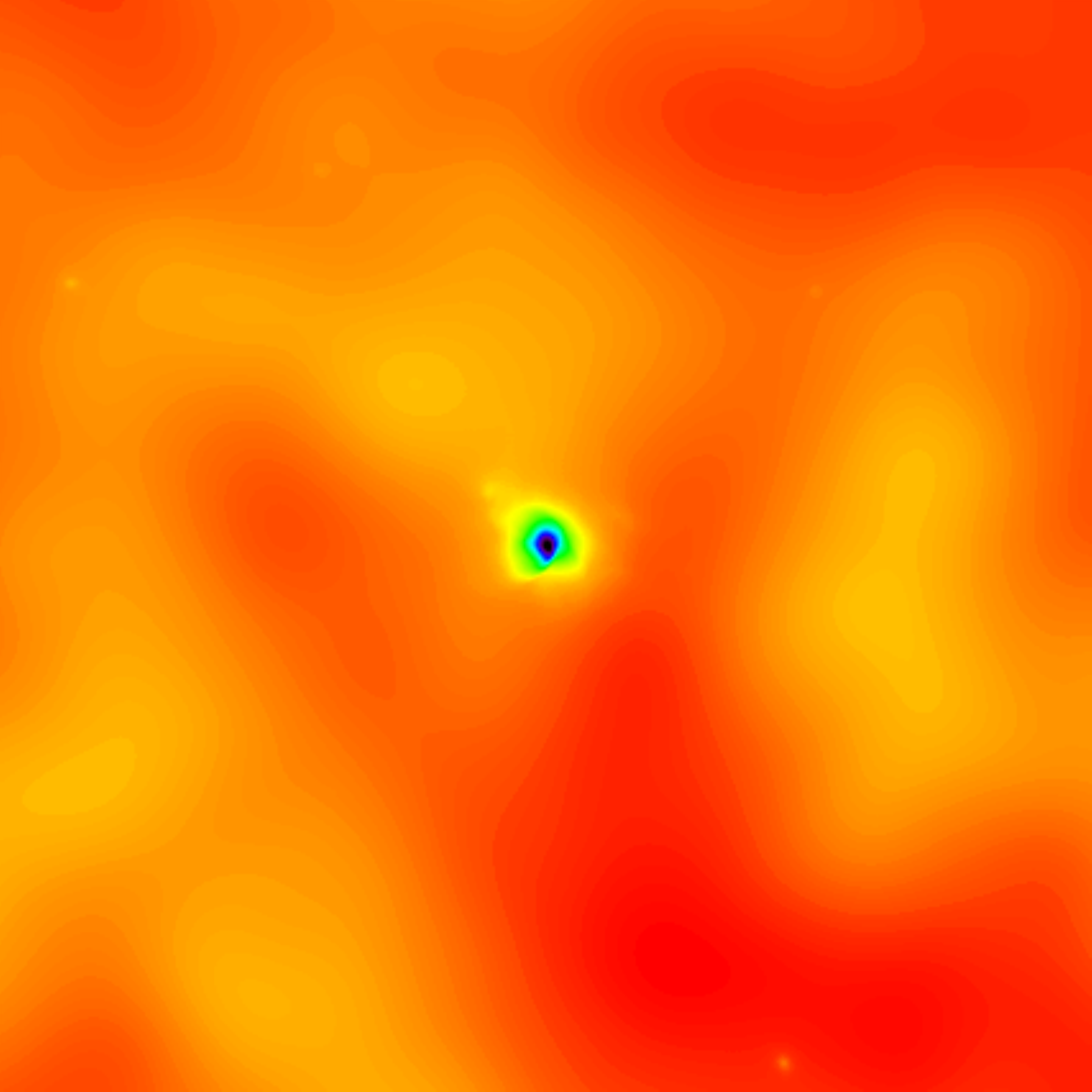}}
     \subfigure{
          \includegraphics[width=.15\textwidth]{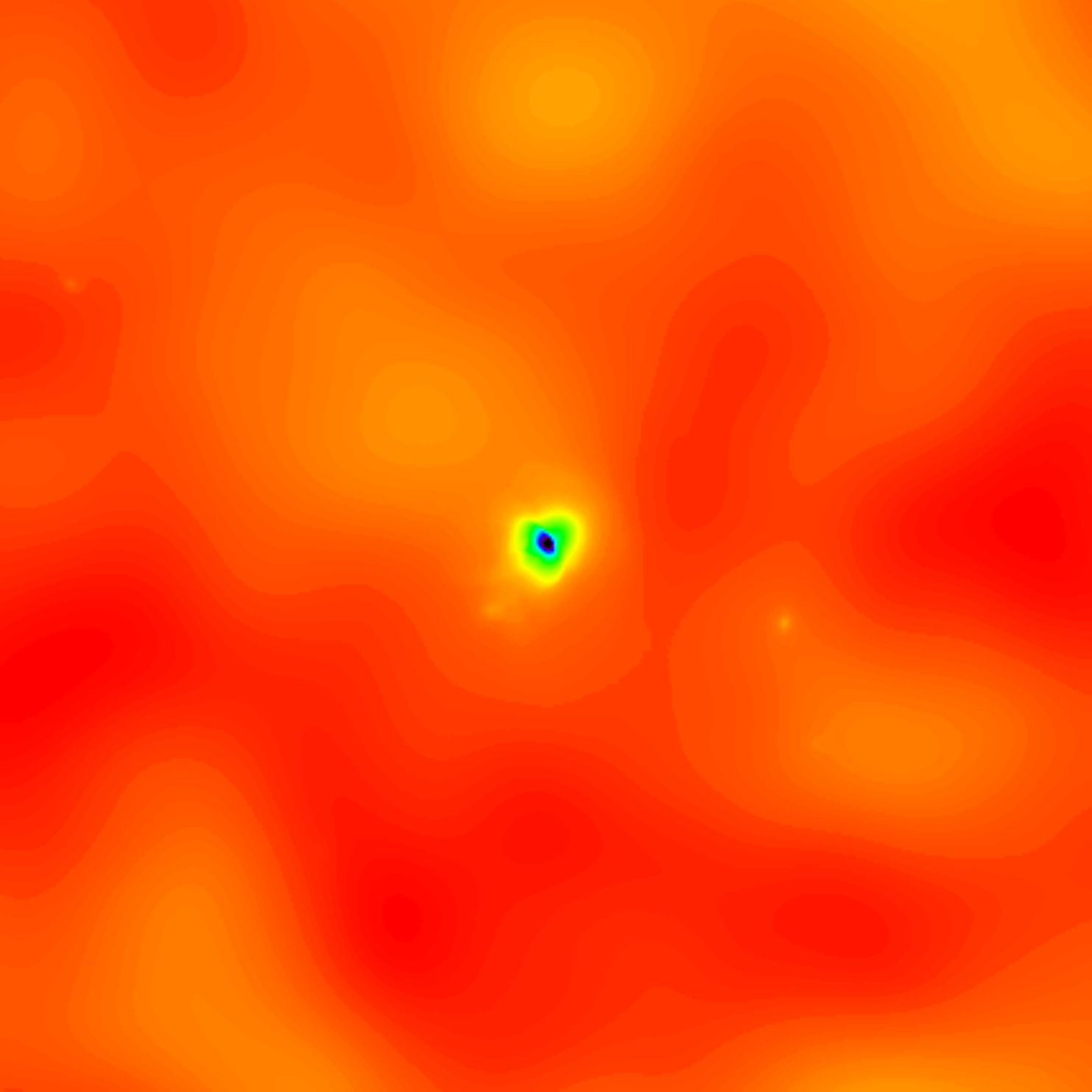}}
      \subfigure{
          \includegraphics[width=.15\textwidth]{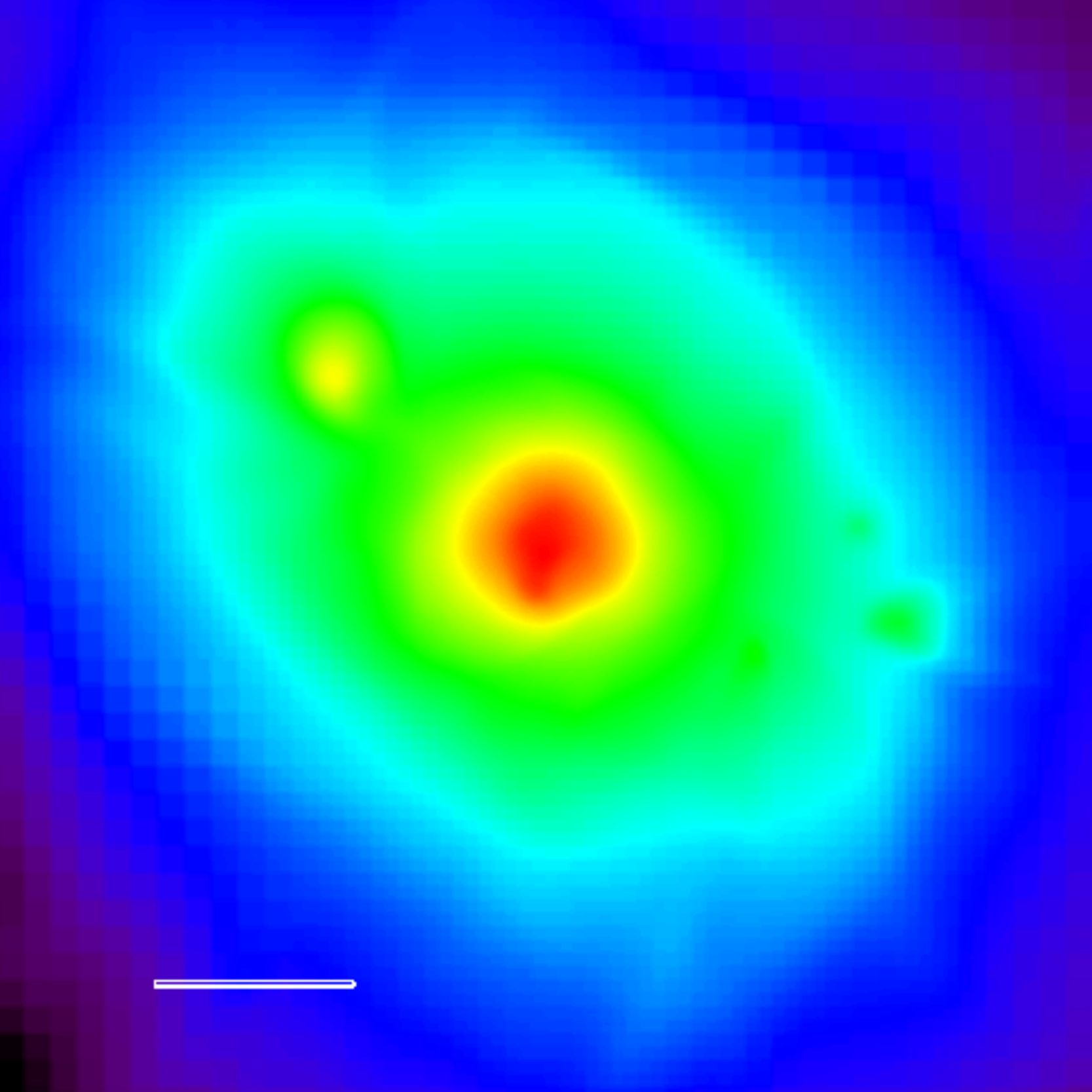}}
     \subfigure{
          \includegraphics[width=.15\textwidth]{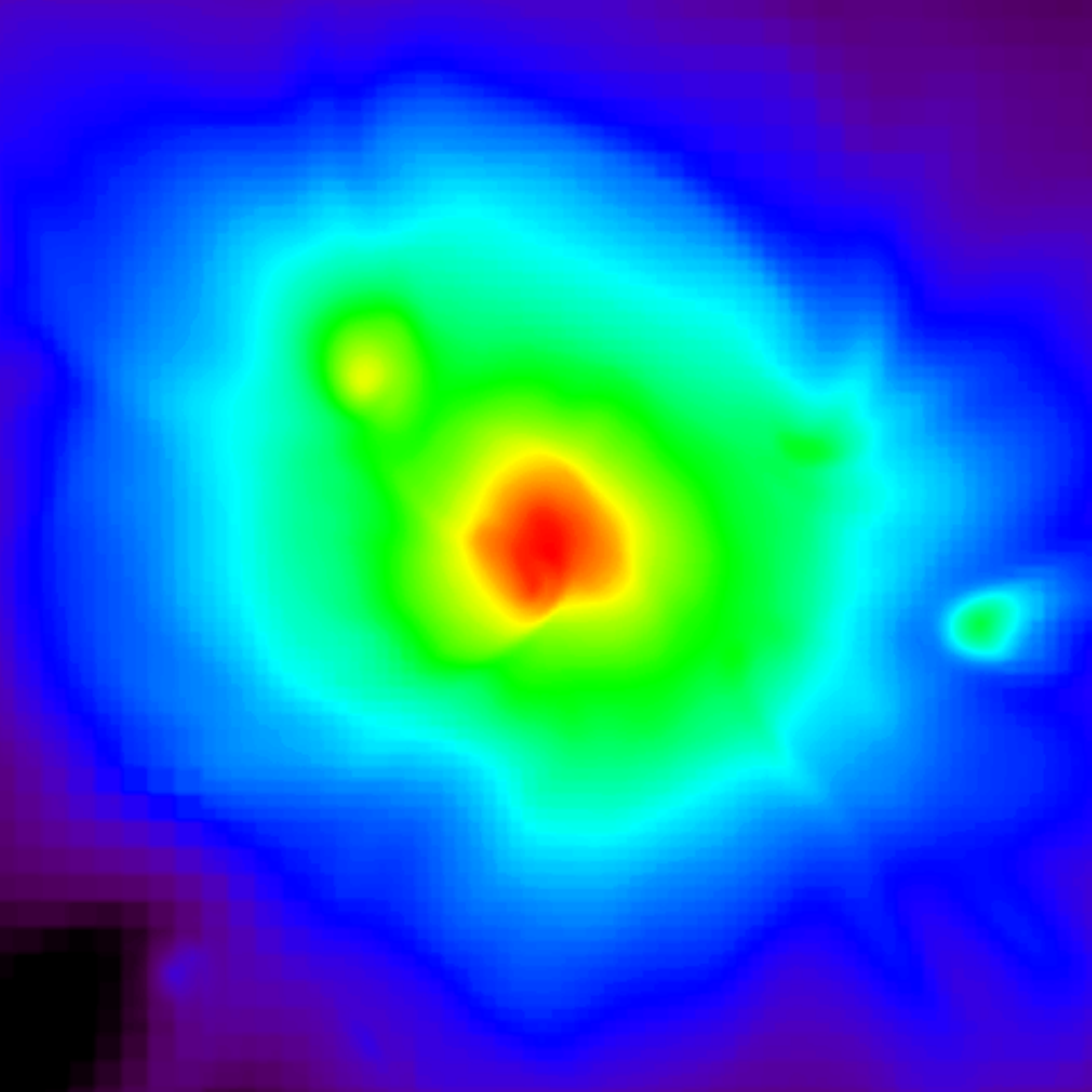}}
     \subfigure{
          \includegraphics[width=.15\textwidth]{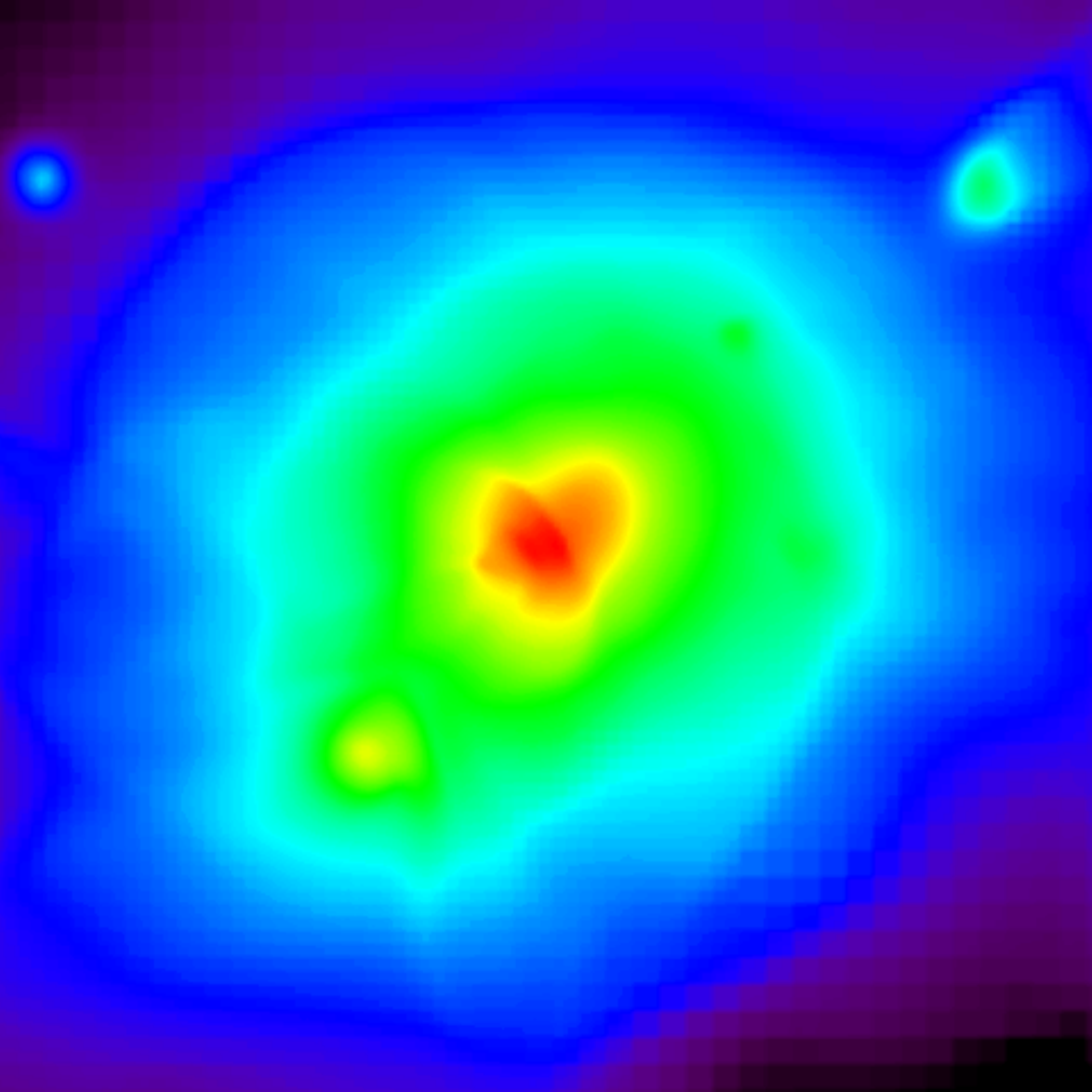}}
     \subfigure{
          \includegraphics[width=.15\textwidth]{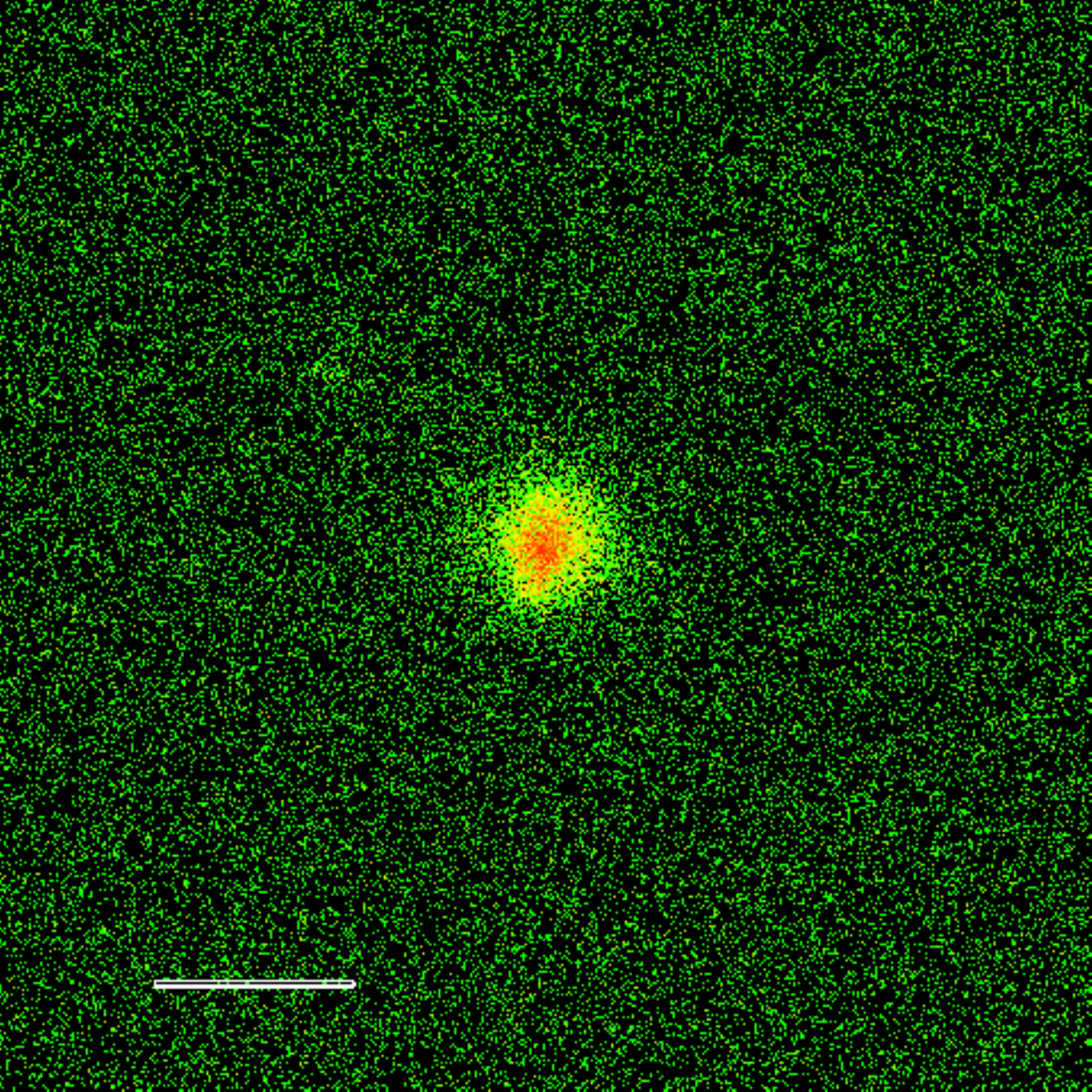}}
     \subfigure{
          \includegraphics[width=.15\textwidth]{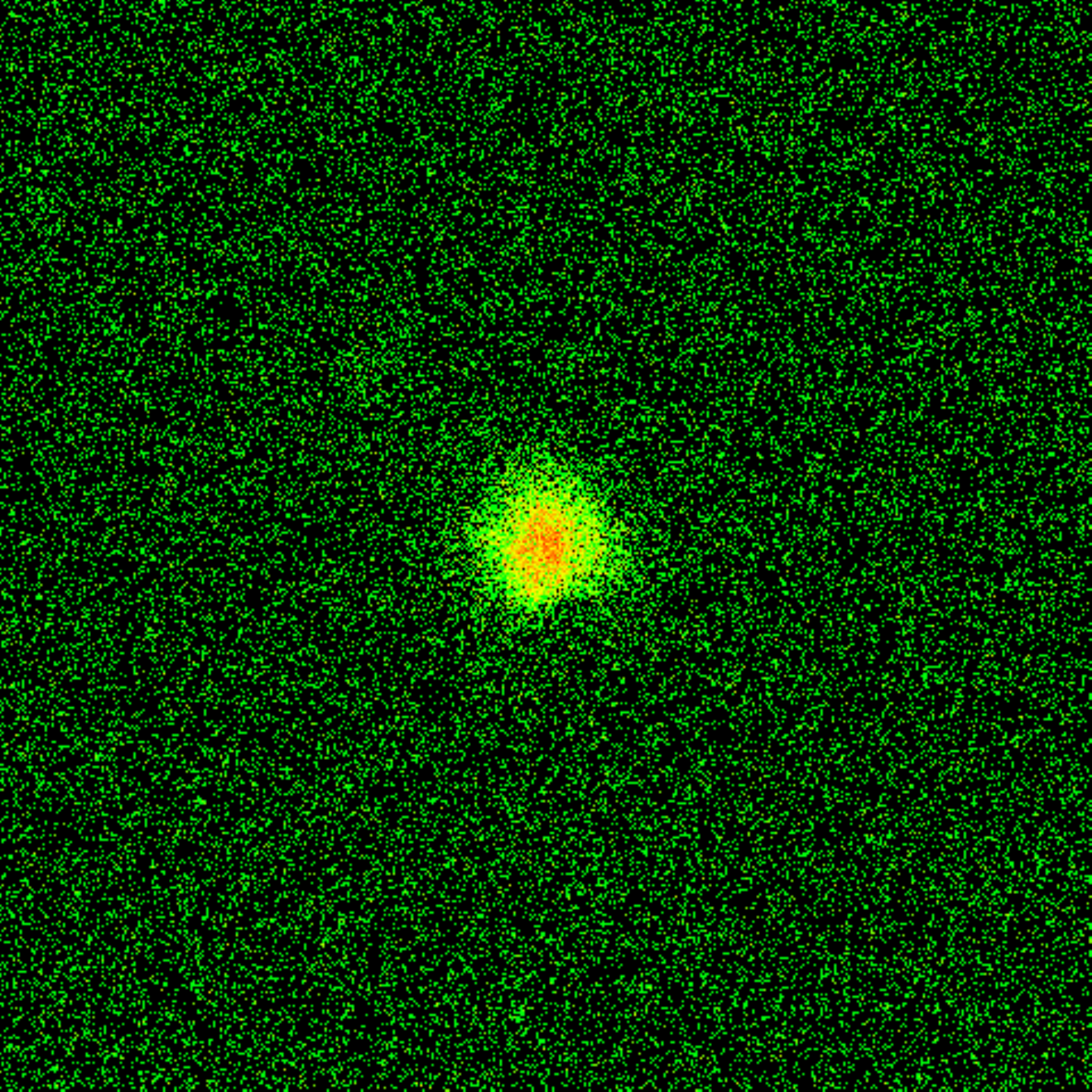}}
     \subfigure{
          \includegraphics[width=.15\textwidth]{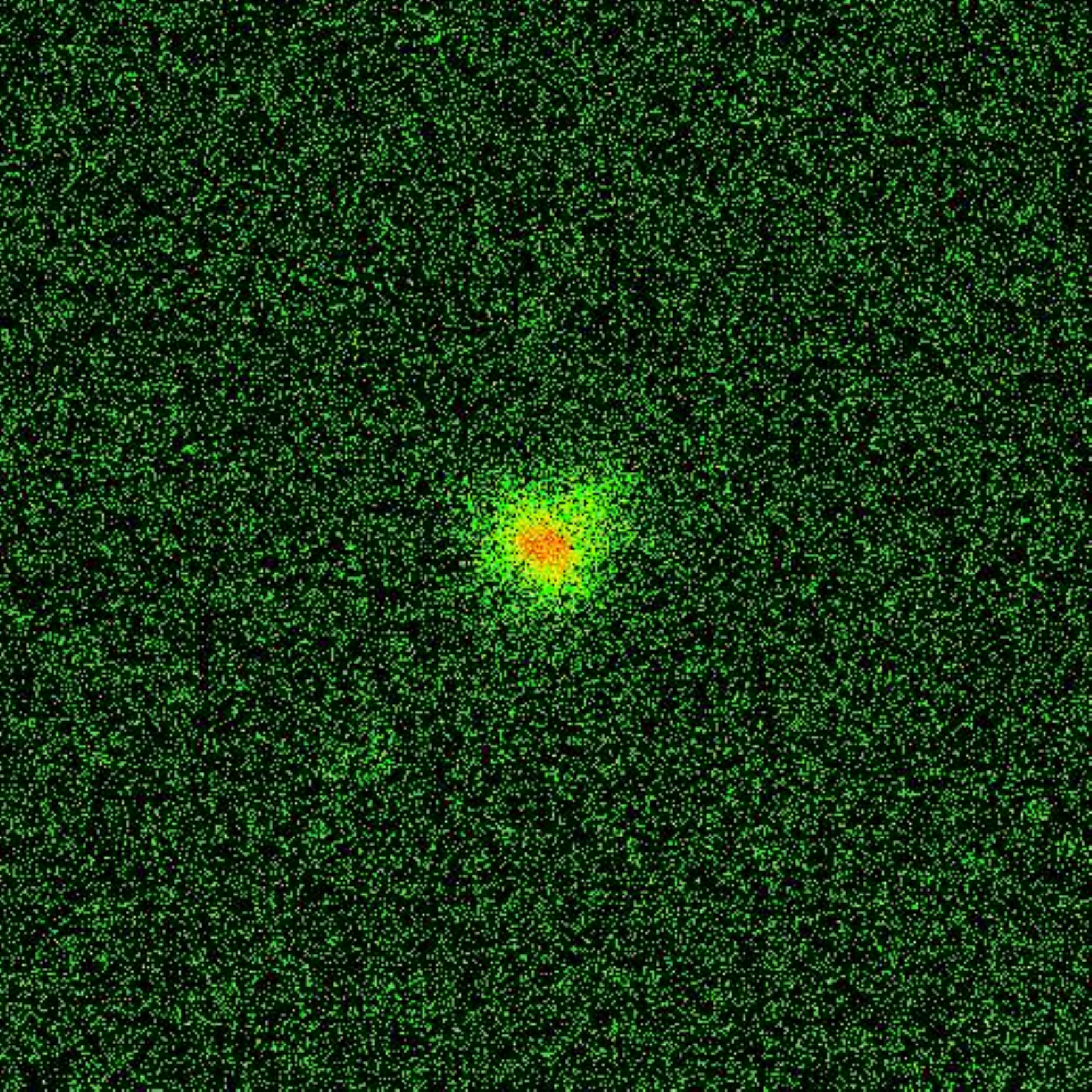}}
\caption{
\label{F:f5}
    Same as Figure~\ref{F:f3} but for simulated images of cluster CL3.
}
\end{figure}

We use only projections of relaxed clusters which seem to show no 
contamination by filaments or distortion due to projection effects 
(Projections X and Y of CL1 and Projection Y of CL2),
and a non-relaxed cluster, CL3, in a most relaxed projection, 
projection X.
We derive simulated visibilities for two-patch observations as

\begin{equation}  \label{E:V2patch}
    V (u_i,v_i) = V_{source} (u_i,v_i) - V_{bkng} (u_i,v_i)
,
\end{equation}

\bigskip
\nop
where the on-source visibilities, \VSOURCE, are

\begin{equation}  \label{E:VSOURCE}
    \VSOURCE (u_i,v_i) = \VCL (u_i,v_i)+ \VCMB (u_i,v_i) + \VNOISE (u_i,v_i)
,
\end{equation}

\bigskip
\nop
and the off-source (background) visibilities, \VBKND, are

\begin{equation}  \label{E:VBKND}
    \VBKND (u_i,v_i) = \VCMBP (u_i,v_i) + \VNOISEP (u_i,v_i)
,
\end{equation}

\bigskip
\nop
where \VSZ\ is the visibility of the SZ signal, 
\VCMB, \VNOISE, and $\VCMBP$, $\VNOISEP$ 
are two sets of visibilities of the CMB fluctuations and noise 
(different for on- and off-source observations).
We assumed that the CMB fluctuation fields for the two patches 
(about one degree apart) are uncorrelated, which is a conservative assumption: 
this way we somewhat overestimate the noise due to the CMB.
Since our models are spherically symmetric, we have
no constraints on them from the imaginary part
of the model visibilities (the imaginary part is identically zero),
therefore we work only with the real part of the visibilities. 
Note that in real applications the imaginary parts can be used 
to check the amplitude of the CMB fluctuations in the field,
assess non-sphericity of the cluster and the pointing accuracy.

As an illustration, in Figure~\ref{F:f7} we show the radial profile 
of the real part, ${\cal R}e(V)$, of the azimuthally averaged simulated visibilities 
of one realization of a two-patch observation for CL1 in Projection~Y
for the 90 and 98~GHz AMiBA channels (Channels A and B, Ho et al. 2009)
as a function of $R_{uv}$.
Since the length scale for the visibility data is in units of the observing wavelength,
the visibilities are represented by two sets of curves. 
The red diamonds and green squares with error bars
represent visibilities for Channel A and B.
The error bars represent the {\it rms} of the azimuthally averaged real part of the
visibilities at the \AMIBAW\ baselines.
Visibilities of other projections of CL1 and projections of CL2 and CL3
with no contamination from filaments are very similar 
due to the structural similarity among the cluster SZ images in these projections
(see Figures~\ref{F:f3}, \ref{F:f4} and \ref{F:f5}).

\smallskip
\section{Model Fitting}
\label{S:AMIBA}

We use non-isothermal double $\beta$ models truncated at the virial radius to describe the ICG.
We determine the best-fit parameters using likelihood functions.
Our model for the SZ surface brightness distributions (Equations~\ref{E:DENS} and \ref{E:TEMP})
has eleven free parameters: ten shape parameters, 
$p = (a_1,r_1,\beta_1,\RCORE,\beta,a_c,r_c,\RT,\delta,\RVIR)$, and one amplitude, $\Delta T_0$.
Unfortunately, due to limited spatial resolution, FOV and receiver noise, 
we do not expect to be able to determine all ten parameters using \AMIBAW.
Thus, similarly to \cite{Mrocet09}, we reduce the number of free parameters in our models. 
We proceed the following way: first we determine the shape parameters for the central part
of the cluster, $a_1$, $r_1$, $\beta_1$, $a_c$, $r_c$, and the core radius for the large scale distribution, 
\RCORE\ from simulated X-ray observations (fixing $\beta=1$, $r_T = 1.0$ \RVIR\ and $\delta = 1.6$ 
based on our results from numerical simulations).
We assume that the X-ray surface density and temperature are determined with acceptable
accuracy only in the central part of the cluster, to 0.5\RVIR\ and 0.2\RVIR.
Therefore \RT\ cannot be determined from fits to the X-ray temperature profile, and 
the X-ray emissivity within 0.5\RVIR\ gives no useful constraints on the large scale
distribution of the temperature.
Having determined the shape parameters for the central part, 
we derive constraints on the temperature scale parameter, \RT,
from simulated \AMIBAW\ visibilities.

%
%
\begin{figure}
\centerline{
\includegraphics[angle=-90,width=.40\textwidth]{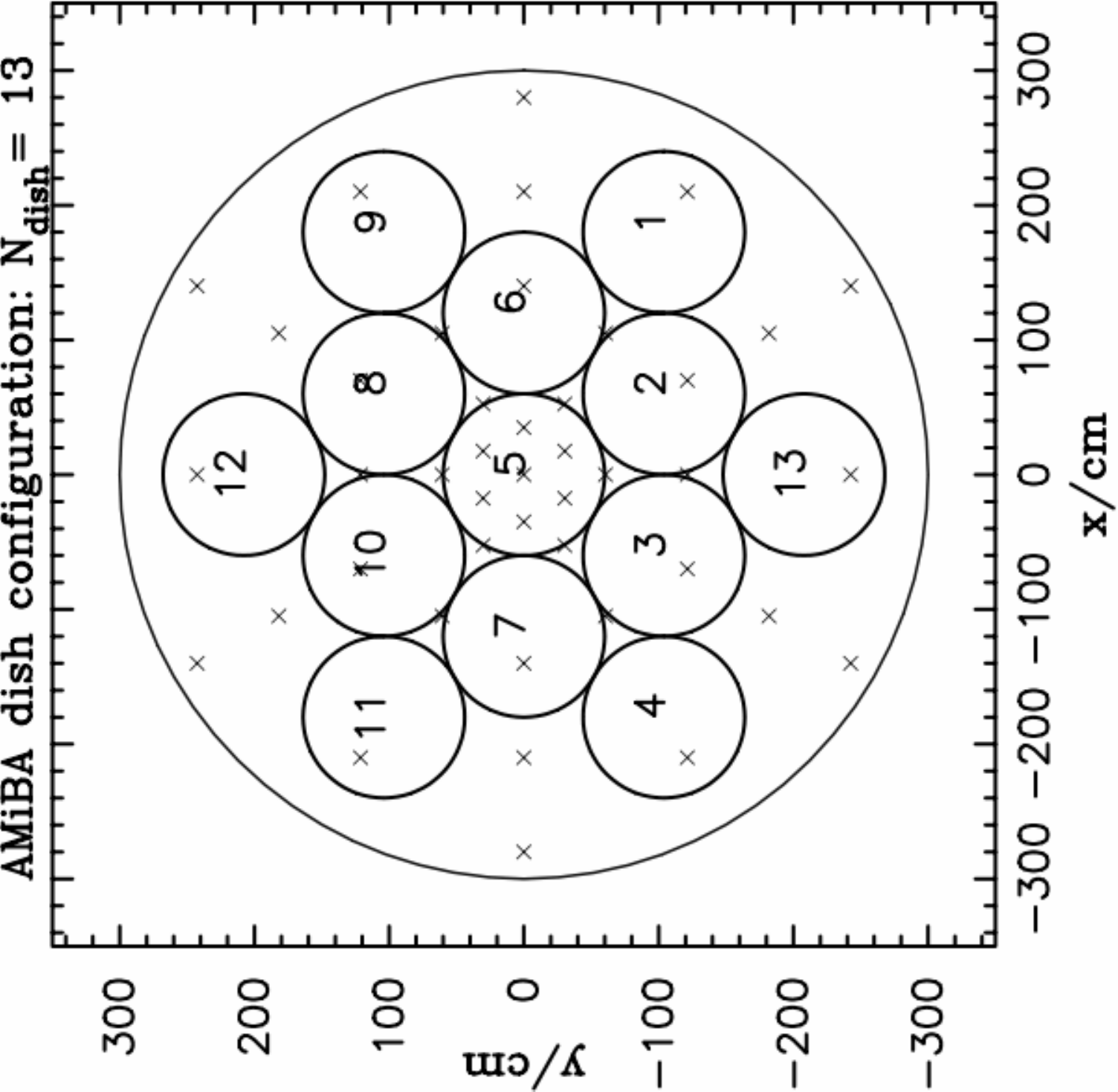}
}
\caption{
 Antenna layout for \AMIBAW\ in the compact configuration
 (13 dishes with 1.2 m diameter) used for simulations in this paper.
 Other possible antenna positions are marked with $\times$-s.
\label{F:f6}
}
\end{figure} 

We calculate the predicted integrated X-ray flux/pixel for our X-ray model as 

\begin{equation}  \label{E:SXMB}
           F_{XM} = F_{CLM} + B_{XM}
,
\end{equation}
where $F_{CLM}$ is our cluster model and $B_{XM}$ is the background.
We calculate $F_{CLM}$ as

\begin{equation}  \label{E:SXM}
  F_{CLM} (x,y, p) = F_1\, N_{X1}\,  {\cal I}_{X1} (x,y, p) +  F_2\, N_{X2}\,  {\cal I}_{X2} (x,y, p)
,
\end{equation}
\nop
where $F_1$ and $F_2$ are the central integrated flux/pixel for the two model components,
the normalizations are 
$N_{X1}^{\;-1} = {\cal I}_{X1} (0,0, p)$ and $N_{X2}^{\;-1} = {\cal I}_{X2} (0,0, p)$, and 

\begin{eqnarray}  \label{E:INTXRAY}
                        {\cal I}_{X1} (x,y,p)   & = &   2\,  \int_0^{\ell_c}  {\cal F}_1^2 \, ( 1 + r/r_T)^{-\delta/2}  d \ell
\nonumber\\     {\cal I}_{X2} (x,y,p)   & = &   2\,  \int_0^{\ell_c}  {\cal F}_2^2 \, ( 1 + r/r_T)^{-\delta/2}  d \ell
,
\end{eqnarray}
where $r^2 = x^2 + y^2 + \ell^2$, and 
the cut off in the LOS is $\ell^2_c = \RVIR^2 - x^2 -y^2$.
We ignore the error in the X-ray background, $B_{XM}$ since 
it is negligible relative to other sources of error, such as,
for example, errors due to azimuthal averaging.

We experimented with the likelihood ratio for
Poisson distribution and Mighell's $\chi_\gamma^2$ statistic 
(see Mighell 1999 for a detailed analysis) in fitting this structure.
We found that the fitted parameters obtained by these two methods are 
virtually identical. Therefore we decided to use the Poisson likelihood ratio test:

\begin{equation} \label{E:LIKEX}
      - \ln {\cal L}_X = \sum_i  M_i - N_i + N_i \, \ln ( N_i/M_i )
,
\end{equation}

\nop
where $N_i$ and $M_i$ are the observed and expected numbers of photons.

We derive spectroscopic-like X-ray temperature profiles, $T_{sp}$, for our simulated
clusters using the weighting scheme of Mazzotta et al (2004), which has been
shown to provide temperature profiles similar to those observed with \CHANDRA\ and XMM
(Nagai et al. 2007),

\begin{equation} \label{E:TSPEC}
       T_{sp} =  { \int w_{sp} \; T dV \over \int w_{sp} \; dV }
,
\end{equation}

\nop
where the weight is $w_{sp} = n^2 / T^{3/4}$.
We determine the best fit parameters for the temperature model
maximizing the likelihood function 

\begin{equation}  \label{E:LIKETEMP}
       -2\,   \ln {\cal L}_{T} =  \sum_{i}  { [ (T_{sp})_i -  (T_M)_i \}]^2  \over \sigma_i^2 }
,
\end{equation}

\nop
where $(T_{sp})_i$ and $(T_M)_i$ are the median values of the observed and model 
spectroscopic temperatures, and $\sigma_i$ is the standard deviation in the $i$th radial bin.

%
%
\begin{figure}[t]
\centerline{
\includegraphics[width=.48\textwidth]{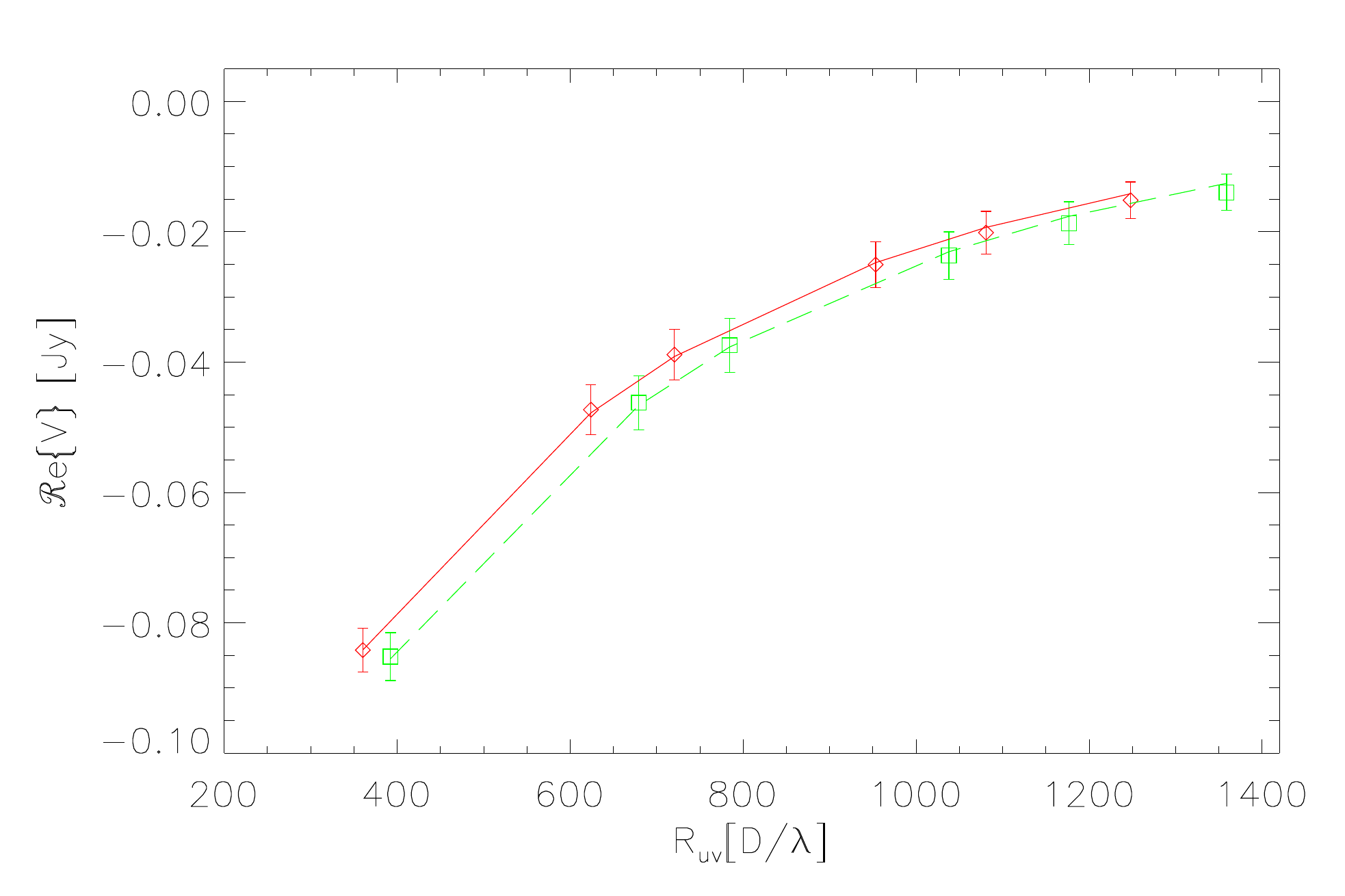}
}
\caption{
         Real part, ${\cal R}e \,\{V\}$ (in Jy), of the azimuthally averaged 
         simulated visibilities as a function of $uv$ radius ($R_{uv}$) in
         Projection Y of the CL1 (plus CMB and noise, one realization) 
         for \AMIBAW\ Channel A and B (red diamonds and green squares)
         in a compact configuration (see Figure~\ref{F:f6}). 
         The error bars represent instrumental errors of 60 hour observations. 
         The best-fit $\beta$ model is also shown for the two channels 
         (red solid and green dashed lines).
\label{F:f7}
}
\end{figure} 

%
%
\begin{figure*}
\centerline{
\includegraphics[width=17.5cm]{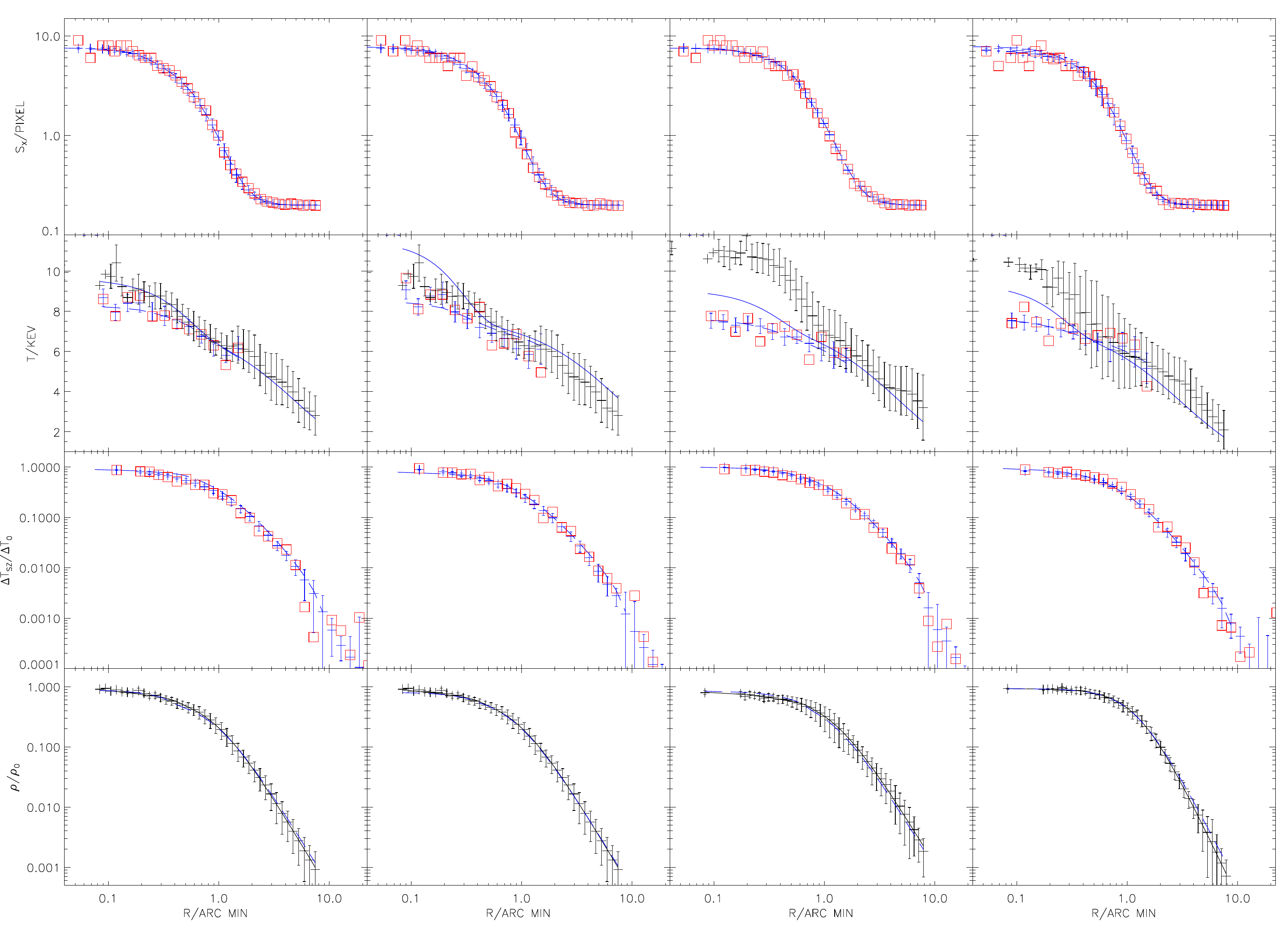}
}
\caption{Radial profiles of X-ray surface brightness, gas temperature, 
              normalized SZ surface brightness and gas density (top to bottom) 
              in projections X and Y of CL1, projection Y of CL2 
              and projection X of CL3 (left to right columns).
              Blue plus signs and error bars: AMR cluster in projection (no noise, no CMB;
              first, second and 3rd rows);
              red squares: simulated data points for one realization including contamination;
              black plus signs and error bars: spherically averaged gas temperature and 
              density distributions in simulated clusters based on their 3D distributions 
              (shown out to the virial radius; second and 4th row);
              blue dashed lines: best fitted models to simulated X-ray and \AMIBAW\ observations
              (first, second and 3rd rows);
              blue solid lines: 3D distribution of the temperature and density 
              in simulated clusters based on fits to the projected (2D) X-ray surface brightness, gas temperature, 
              and SZ surface brightness profiles (second and 4th rows). 
}
\label{F:f8}
\end{figure*}

Since the shape parameters of the inner part of clusters, 
$r_1$, $\beta_1$, \RCORE, $a_c$, and $r_c$, 
are not sensitive to the large-scale distribution of the temperature 
(heavily weighted towards the center of the cluster),
we determine them fixing $\beta = 1$, $r_T = 1$ and $\delta = 1.6$ 
(based on our results from fitting to the density and temperature distributions 
of our simulated clusters).
In our case, since $\beta = 1$, the X-ray emissivity, which is proportional to 
$(1 + r^2 / \RCORE^2)^{-3}$ in the outer parts of the cluster, 
drops about six orders of magnitude from the cluster center to 
the virial radius. Therefore a moderate change in \RVIR\ (say 20\%) causes only an insignificant 
change in the X-ray signal (Equation~\ref{E:INTXRAY}) due to a change in the upper limit, $\ell_c$
(except around \RVIR\, where the X-ray signal is negligible; 
see Figures~\ref{F:f3}, \ref{F:f4} and \ref{F:f5}).
For our purposes, therefore, we fix the value of \RVIR\ assuming that an estimate 
for its value with a 20\% accuracy is available from other measurements.
We determine the shape parameters, $r_1$, $\beta_1$, \RCORE, $a_c$, and $r_c$ 
by maximizing the likelihood functions, Equations~\ref{E:LIKEX} and \ref{E:LIKETEMP}.
In practice, since the spectroscopic-like temperature is not sensitive to a few tens of percent 
change in the $\beta$ model parameters, we can determine the shape parameters using 
iteration. We use the likelihood function for the X-ray emission 
(Equation~\ref{E:LIKEX}) to determine $r_1$, $\beta_1$ and \RCORE, and then the 
likelihood function for the X-ray temperature (Equation~\ref{E:LIKETEMP})
to determine $a_c$, and $r_c$, then we iterate
over these steps (usually only two steps are needed).
This method proved to be faster than a search for the maximum likelihood 
in the five dimensional parameter space.

In Figure~\ref{F:f8} we show the radial profiles of the X-ray surface brightness distribution,
the spectroscopic temperature and the SZ signal for Projections X and Y of CL1, 
Projection Y of CL2, and Projection X of CL3
(blue plus signs with error bars representing dispersion in azimuthal averaging). 
For each projection, we also show one realization of our Monte Carlo simulations (red squares),
and, for comparison, the temperature and density profiles derived from their respective 3D 
distributions (black plus signs with error bars representing dispersion in spherical averaging, 
solid black lines represent best fit models; same as in Figure~\ref{F:f1}). 
We also show the best fit X-ray surface brightness models 
(blue dashed lines) derived as described in the previous paragraph.
The other dashed and solid blue lines, explained later in this section, 
are included here for later convenience. 
Based on the first row in this figure, we conclude that 
our non-isothermal double $\beta$ models provide good fits to the 
X-ray surface brightness profiles in our relaxed clusters (CL1 and CL2), and
even in our cluster with a non-relaxed core (CL3).

The best fit core radii of the outer $\beta$ model from fitting our non-isothermal double $\beta$ 
models to the X-ray surface brightness distributions and temperature profiles (with $\beta = 1$ fixed) 
in different projections are shown in Figure~\ref{F:f2} (squares and triangles).
These values are within the 68\% CL of the best fit values obtained
from fitting double $\beta$ models directly to the 3D distribution of the density 
(solid red lines).
We conclude that the core radius for the large scale distribution can be determined
accurately, with no bias, from X-ray data using our ICG models.

%
%
\begin{figure}
\centerline{
\includegraphics[width=.40\textwidth]{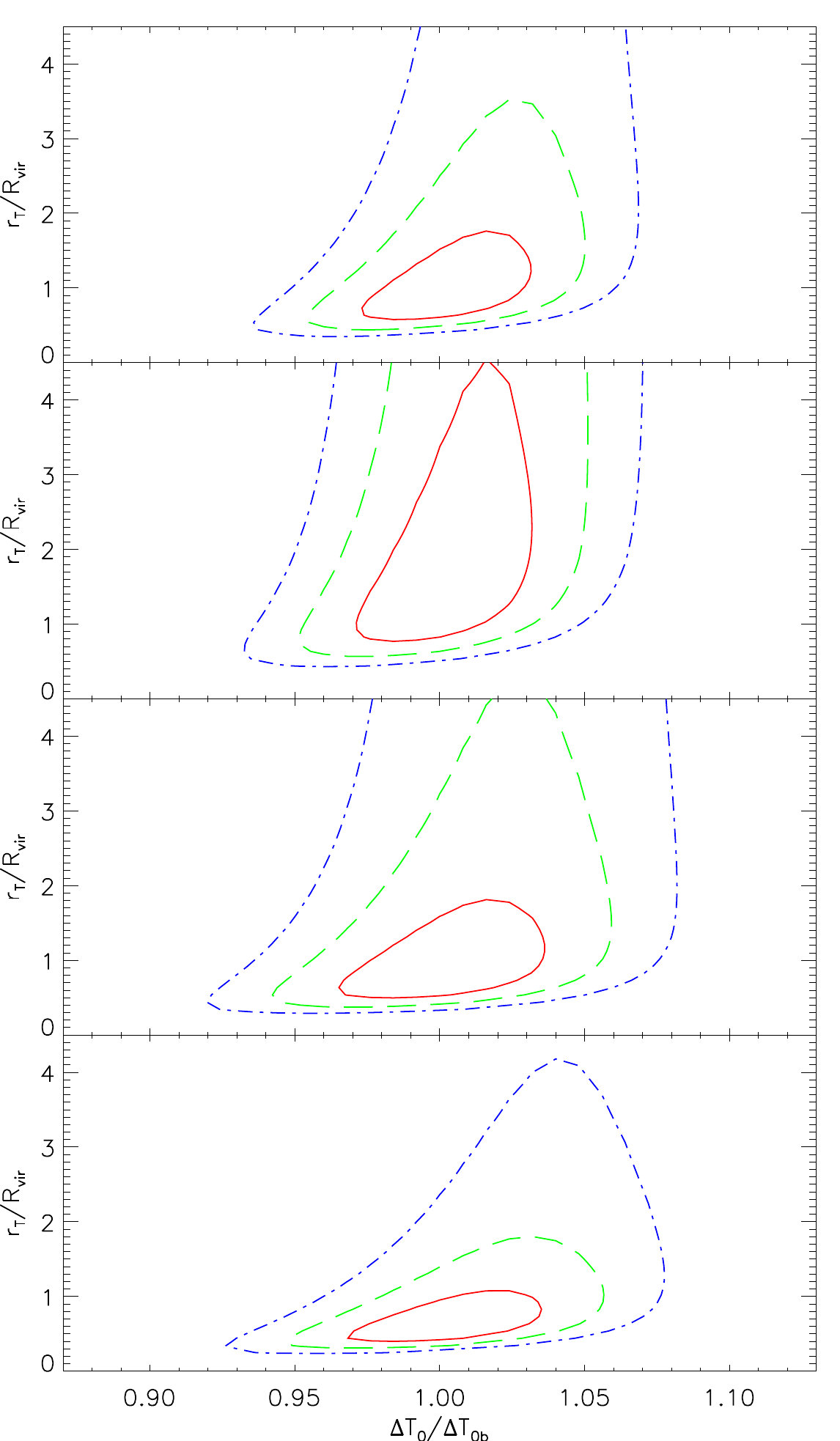}
}
\caption{Likelihood contours (68\%, 95.4\% and 99.7\% CLs, solid red, dashed green, and dash-dotted blue lines)
              for fitting for the normalized SZ amplitude ($\Delta T_0 / \Delta T_{0b}$) and 
              the temperature scale radius $r_T$ 
              in projections X and Y of CL1, projection Y of CL2 and projection X of CL3 (from top to bottom).
\smallskip
}
\label{F:f9}
\end{figure}

As a second and final step, 
we calculate the visibilities for our models, $V_M$, at frequency $\nu$, 
using Equations~\ref{E:VISIB} and \ref{E:DELI}, as

\begin{equation} \label{E:DELTATM}
           \Delta T_{CLM} (x,y) =  \Delta T_0 \, N_{SZ}\, {\cal I}_{SZ} \,(x,y, p)
,
\end{equation}

\bigskip
\nop
where $\Delta T_0$ is the central SZ amplitude, the normalization is 
$N^{\,-1}_{SZ} = {\cal I}_{SZ} (0,0, p)$, 
and

\begin{equation} \label{E:INTSZ}
   {\cal I}_{SZ} (x,y, p) = 2\, \int_0^{\ell_c} {   \rho_1 \, {\cal F}_1  + \rho_2 \, {\cal F}_2 \over \rho_1 + \rho_2} \, ( 1 + r/r_T)^{-\delta} d \ell
.
\end{equation}

\nop
Note, that since $r/r_{\rm core}$ and $r/r_T$ are both dimensionless, 
$\Delta T_{CLM} (x,y)$ depends on \RVIR\ only through $\ell_c$.
We use Equation~\ref{E:V2patch} to derive simulated observed visibilities, $V_O$, 
for each realization as described in Section~\S\ref{S:AMIBAVIS}.

We determine the best fit parameters for our models by maximizing the SZ likelihood function 
defined as

\begin{equation}  \label{E:LIKESZ}
       -2\,   \ln {\cal L}_{SZ} =  \sum_{i,j}  { [ { \cal R}e \{ V_O(R_{uv}^{\,i};\nu_j)\}  -  
                                                                         { \cal R}e \{ V_M(R_{uv}^{\,i};\nu_j) \}]^2 \over \sigma_{ij}^2 }
,
\end{equation}

\bigskip
\nop
where ${\cal R}e \{V_O \}$ and ${\cal R}e \{V_M\}$ are the azimuthally averaged real part of 
the observed and model visibilities at the $i$th $uv$ radius, $R_{uv}^{\,i}$, at frequency $\nu_j$, 
where $j=1,2$ for the two \AMIBA\ frequency channels, and $\sigma_{ij}$ is the 
Gaussian noise, which is assumed to be the same for all antenna pairs and frequencies.

Similarly to the X-ray signal, a moderate change in \RVIR\ 
(all the other parameters fixed) causes only an insignificant change in the SZ signal 
(Equation~\ref{E:INTSZ}) because the pressure is about four orders of magnitude 
smaller at the virial radius than at the center of the cluster. 
In this final step, we assumed that we determined the model parameters 
for the central region of clusters ($r_1$, $\beta_1$,  $a_c$, $r_c$) and \RCORE\ 
from fitting our models to the X-ray surface brightness and temperature 
distribution (as above), therefore, among the shape parameters 
of our non-isothermal double $\beta$ models,  
$p = (r_1, \beta_1, \RCORE, 1, a_c, r_c, \RT,1.6,\RVIR)$, 
there is only one unknown parameter, \RT.
We determine the best fit for \RT\ and the normalized SZ amplitude ($\Delta T_{SZ} / \Delta T_0$) 
by maximizing the SZ likelihood function, Equation~\ref{E:LIKESZ}.

In Figure~\ref{F:f8} we show the 2D projected temperature and SZ profiles for the best fits
(dashed blue lines, 2nd and 3rd row). Solid blue lines represent the 3D best fit radial profiles 
for the temperature and density (2nd and 4th row). 
In some cases the deprojected temperature profiles are underestimated by 10--20\% 
close to the cluster center because cold substructures in the LOS reduce the 
spectroscopic temperature and the weighting suggested by Mazzotta et al (2004) does 
not always correct for this effect accurately.
However, these deviations in the central temperature do not affect the fits to the 
large scale SZ profiles as observed by \AMIBAW\ because of the low resolution of the 
instrument.

In Figure~\ref{F:f9} we show the CLs for
the temperature scale radius $r_T$ and the normalized SZ amplitude
in different projections of X and Y of CL1, Projection Y of CL2 and Projection X of CL3.
The normalization, $\Delta T_0 / \Delta T_{0b}$, is the ratio of the best-fit $\Delta T_0$ 
for each Monte Carlo simulation to the best-fit value determined from all Monte Carlo simulations.
The confidence levels have been determined using Monte Carlo simulations. 
The contours for the 68\% CL, for example, are the smoothed version of the 
contours containing 68\% of the simulated best fit points based on simulations.

\bigskip
\bigskip
\bigskip
\section{Discussion}
\label{S:Discussion}

We have simulated \AMIBAW\ observations of massive relaxed clusters of galaxies 
including CMB contamination and receiver noise using clusters drawn from 
cosmological numerical simulations to assess how well we should be able to constrain the
large-scale distribution of the ICG.
Our simulated SZ images (rows 2 in Figures~\ref{F:f3}, \ref{F:f4} and \ref{F:f5}) show that
at 94 GHz, on the scale of the cluster cores
(few arc minutes), the cluster SZ signal dominates the CMB fluctuations, so
  that CMB contamination is not important in surveys searching for rich clusters. 
  On a scale of ten arc minutes, corresponding to the extent of the ICG in massive clusters
  at $z \approx 0.3$, the CMB contamination is at a comparable level to the SZ signal. 
  Contamination from CMB fluctuations is also important in the regions in galaxy clusters close 
  to their virial radius. 
 Spectral separation of the cluster SZ and the CMB signals based on multi-frequency observations
 seems to be essential for studying the outskirts of galaxy clusters and the 
 SZ signature of the large-scale structure.

Using our AMR simulations, we showed that a spherical non-isothermal double $\beta$ 
model with a temperature distribution described by Equation~\ref{E:TEMP}
provides good fits to the radial distributions of the ICG in our selected
massive clusters. 
We generated X-ray and SZ images of our clusters drawn from numerical simulations
assuming that the clusters are at a redshift of 0.3.

We used the simulated X-ray data to determine the shape parameters, $r_1$, $\beta_1$, 
\RCORE, of the double $\beta$ model for the density distribution 
and the central model parameters, $a_c$ and $r_c$ of the temperature model
by minimizing the likelihood functions Equations~\ref{E:LIKEX} and \ref{E:LIKETEMP}.
We assumed that \RVIR\ is constrained to $\pm 20$\% by other measurements,
for example, from gravitational lensing, as done by Broadhurst and Barkana (2008) 
and  Umetsu et al. (2009). This is a conservative estimate since combining weak and 
strong lensing data, one can determine \RVIR\ with about 5\% accuracy \citep{Umetet2010ApJ714}.
The values of the likelihood functions corresponding to the 
68\%, 95.4\% and 99.7\% CLs for fitting for \RCORE\ and $\beta$ to the 3D density
distribution of simulated clusters are shown in Figure~\ref{F:f2}.
The elongated shape of the confidence levels in this figure shows a degeneracy 
between the two shape parameters of the $\beta$ model: \RCORE\ and $\beta$. 
 Combinations of small ($r_{core},\beta$) or large ($r_{core},\beta$) both give good fits
 to the density distributions.
A similar degeneracy has been reported for fitting $\beta$ models to X-ray observations
(e.g., \citealt{Greget01,Reeset00,Birket94}).

As a final step, we used simulated \AMIBAW\ visibilities to determine the temperature scale radius,
\RT, and the SZ normalization (fixing all other parameters)
by maximizing the SZ likelihood function, ${\cal L}_{SZ}$ (Equation~\ref{E:LIKESZ}).
Our results are shown in Figure~\ref{F:f9}.
For model fits to Projections X and Y of CL1, 
we obtain $r_T = 1.023_{-0.4}^{+0.5}$ \RVIR, and $r_T = 1.60_{-0.5}^{+4.0}$ \RVIR.
Fits to Projection Y of CL2 yield $r_T = 0.927_{-0.4}^{+0.5}$ \RVIR.
We obtain $r_T = 0.631_{-0.2}^{+0.3}$ \RVIR\ for our cluster with
a non-relaxed core, CL3, in Projection X.

All results for the temperature scale radius, \RT, are within 68\% of the
best-fit values based on fitting the 3D distributions (Table~\ref{tab:Table1}).
The 68\% CLs for all clusters with circular projected
X-ray distribution (Projection X of CL1 and Projection Y of CL2)
and the cluster with a non-relaxed core (Projection X of CL3)
are within 50\% of the best fit values. 
Therefore our results suggest that, using relaxed clusters with
circular morphology, we should be able to use \AMIBAW\ along with X-ray
observations to obtain unbiased parameters for our 
non-isothermal double $\beta$ models
(even for clusters with a non-relaxed core, such as CL3), 
and constrain \RT\ within 50\%. 
Our model fits to Projection Y of CL1, a relaxed cluster with
elliptical morphology, although returning \RT\ within 68\% of the
3D fitted value, shows a large error due to asphericity.
We expect that using a more accurate elliptical model would 
result in better constraints on \RT.
The SZ amplitude, which is important for the determination of the Hubble
constant for example, can be determined with 3-4\%, 
which is better than the expected accuracy of the absolute calibration of \AMIBAW.

We carried out simulations at different redshifts between 0.1 and 0.4.
For a distant cluster the beam dilution reduces the signal;
if the cluster is too close, the outer parts of the cluster fall outside of the FOV. 
We have found that the optimal redshift for determining the large-scale
distribution of the ICG with \AMIBAW\ is $z \approx 0.3$.

We conclude that we should be able to use \AMIBAW\ to determine the
large-scale distribution of the ICG in massive relaxed clusters of galaxies
located at a redshift of 0.3 by determining the temperature
scale radius with an about 50\% statistical accuracy.
\AMIBA, as upgraded to 13 dishes with 1.2-m in diameter, 
will be a powerful tool for constraining the large-scale 
distribution of the ICG.

The degeneracy between density and X-ray temperature can be
broken and determined out to the virial radius using current
X-ray telescopes. However, an accurate determination of the temperature
profile out to \RVIR\ requires a long exposure time (due to the low photon 
count rates at the outer parts of clusters) and mosaicing (due to the limited FOV).
This is the reason why only a few attempts have been carried out
to map the outer regions of clusters (see \S\ref{S:LargeScale}).
Constraints on the large scale distribution of the ICG similar to AMIBA
can be derived using bolometer cameras.
In principle, bolometer cameras with arc minute resolution can cover clusters
with the necessary sensitivity out to \RVIR\ (ACT: Hinks et al. 2009; SPT: Plagge et al. 2009),
although this has not yet been done.

In this paper we considered spherically symmetric galaxy cluster models, and
focused on statistical errors due to the \AMIBAW\ telescope and receiver system.
We should be able to reduce the observation time needed to measure \RT\ by 
using a more sophisticated observational strategy. 
A single pointing to the center of the cluster, although it simplifies the data analysis, 
has the disadvantage that it has a reduced sensitivity at the outer parts of the cluster, 
where the signal is weaker. 
We expect that mosaic observations including pointings towards the outer regions in clusters,
although more difficult to analyze, would enable us to reach our goal
with an exposure time shorter than 60 hrs.
We leave simulations to quantify the effects of mosaic observations and
more sophisticated ICG models on parameter determination for future work,
as well as a detailed study of parameter estimation from a large set of simulated
relaxed clusters with a wider range of mass and redshift.

\acknowledgements
We thank the anonymous referee for constructive comments which helped to improve our 
paper substantially.
Our special thanks go the the \AMIBA\ team for their many years of dedicated 
work which made this project possible.
KU is partially supported by the National Science Council of Taiwan under the
grant NSC97-2112-M-001-020-MY3.
MB acknowledges support from the STFC.
GB acknowledges support from NSF grants AST-05-07161, 
AST-05-47823 and supercomputing resources from the National Center for 
Supercomputing Applications.
ZH was supported by the NSF grant AST-05-07161 and by the Pol\'anyi Program 
of the Hungarian National Office for Research and Technology (NKTH).
NH acknowledges support from the ASC Academic Alliances Flash Center 
at the University of Chicago, which is supported in part by the U.S. Department of Energy, 
contract B523820.

%
%
\bibliographystyle{apj}

 
\end{document}